\documentclass[twocolumn,times,tighten]{aastex63}
\usepackage[utf8]{inputenc}
\usepackage{enumerate}
\usepackage{amsmath, hyperref, enumitem, graphicx, amssymb}
\usepackage[caption=false]{subfig}

\newcommand{\sgg}[1]{\textcolor{green}{[{\bf SG}: #1]}}

\begin{document}

\title{Reorientation Rates of Structural and Kinematic Axes in Simulated Massive Galaxies \\ and the Origins of Prolate Rotation}

\author[0000-0002-9370-8061]{Sahil Hegde}
\affiliation{Department of Physics, Columbia University, 550 West 120th Street, New York, NY 10027, USA}
\affiliation{Columbia Astrophysics Laboratory, Columbia University, 550 West 120th Street, New York, NY 10027, USA}
\affiliation{Department of Physics \& Astronomy, University of California, Los Angeles, 475 Portola Plaza, Los Angeles, CA 90095, USA}

\author[0000-0003-2630-9228]{Greg L. Bryan}
\affiliation{Department of Astronomy, Columbia University, 550 West 120th Street, New York, NY 10027, USA}
\affiliation{Center for Computational Astrophysics, Flatiron Institute, 162 Fifth Avenue, New York, NY 10010, USA}

\author[0000-0002-3185-1540]{Shy Genel}
\affiliation{Center for Computational Astrophysics, Flatiron Institute, 162 Fifth Avenue, New York, NY 10010, USA}
\affiliation{Columbia Astrophysics Laboratory, Columbia University, 550 West 120th Street, New York, NY 10027, USA}

\begin{abstract}
In this work, we analyze a sample of $\sim$4000 massive ($M_*\geq 10^{11} M_\odot$ at $z=0$) galaxies in TNG300, the $(300 \mathrm{Mpc})^3$ box of the IllustrisTNG simulation suite. We characterize the shape and kinematics of these galaxies with a focus on the kinematic misalignment ($\Psi_\mathrm{int}$) between the angular momentum (AM) and morphological major axis. We find that the traditional purely shape- or kinematics-based classifications are insufficient to characterize the diversity of our sample and define a new set of classes based on the rates of change of the galaxies' morphological and kinematic axes. We show that these classes are mostly stable over time and correspond to six distinct populations of galaxies: the rapid AM reorienters (58\% of our sample), unsettled galaxies (10\%), spinning disks (10\%), twirling cigars (16\%), misaligned slow reorienters (3\%), and regular prolate rotators (galaxies that display major axis rotation; 2\%). We demonstrate that the most-recent significant (mass-ratio $\mu>1/10$) mergers of these galaxies are the primary cause for their present-day properties and find that these mergers are best characterized at the point of the satellite's final infall – that is, much closer to the final coalescence than has been previously thought. We show that regular prolate rotators evolve from spinning disk progenitors that experience a radial merger along their internal AM direction. Finally, we argue that these regular prolate rotators are distinct from the similarly-sized population of rapid AM reorienters with large $\Psi_\mathrm{int}$, implying that a large $\Psi_\mathrm{int}$ is not a sufficient condition for major axis rotation.
\end{abstract}


\section{Introduction}
Studies of both real and simulated early type galaxies (ETGs) – bulge-dominated, older systems – have been used to develop an understanding of their kinematic and morphological properties. These in turn yield insight into the formation processes of these galaxies, such as their merger history (e.g.~\citealp{EL2017}, \citealp{BF2019}, \citealp{Lagos2020}, \citealp{Dolfi2021}).

Prior to the advent of integral field spectroscopy (IFS), early work focused on describing galactic structure via photometry, such as through the parametrization of global intensity profiles \citep{Sersic1968} or the classification of galaxies by isophotal shape (e.g.~\citealp{Bender1989}, \citealp{Kormendy1996}). Ultimately, this yielded two broad classes of ETGs: the giant ellipticals and the normal-luminosity ellipticals (\citealp{Kormendy1996}, \citealp{Faber1997}, \citealp{Ferrarese2006}). 

The turn of the 21st century brought with it advancements in instrument data quality, specifically with the \texttt{SAURON} (\citealp{Bacon2001}, \citealp{deZeeuw2002}) IFS survey, which enabled analysis of two-dimensional maps of stellar kinematics. From this, the field of kinemetry, or the generalization of photometry to stellar kinematic maps, was born (\citealp{Krajinovic2006}, \citealp{Krajinovic2008}, \citealp{Krajinovic2011}), and, shortly afterwards, this was followed up by the ATLAS$^\mathrm{3D}$ project (\citealp{Cappellari2011}). Measurements of correlations between quantities, such as the spin parameter $\lambda_R$ (\citealp{Emsellem2007}, \citealp{Emsellem2011}) and the ellipticity $\varepsilon$, led to a kinematic distinction between \textit{fast} and \textit{slow} rotating ETGs (\citealp{Emsellem2011}, \citealp{Graham2018}). These and further IFS surveys, e.g.~MASSIVE (\citealp{Greene2015}) and MaNGA (\citealp{Wilkinson2015}), have greatly expanded our ability to study ETGs and have produced evidence of a wide range of kinematic characteristics beyond the simple fast-slow rotator dichotomy, such as the presence of a kinematically decoupled core (KDC), counter-rotating component, or misalignment of the photometric and kinematic axes (e.g.~\citealp{Krajinovic2011}).

Based in large part on work using cosmological simulations, it is now believed that fast and slow rotators evolve broadly following a two-phase evolutionary process (e.g.~\citealp{Oser2010}). These galaxies are born in rapidly-growing dark matter (DM) halos and experience a strong starburst in the initial formation phase, followed by hierarchical assembly through mergers, wherein the details of the merger (e.g.~mass ratio, gas fractions, etc.) dictate the subsequent formation pathways of the two main kinematic classes (\citealp{Naab2014}, \citealp{Penoyre2017}, \citealp{Pulsoni2020}). With this in mind, analysis of simulated ETGs has been used to demonstrate that these are layered structures that have kinematic and morphological properties that change with radius (e.g.~one can identify a kinematic `twist' radius; see \citealp{Pulsoni2020}). In addition, it is believed that the aforementioned kinematic anomalies are signatures of mergers with peculiar properties (e.g.~\citealp{barreraballesteros2015}, \citealp{Ebrova2015}, \citealp{Tsatsi2017}, \citealp{Nevin2021}).

Of special interest are classes of galaxies derived from measurements of the intrinsic kinematic misalignment ($\Psi_\mathrm{int}$; the complement to $90^\circ$ of the angle between the angular momentum (AM) and major axis) and these have been studied in a variety of contexts. While diskier/more oblate galaxies (which have near orthogonal AM and major axes; $\Psi_\mathrm{int}\sim 0^\circ$) are physically better understood (see e.g.~\citealp{BF2019}, \citealp{Zeng2021}), describing the phenomenon known as prolate rotation (alignment of the AM and major axes in elongated, prolate systems; $\Psi_\mathrm{int}\sim 90^\circ$) is more complex. Prolate shapes are more prevalent at higher stellar masses (e.g.~\citealp{Pulsoni2020}) and, in kind, the latter kinematic phenomenon is increasingly more common as IFS observations of ETGs extend to higher stellar masses (e.g.~\citealp{Tsatsi2017}, \citealp{Krajinovic2018}). In fact, the utility of the fast-slow rotator dichotomy begins to break down at higher masses, wherein the majority of observed systems are slowly rotating with prolate shapes (e.g.~\citealp{Cappellari2016}, \citealp{Lagos2020}) and further work is needed to better understand the nuance of the joint shape and kinematic distribution. In addition, not only are the most massive galaxies the most clearly resolved systems, making comparison samples more readily available, but also with our picture of hierarchical merger-driven growth, they experience the most significant mergers. Therefore, developing an understanding of these systems lends insight into galaxy evolution on the largest scales. Taking these broad components – the utility of kinematic anomalies as tell-tale signs of merger episodes, the knowledge that massive galaxies experience complex merger histories, and the observational relevance of the most massive galaxies – in conjunction, the need for further investigation into the origins of the diverse range of observed behaviors is clear. 

Given the limited observational evidence currently available – a matter of small samples – to constrain the nature of such systems, especially at masses $M_*>10^{11} M_\odot$, and with the utility of simulations to directly trace galaxy evolution over cosmic time, in this paper, we explore the distribution of kinematic and morphological properties of the most massive \textit{simulated} galaxies. We begin by analyzing the galaxies as has been done in the literature and testing the utility of traditional shape-based classifications of ETGs at the high-mass end of the spectrum. We demonstrate that a simple shape division is not sufficient to capture the true diversity of physical behaviors exhibited by these systems and propose a new classification system that leverages the rate-of-change of the kinematic and morphological axes associated with the galaxies to better describe their properties. We evaluate the stability and robustness of this classification over cosmic time and trace back the merger trees of our sample to identify properties of the mergers that correspond to different observed $z=0$ behaviors. Finally, we isolate a subset of our galaxies – the prolate rotators – and argue that dynamical and evolutionary properties can be used to distinguish them from galaxies that are simply instantaneously misaligned. We note that while the results of our time-averaged analysis are not directly observable quantities, the focus of this study is on identifying the physical source of observed galaxy behaviors and, in future work, these results can be extended to connect with direct, instantaneously measured observables. 

This paper is organized as follows. In Section \ref{sec:methods}, we describe our sample selection and methodology of characterizing the galaxies, summarizing the results of a shape-based analysis. Section~\ref{sec:new_classes} introduces the new classification system, which is done by first providing sample galaxies chosen to illustrate the behaviors representative of each class (Section~\ref{ssec:class_exemplars}), before going on to formally define the classes (Section~\ref{ssec:def_classes}), characterize the distribution of galaxies in the classes (section~\ref{ssec:characterizingclasses}), and then finally to examine class stability over cosmological time (Section~\ref{sec:class_stability}). Then, in Section~\ref{sec:merg_history}, we tackle the origins of the various classes. To do this, we first demonstrate in Section~\ref{ssec:merg_overview} that mergers are the primary driver of class change, before going on to demonstrate that it is not until the final phase of the merger that the class membership of the descendant is decided (Section~\ref{ssec:final_infall}).  In the discussion section, we briefly explore the origin of the angular momentum reorientation seen in our galaxies (Section~\ref{ssec:tidaltorques}), arguing that mass growth, rather than external torques, plays a dominant role, while in Section~\ref{sec:misaligned_comp}, we analyze properties of misaligned galaxies and identify prolate rotation as a distinct subclass of the misaligned population. In Section~\ref{sec:prev_work} we discuss connections to previous works and finally, Section~\ref{sec:conclusions} summarizes the conclusions we draw from this analysis.

\section{Methods}\label{sec:methods}
Our methodology can be summarized as as follows: first, we select a sample of simulated, high-mass galaxies for our analysis; then, we characterize this sample in the present-day universe ($z=0$); next, we develop a new classification system that offers a distinct perspective on understanding the behavior of these galaxies over time; and finally, we trace back the merger trees for these systems and analyze their merger histories.

\subsection{Selecting the Sample}
For the purposes of this investigation, we sought to analyze the \textit{most massive} galaxies in the (simulated) universe. Our simulated galaxies come from the `Illustris: The Next Generation' cosmological simulation suite (hereafter referred to as TNG; \citealp{TNG1}, \citealp{TNG2}, \citealp{TNG3}, \citealp{TNG4}, \citealp{TNG5}). TNG is a suite of hydrodynamical galaxy formation simulations that use the moving mesh code \texttt{AREPO} (\citealp{Springel2010}) in three cosmological volumes, each of which highlights a different scale of galactic physics. Building from the original `Illustris' simulation, the sample produced by the effective galaxy formation model in TNG has been shown to agree satisfactorily with observed relations of size and mass (\citealp{Genel2018}) and reflects a reasonable distribution of morphological types (\citealp{RodriguezGomez2019}). 

The most massive galaxies experience complex growth histories and are often found in dense cluster environments. Therefore, the largest volume realization of the TNG simulations, the $300\ \mathrm{Mpc}^3$ box (denoted TNG300), which captures the evolution of thousands of massive galaxies in a variety of environments, including clusters, is the best suited for such analysis. We select galaxies from the final snapshot of TNG300, corresponding to $z=0$, using a single selection cut – the stellar mass – to select all galaxies that have $\mathbf{M_* \geq 10^{11} M_\odot}$. With an approximate mass resolution of $10^7 M_\odot$ for baryonic particles, this roughly corresponds to a lower bound of $\gtrsim 10^{4}$ particles. This selection criterion focuses our analysis on a relatively understudied end of the mass spectrum – in favor of studying smaller structures at higher resolutions, most previous works have limited their analysis to galaxies between $10^{10}-10^{11}M_\odot$ (see e.g.~\citealp{EL2017}, \citealp{BF2019}, \citealp{Pulsoni2020}). In the highest resolution  TNG300 box (TNG300-1), we have a sample of 3890 galaxies with the aforementioned selection criterion. 

\subsection{Characterizing the Sample}\label{ssec:characterizingsample}
Following the approach of \citet{BF2019} and \citet{Pulsoni2020}, we characterize the shapes and kinematics of the galaxies in our sample assuming that the stellar particle distributions are reasonably well modeled by a three-dimensional ellipsoid. With the effects of interactions and other galactic evolution processes, asymmetries in the stellar distributions reflect deviations from this simple picture at large distances from the galactic center and measured morphological quantities can be quite sensitive to such variations. Additionally, it becomes observationally difficult to resolve stellar distributions at large radii. Therefore, we limit our analysis to the stellar particles within the ellipsoidal half-mass radius of the galaxy, $R_{1/2}$, where the distribution is relatively symmetric and can be more readily observed. Unless explicitly stated otherwise, every quantity we discuss has been measured within $R_{1/2}$. Then, the first step is to compute the three dimensional shape of these stellar distributions for each galaxy in our sample, and to do this, we follow the iterative algorithm in the literature (\citealp{BF2019}), which can be broken down as follows: 
\begin{enumerate}
    \item Beginning with an initial guess at axis ratios of $p = q = 1$ (spherical) for the shape (only for initialization), we compute the half-mass radius for the galaxy, $R_{1/2, \mathrm{spherical}}$.
    \item For the particles within $R_{1/2}$, we compute the reduced inertia tensor given by:
    \begin{equation}\label{eq:inertia}
    I_{ij} = \frac{\sum_n \frac{m_n x_{n, i} x_{n,j}}{r_n^2}}{\sum_n m_n}
    \end{equation}
    where $m_n$ is the mass, $x_{n,i}$ is the $i$th component of the position, and $r_n = \sqrt{x_n^2 + (y_n/p)^2 + (z_n/q)^2}$ is the ellipsoidal galacto-centric radius of the $n$th particle. Diagonalizing this tensor yields the eigenvectors and eigenvalues corresponding to the principal axes of the best-fitting ellipsoid. The ratio of the second largest and smallest eigenvalues to the largest gives the square of the axis ratios $p$ and $q$, respectively. The eigenvectors corresponding to the largest, middle, and smallest eigenvalue indicate the direction of the major, intermediate, and minor axes as well. 
    \item We rotate our system into the eigenbasis of the inertia tensor such that the new $x$, $y$, and $z$ coordinates correspond to the major, intermediate, and minor axes of that best fitting ellipsoid. Having updated these ratios and projected our system into this new basis, we calculate a new ellipsoidal radial distance $r_n$ to each particle and begin with step 1 again – now using the updated axis ratios that result from this iteration – until the resulting axis ratios ($p$ and $q$) vary less than $1\%$ between subsequent iterations.
\end{enumerate} 
The eigenbasis from the final iteration of this procedure yields the principal axis projection (i.e.~the $\mathbf{e}_1, \mathbf{e}_2, \mathbf{e}_3$ projection) for the stellar particles within the half-mass radius of the galaxy and identifies the shape with the axis ratios, $p$ and $q$, which correspond to the intermediate-to-major and minor-to-major axis ratio, respectively. 

Again for the stellar particles within $R_{1/2}$, the ellipsoidal half-mass radius that results from the converged process detailed above, we compute the angular momentum (AM) direction, given by:
\begin{equation}
\Vec{\mathbf{L}} \equiv \sum_n m_n \Vec{\mathbf{r}}_n \times \Vec{\mathbf{v}}_n
\end{equation}
where $\Vec{\mathbf{r}}_n$ is the spherical radius to the $n$th stellar particle and $ \Vec{\mathbf{v}}_n$ and $m_n$ are the corresponding particle velocity and mass. Then, as defined in \citealp{BF2019}, the intrinsic kinematic misalignment is defined as
\begin{equation}
\Psi_{\mathrm{int}} \equiv 90^\circ - \cos^{-1}\left(\frac{\Vec{\mathbf{L}} \cdot \hat{\mathbf{e}_1}}{\|\Vec{\mathbf{L}}\|}\right)
\end{equation}
Note that this is defined as the magnitude of the angular deviation between the AM and major axis directions relative to  $90^\circ$ – consistent with the definition in the literature – so that small values of misalignment (e.g.~$\Psi_\mathrm{int} < 10^\circ$) correspond to a rotation around an axis almost perpendicular to the major axis, and large values ($\Psi_\mathrm{int} \sim 90^\circ$) correspond to rotation (almost) around the major axis. In the case of prolate objects, where the major axis is significantly larger than the other two axes, rotation that is approximately around that axis ($\Psi_\mathrm{int} \sim 90^\circ$) is often in the literature called 'prolate rotation'.

As a measure of the rotational support, or strength of rotation, we also compute the `stellar spin parameter', which is given by a normalization of the specific angular momentum of the stellar particles within $R_{1/2}$:
\begin{equation}
    \lambda_* \equiv \frac{\|\Vec{\mathbf{L}}_\mathrm{spec}\|}{\sqrt{2GM_*R_{1/2}}}
\end{equation}
where $\Vec{\mathbf{L}}_\mathrm{spec}$ is the specific angular momentum of the stellar distribution (i.e.~$\Vec{\mathbf{L}}/M_*$). This takes on values between 0 and 1, with values closer to 1 corresponding to strong, ordered rotation.

Finally, we measure the triaxiality of the galaxy, defined as:
\begin{equation}
T \equiv \frac{1-p^2}{1-q^2}
\end{equation}
where $p$ and $q$ are the aforementioned axis ratios. Therefore, $T$ takes on values between 0 and 1, with $T<1/3$ roughly corresponding to an oblate shape and $T>2/3$ to a prolate shape, with intermediate values being denoted triaxial.

\begin{figure}%
    \centering
        \subfloat[The $p$ vs $q$ distribution, colored by the intrinsic kinematic misalignment, $\Psi_\mathrm{int}$. The dashed lines indicate the morphological definitions of spherical, oblate, prolate, and triaxial taken from (\citealp{Li2018} see \citealp{BF2019}).]{{\includegraphics[scale=0.43]{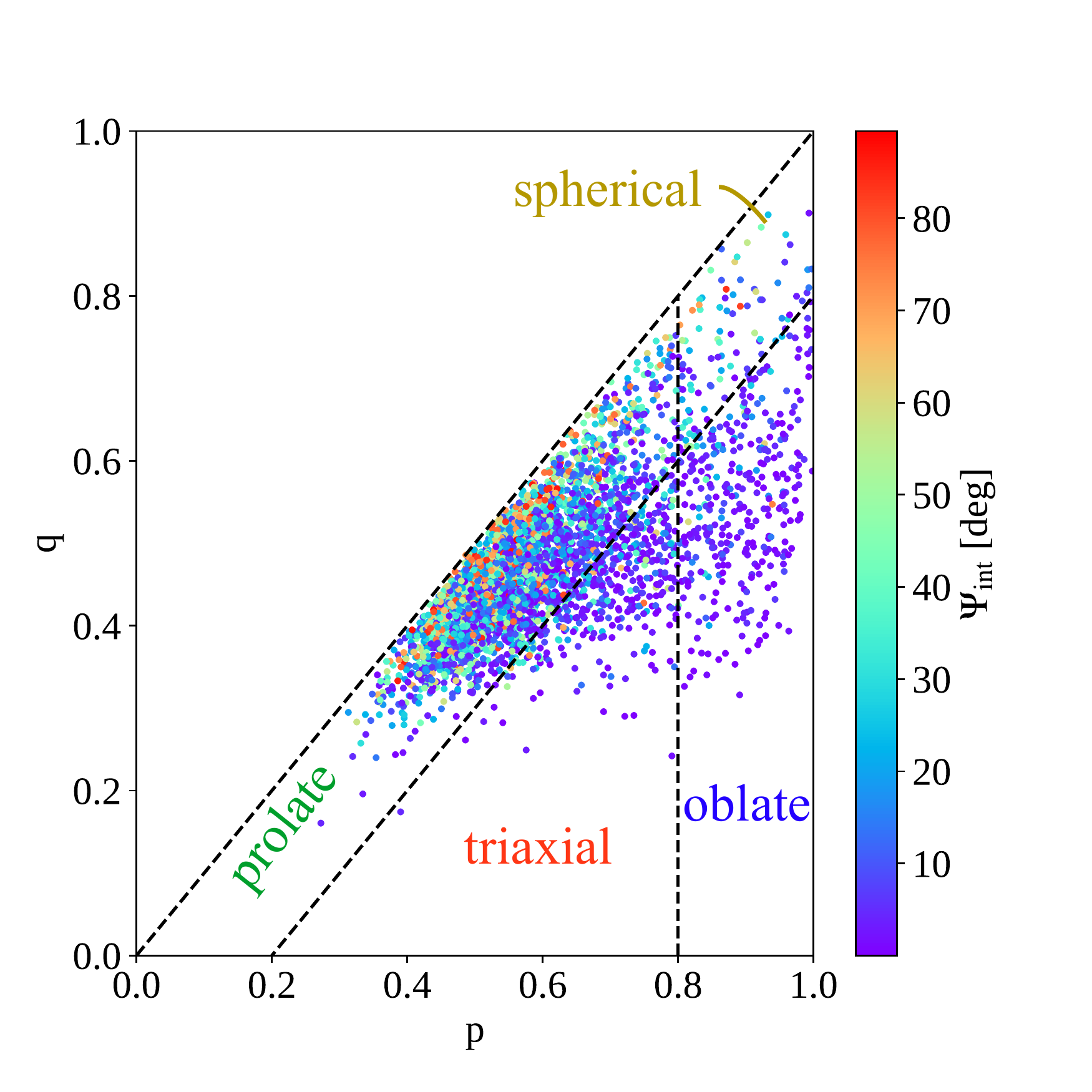} }}%
        \qquad
        \subfloat[The triaxiality ($T = \frac{1-p^2}{1-q^2}$) vs the kinematic misalignment. The points are colored by their morphological definitions.]{{\includegraphics[scale=0.43]{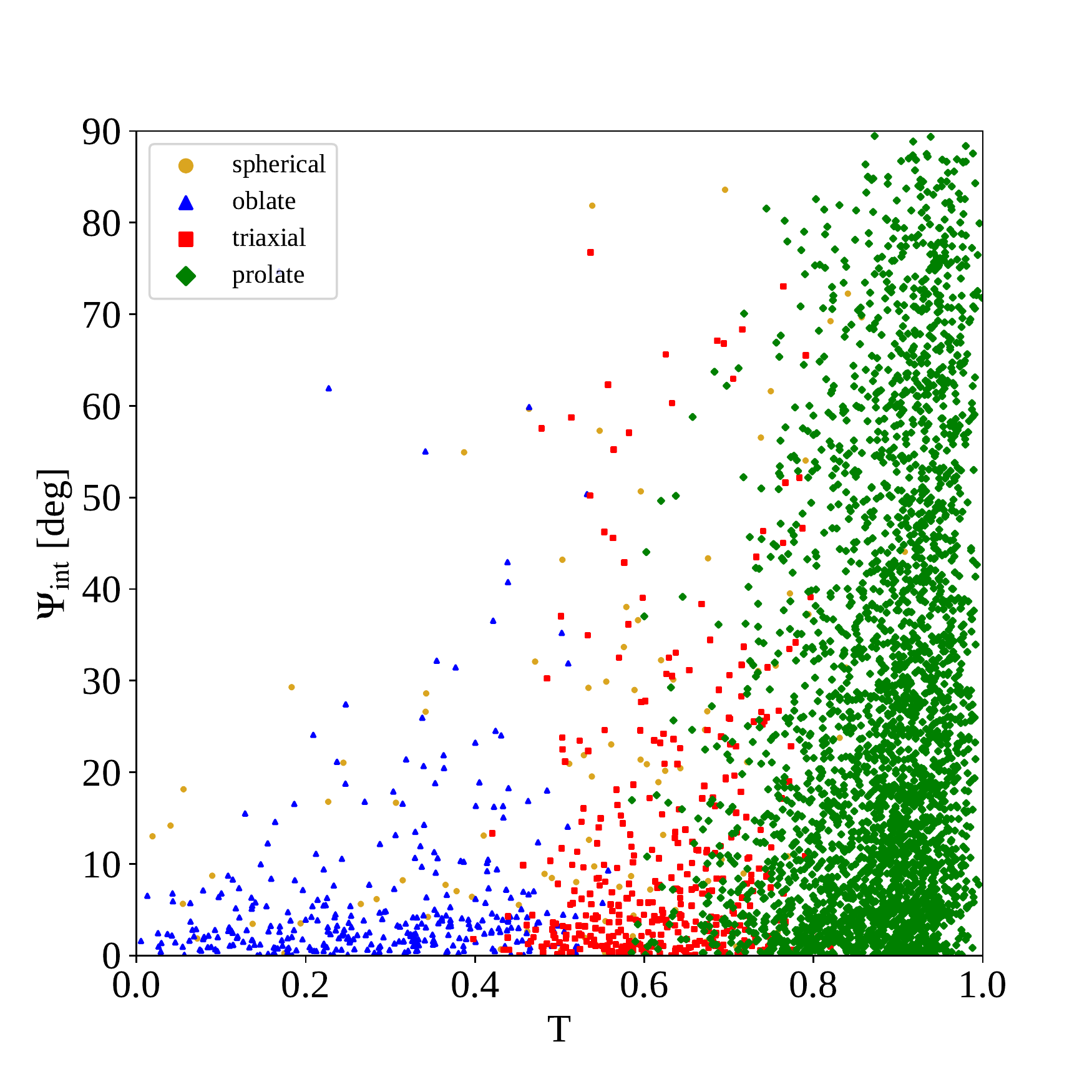} }}%
        \caption{The distribution of a few basic morphological and kinematic properties of the galaxies in our sample.}
        \label{fig:morph_distribution}%
\end{figure}

With these preliminary morphological and kinematic descriptors in hand, we survey the distribution of shapes of galaxies in our sample, using the designation from \citet{Li2018} (see \citealp{BF2019}):
\begin{itemize}
    \item Spherical: $p-q < 0.2$ and $p\geq 0.8$
    \item Oblate: $p-q \geq 0.2$ and $p\geq 0.8$
    \item Prolate: $p-q < 0.2$ and $p < 0.8$
    \item Triaxial: $p-q \geq 0.2$ and $p < 0.8$
\end{itemize}
From this definition, the intuition is clear – oblate galaxies tend to be more disky (comparable intermediate and major axis lengths with a relatively smaller minor axis) and prolate galaxies are more oblong, or `cigar' shaped (one very long, major axis, with small, comparable intermediate and minor axes). With this designation, we can broadly sketch out the distribution of the different shapes (Figure \ref{fig:morph_distribution}). In the first panel, we see the distribution of galaxies colored by their misalignment, with the shape distinctions overlaid as dashed lines. In the second panel, we display the kinematic misalignment against the triaxiality, with symbols indicating our morphological classification. Comparing this against a similar investigation conducted with a sample of lower mass galaxies in the 100 $\mathrm{Mpc}^3$ Illustris box (see \citealp{BF2019}), we see an extension of the same trends, with a larger proportion of prolate and triaxial galaxies. From this, we can summarize the properties of each morphological class. 

First, the \textit{prolate galaxies} (green diamonds) comprise the majority of our sample (79\%) and have the largest triaxialities. These galaxies display the largest range of misalignments, with a near uniform spread of galaxies with $\Psi_\mathrm{int}$ between $0^\circ$ and $90^\circ$ in Figure \ref{fig:morph_distribution}b. Analysis of the radial dependence of the measured principal axis and AM directions for these galaxies (not shown) demonstrates that the major axis is very well defined whereas the AM direction is fairly radius-dependent. In Figure \ref{fig:morph_distribution}a, we can see that the prolate region (the lower-left segment between the dashed lines) reflects this spread of misalignments and that the galaxies tend to be concentrated around axis ratios of $p\sim q \sim 0.5$, so their prolateness is not extreme. This subset tends to have the highest fraction of ex-situ stars, namely those that accreted via mergers rather than formed in-situ (not shown).

Next, the \textit{oblate galaxies} (blue triangles) make up 8.6\% of our sample and have the smallest triaxialities. The vast majority of these galaxies have misalignments less than $5^\circ$. The relative abundance of oblate galaxies (compared to prolate systems) within our sample is a stark contrast to that observed in the lower mass sample analyzed in \citet{BF2019}, for example. These galaxies tend to have the largest fraction of in-situ stars.

The \textit{triaxial galaxies} (red squares) make up a similar fraction (9.8\%) to the oblate galaxies and occupy intermediate triaxialities and misalignments between the oblate and prolate galaxies. These galaxies have smaller in-situ fractions than oblate galaxies.

Finally, the \textit{spherical galaxies} (2.5\% of the sample) are the least well-defined in this shape and kinematic description because the three principal axes are all comparable and the axis ratios are close to 1. Therefore, the measured values (e.g.~misalignment, triaxiality, etc.) for these galaxies are generally unreliable and highly variable. 

We note that the median masses for all the morphological types are almost identical, roughly $\log_{10}(M_*/M_\odot)\sim 11.20$.

\section{A New Classification from Rates of Change of the Axes}\label{sec:new_classes}
It is clear that the simple morphological classifications outlined in Section \ref{ssec:characterizingsample} are insufficient to capture the detailed variations of the behaviors observed in these high-mass galaxies, as almost 80\% of the sample is prolate. In addition, we generally describe galaxies using their instantaneous properties, often focusing on the orientation of their kinematic and shape axes, so it is worth asking how robust these measurements actually are. In other words, would these observations be consistent over time? Are the axes changing direction rapidly and, if so, how are they changing? Looking to Figure~\ref{fig:morph_distribution}b for example, it is interesting to consider whether all the galaxies in the high-triaxiality, large-misalignment range truly are in a state of regular rotation around their major axis, and whether the combination of these two properties are sufficient for a designation as a prolate rotator, as is often done in the literature?

To investigate this, and to begin thinking about how these galaxies achieve their current morphological and kinematic states, it is instructive to inspect how they change over time. Using such measurements, we introduce a new system of classes, determined by the rate-of-change of the AM and principal axis directions, that offers a new perspective for analyzing the physical behavior of these simulated systems. In fact, using this classification, as we will discuss in Section~\ref{sec:discussion}, we are able to identify distinct physical and evolutionary behaviors for each of our classes, one of which is prolate rotation.

We define the classes using a new quantity $\dot{\Phi}$ which we term the `reorientation rate'. This is calculated by determining the angular change of a vector measured in consecutive simulation snapshots divided by the time separation of the snapshots,
yielding a quantity measured in units of degrees per gigayear ($\mathrm{deg}/\mathrm{Gyr}$). In particular, we focus on the AM reorientation rate $\dot{\Phi}_L$ and major axis reorientation rate $\dot{\Phi}_{\mathbf{e}_1}$. Averaging each of these measured quantities over a window of five snapshots (0.81 Gyr at $z=0$ and 0.65-0.8 Gyr for earlier times) results in a quantity that is relatively resistant to small-scale fluctuations in the distribution of the stellar particles and in the measured direction of the AM or major axis – that is, our reorientation rate provides a robust measure of the consistency (or lack thereof) in the direction of the AM or major axis.

\subsection{Exemplars of the classes}\label{ssec:class_exemplars}
Before we formally introduce the classes, it is useful to build intuition using a sample galaxy drawn from each class. 
In the upper panels of Figures~\ref{fig:prolate_slow_rotator_unsettled_map}-\ref{fig:prolate_rotator_PSR_mis_map}, we present smoothed velocity maps of a set of exemplar galaxies at four stages of their evolution, beginning with the upper left panel at the time immediately following the most recent major merger (mass ratio $\mu > 1/4$), namely the immediate descendant of that merger. In the bottom right panel, we display the galaxy in the present universe ($z=0$), and the other two panels capture the state of the galaxy at two intermediate snapshots. 

The lower panels of Figures~\ref{fig:prolate_slow_rotator_unsettled_map}-\ref{fig:prolate_rotator_PSR_mis_map} contain the evolution of certain descriptive quantities for each galaxy over time, from the point of most recent major merger to the present day. In order, we present the triaxiality $T$, intermediate-to-major axis ratio $p$, minor-to-major axis ratio $q$, AM and major axis reorientation rates as defined above, the stellar spin parameter $\lambda_*$, the intrinsic kinematic misalignment $\Psi_\mathrm{int}$, and the cumulative AM reorientation rate between the descendant galaxy of the merger and subsequent descendants galaxies (and the analogous quantity for the major axis). The latter two quantities give an indication of the net angular change of the kinematic (AM) and morphological (major axis) axes of the galaxy in contrast with the reorientation rate, which provides an instantaneous measure of these changes. In the panel with the reorientation rate, we have included a dotted line to indicate our selected boundary between `fast' and `slow' reorienters of 75 deg/Gyr (see Section~\ref{ssec:def_classes}). In sum, these panels capture the evolution of a few specially selected galaxies both visually and with respect to several key descriptors of their kinematics and shape.

In Figure \ref{fig:class_definition}, we present the distribution of parameters we ultimately use to define the classes. We classify each galaxy by its AM reorientation rate, major axis reorientation rate, triaxiality, and kinematic misalignment. Specifically, we begin by segmenting the distribution in the AM reorientation rate-major axis reorientation rate plane based on visual separations in the distribution (the aforementioned 75 deg/Gyr cutoff) and further subclassify the galaxies by their kinematic misalignments and triaxialities. The forthcoming descriptions of exemplar galaxies are based on the quantities displayed in this figure. The full classification is outlined in Section \ref{ssec:def_classes}.

\subsubsection{Rapid AM Reorienters}\label{sssec:psr_ex}
First (Figure \ref{fig:prolate_slow_rotator_unsettled_map}a), we display a galaxy selected from the large AM and small major axis reorientation rate arm of the distribution given in Figure \ref{fig:class_definition}a (i.e.~the yellow points). In this projection, the elongated prolate shape is evident at all times and its orientation is fairly fixed. Next, we note that the galaxy has very little coherence in the individual velocity maps; in other words, there are small regions of particles collectively moving in different directions and no clear sense of rotation in the galaxy. This is reaffirmed by the observation that the galaxy has a vanishingly small stellar spin parameter – there is no strong rotational support. In the velocity maps, we also observe that the AM vector (black line) is quite variable and does not hold a fixed direction relative to the major axis, a direct consequence of the rapid AM reorientation rate, one of the definition criteria of this class. The misalignment angle (between AM and major axis) reflects this variability and shows significant fluctuations as well. We note that at some moments in its evolution (such as 4 Gyr or 1.5 Gyr), the galaxy obtains a state of significant misalignment ($\sim 90^\circ$) that according to some previous studies could earn it the designation of a `prolate rotator'; however, we argue that the instability of this measurement and the incoherence of the velocity maps indicate that a different characterization is more appropriate. In kind, we see that there is significant variation in the AM direction and a gradual drift to the major axis direction. Despite all this variability, we see that our classification remains consistent throughout the galaxy's history (the yellow background persists) so this is a clear example of what we suggestively call a \textit{rapid AM reorienter}.

\subsubsection{Unsettled galaxies}\label{sssec:unsettled_ex}
Next, in Figure \ref{fig:prolate_slow_rotator_unsettled_map}b, we present a galaxy selected from the fast reorientation rate region of Figure \ref{fig:class_definition}a in terms of both AM and major axis (i.e.~red-orange points). We see that the galaxy not only grows significantly over time (the half mass radius quadruples and the mass increases by nearly 50\%), but the velocity map also becomes increasingly incoherent. That is, in the earlier times, we see a clear sense of rotation (the red-blue division), but by the present day, this is diminished. The lower panels help clarify the story: this galaxy experiences significant class changes over time. For a large portion of its evolution, this galaxy holds a certain classification (blue) but very close to $z=0$, we see that it experiences a sharp growth in mass of approximately 7.5\% (identified by the black dashed line). This is indicative of a mini merger that has moved the galaxy from one state to a position of disequilibrium. Note that in the bottom left panel the displayed AM vector looks to be pointing in a different direction than the ostensible red-blue divide in the velocity because the true region within which the AM is computed is obscured by high velocity features at large radii (cf. $R_{1/2}$ at this time is roughly 6 kpc). In fact, we can see when the merging progenitor galaxy begins to interact with the parent system. At lookback times of approximately 2.5 and 1.5 Gyrs, we see spikes in the misalignment and AM reorientation, as well as in the cumulative angular change of the AM vector. Prior to the first of these spikes, the galaxy had a well-defined class and its behavior was one of steady, coherent rotation, evidenced by strong rotational support and periodic variation in the cumulative angular change of the major axis. Therefore, these two spikes are flybys of the satellite galaxy that are upsetting the equilibrium state of the system. In the presence of such interactions, and immediately following the merger, we see that the galaxy achieves the same classification as just before the present day (red-orange), so identifying this as an exemplar of the \textit{unsettled} class of galaxies is quite straight-forward. 

\subsubsection{Spinning disks}\label{sssec:sd_ex}
Following this, in Figure \ref{fig:spinning_disk_twirling_cigar_map}a, we present a galaxy selected from the slow AM reorientation region of Figure \ref{fig:class_definition}a (light blue points) and small misalignment region of panel b. Within this designation, however, we split the galaxies by their triaxiality, so this is one that has $T < 0.6$ (panel c), yielding a selection of galaxies that tend to be more triaxial or oblate. In the velocity maps, we see that this galaxy has a clear sense of rotation at all times through its evolutionary history and a consistent AM direction that is orthogonal to the horizontal/major axis. With this, one can imagine rotation of the major and intermediate axes in the plane of the oblate disk of this galaxy. The lower panels confirm this assessment – we see a significant degree of rotational support, a misalignment that is very nearly $0^\circ$ (indicating consistent orthogonality between the AM and major axis), a well defined AM direction, and periodic variation of the major axis (corresponding to rotation in the plane of a disk). These are all characteristic behaviors of a disky galaxy – earning such systems the designation of \textit{spinning disk}.

\subsubsection{Twirling cigars}\label{sssec:tc_ex}
Next, in Figure \ref{fig:spinning_disk_twirling_cigar_map}b, we display a galaxy selected from the same region in Figure~\ref{fig:class_definition}a as the galaxy shown in Figure \ref{fig:spinning_disk_twirling_cigar_map}a, but with a triaxiality larger than 0.6 (panel c). Such galaxies will be, by definition, more elongated or prolate. Inspecting the maps and panels associated with this galaxy, we can see that it exhibits many of the same behaviors as the spinning disk, with the primary difference being the triaxiality. In fact, these two classes have extremely similar behaviors and are only distinguishable by shape, so we term these galaxies the \textit{twirling cigars}.

\subsubsection{Regular prolate rotators}\label{sssec:pr_ex}
The final class that we identify is given in Figure \ref{fig:prolate_rotator_PSR_mis_map}b. This galaxy is selected from the slow reorientation corner of Figure \ref{fig:class_definition}a under a secondary selection cut by misalignment of $\Psi_\mathrm{int}>60^\circ$ (purple points; see panel b). In the velocity maps, we see that the galaxy shows a distinct sense of rotation but in a manner that is a contrast to the spinning disks or twirling cigars. That is, these galaxies are rotating around their major axes and the AM alignment with the major axis indicates just that. In the lower panels, we see that the galaxy has a reasonable degree of rotational support, is consistently misaligned, and maintains a well-defined AM and major axis direction. Over time, it experiences a relatively consistent classification as a \textit{regular prolate rotator}. 

The unique behavior of the regular prolate rotators is clearer when taken in contrast with another galaxy that is similarly misaligned but differs in its AM reorientation (i.e.~a misaligned rapid AM reorienter). An example of one such galaxy is given in Figure \ref{fig:prolate_rotator_PSR_mis_map}a. While the roughly consistent misalignment is a shared characteristic of these and the regular prolate rotators, we see that there is little coherence in the velocity map and the degree of rotational support reaffirms that. The misalignment also shows more susceptibility to variation and dips further than that of the regular prolate rotators. We can begin to see that regular prolate rotators demonstrate certain unique characteristics relative to other galaxies that may be instantaneously or even consistently misaligned and will explore this dichotomy more deeply in Section \ref{sec:misaligned_comp}.

\begin{figure*}
    \centering
        \subfloat[The rapid AM reorienter map.]{{\includegraphics[scale=0.35]{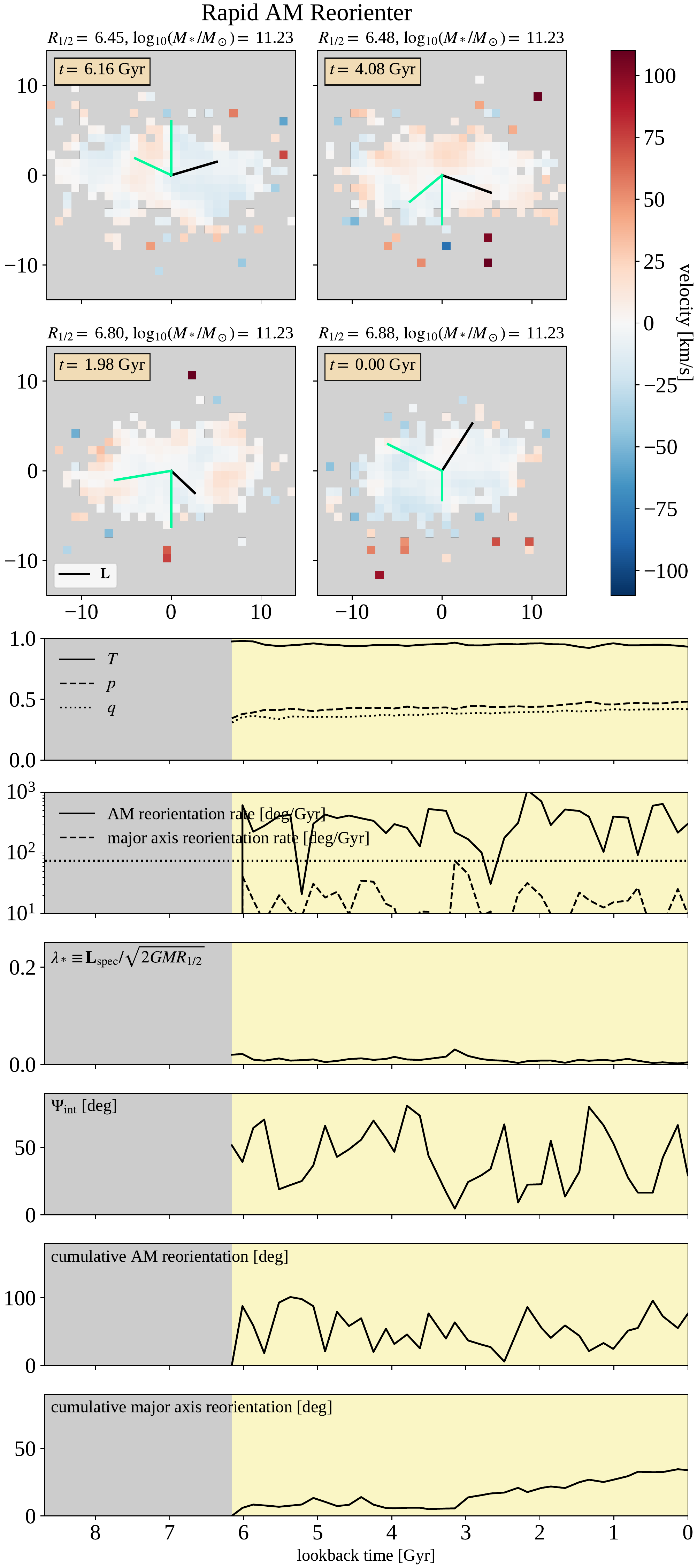} }}%
        \qquad
        \subfloat[The unsettled galaxy map.]{{\includegraphics[scale=0.35]{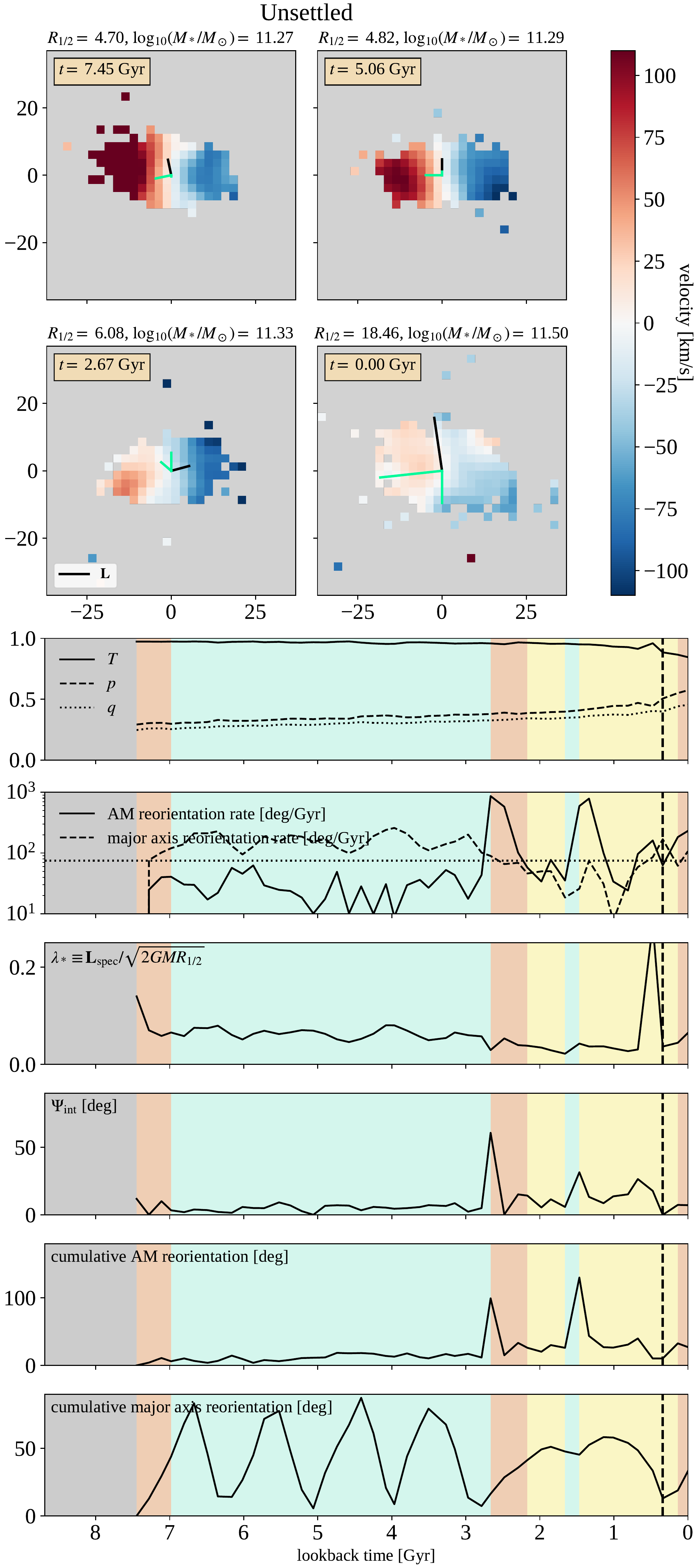} }}%
    \caption{The velocity map and associated panels for a characteristic \textbf{rapid AM reorienter} (left) and \textbf{unsettled} (right) galaxy. The top four panels are maps of the mass-weighted average line-of-sight velocity of the stellar particles in each pixel at a variety of lookback times – the first panel is the snapshot right after the merger (descendant galaxy) and the fourth is at $z=0$. These maps are oriented with the major axis as the abscissa and the minor axis as the ordinate (taking the principal axes from the second displayed snapshot), and the black solid line shows the \textbf{direction} of the AM (note that these are all scaled to be the length of $R_{1/2}$) – the two green lines are provided for visual purposes. The colormap is normalized to the maximum velocities of pixels within $1.5 R_{1/2}$ and pixels with fewer than 5 particles or standard errors $\sigma_v/\sqrt{n-1}> 35\ \mathrm{km/s}$ are displayed in gray. The lower panels indicate the evolution of several morphological and kinematic quantities measured for the galaxy at every snapshot \textit{after} its most recent major merger ($\mu > 1/4$). The lowest two panels indicate the angle between the vector direction (AM or major axis) at a given snapshot and that same vector at the snapshot immediately after the most recent major merger (i.e.~descendant snapshot; the snapshot corresponding to the upper left pixel panel). A vertical dashed line (if present) in all the panels indicates a time at which the mass of the galaxy increased by more than 5\% between subsequent snapshots. The colors in the background of the panels reflect the classification assigned to the galaxy at that time (see Figure \ref{fig:class_definition} for a reference of the colors), with the gray region denoting the time \textit{before} the last major merger.}
    \label{fig:prolate_slow_rotator_unsettled_map}
\end{figure*}

\begin{figure*}
    \centering
        \subfloat[The spinning disk map.]{{\includegraphics[scale=0.35]{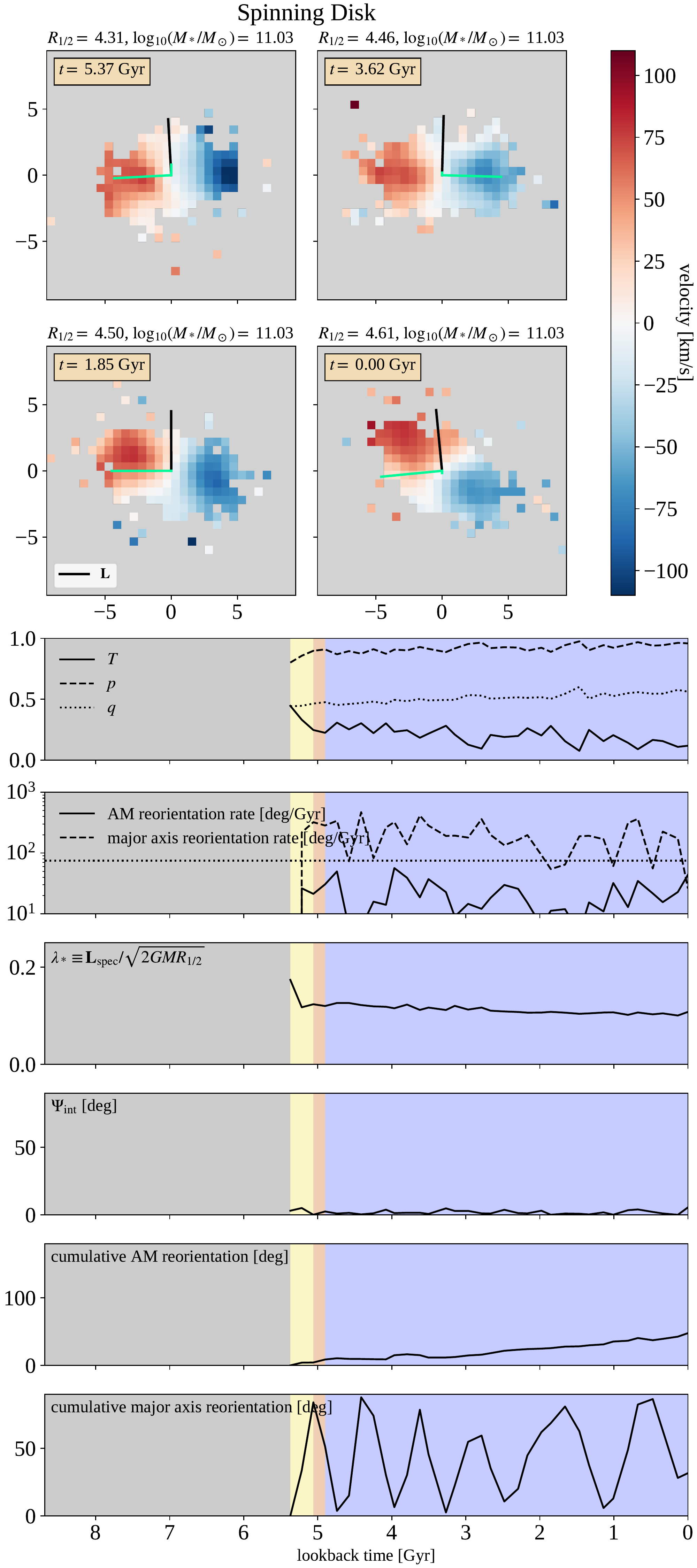} }}%
        \qquad
        \subfloat[The twirling cigar map.]{{\includegraphics[scale=0.35]{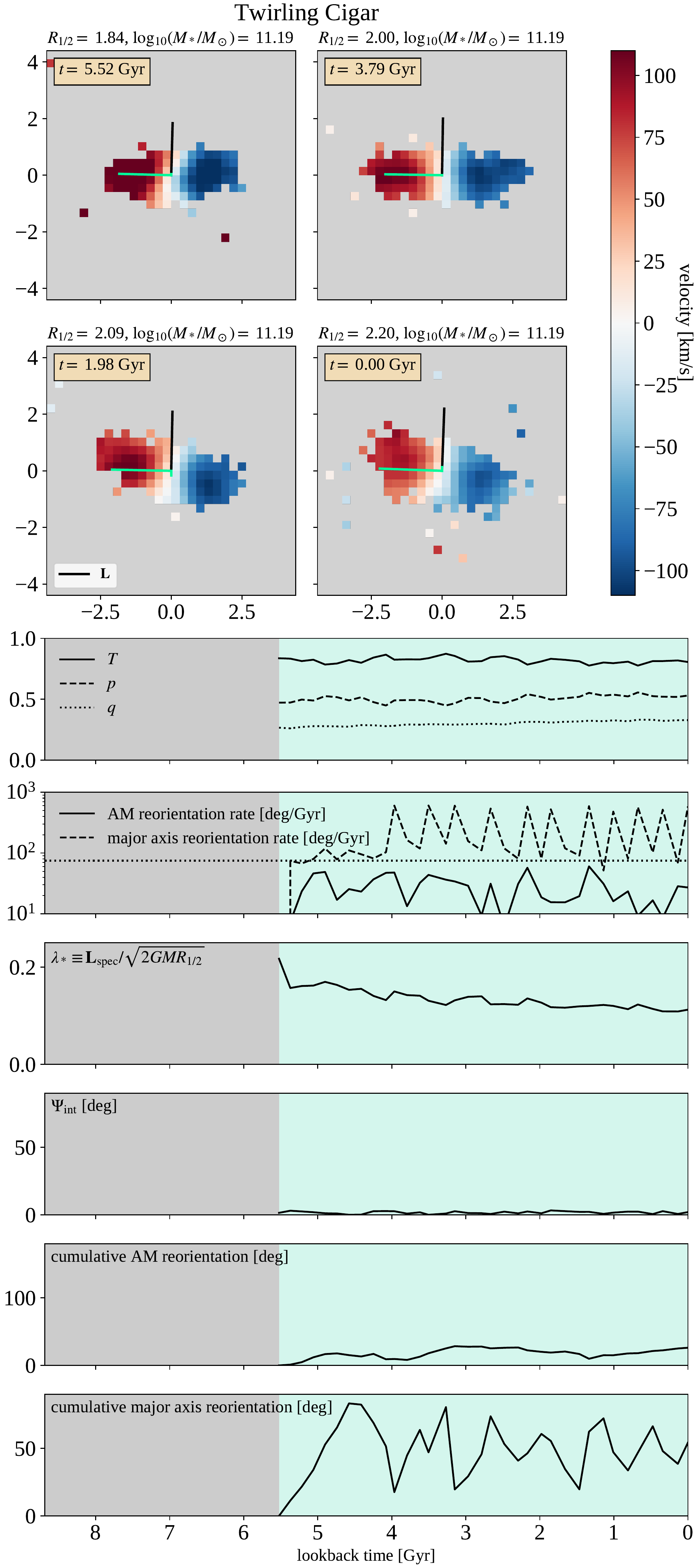} }}%
    \caption{The velocity map and associated panels for a characteristic \textbf{spinning disk} (left) and \textbf{twirling cigar} (right) galaxy. See Figure \ref{fig:prolate_slow_rotator_unsettled_map} for a description of what is presented in each panel.}
    \label{fig:spinning_disk_twirling_cigar_map}
\end{figure*}

\begin{figure*}
    \centering
        \subfloat[A misaligned rapid AM reorienter.]{{\includegraphics[scale=0.35]{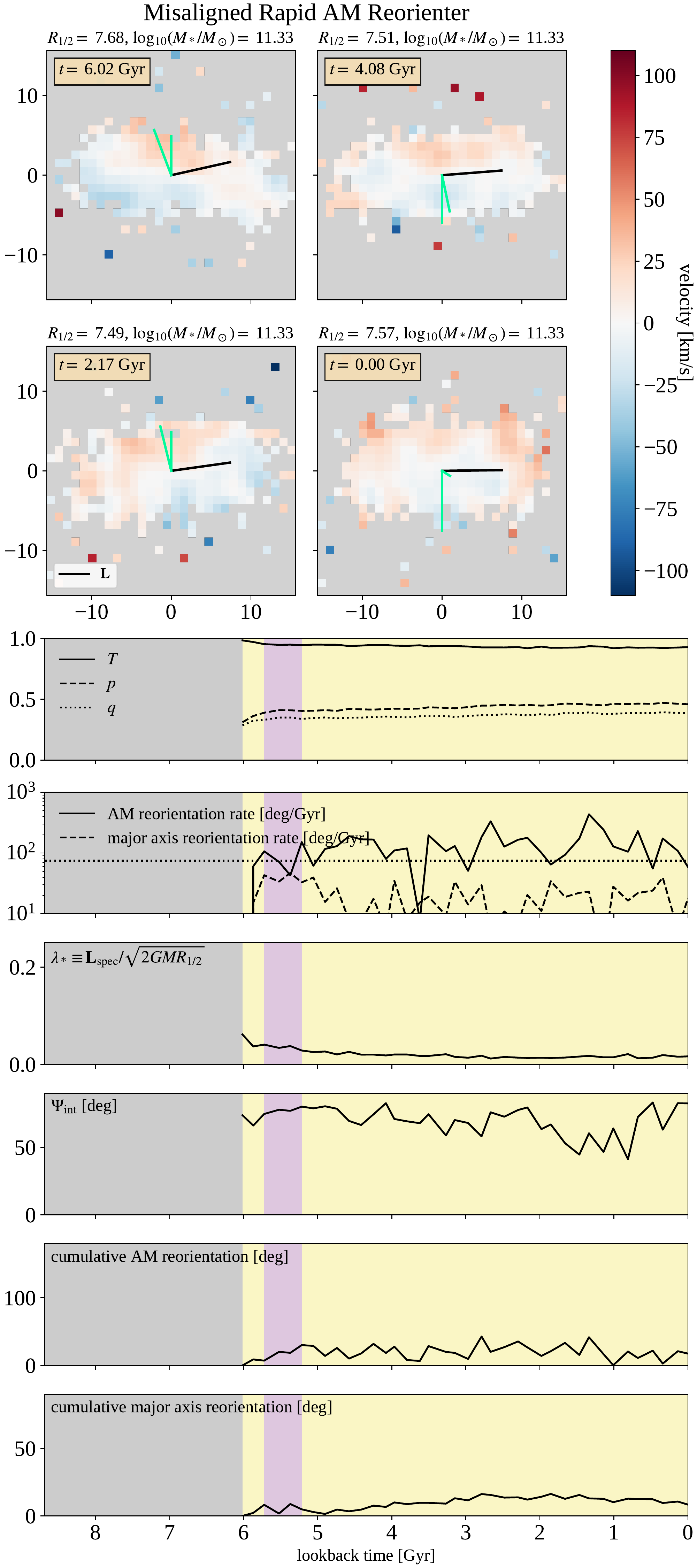} }}%
        \qquad
        \subfloat[The regular prolate rotator map.]{{\includegraphics[scale=0.35]{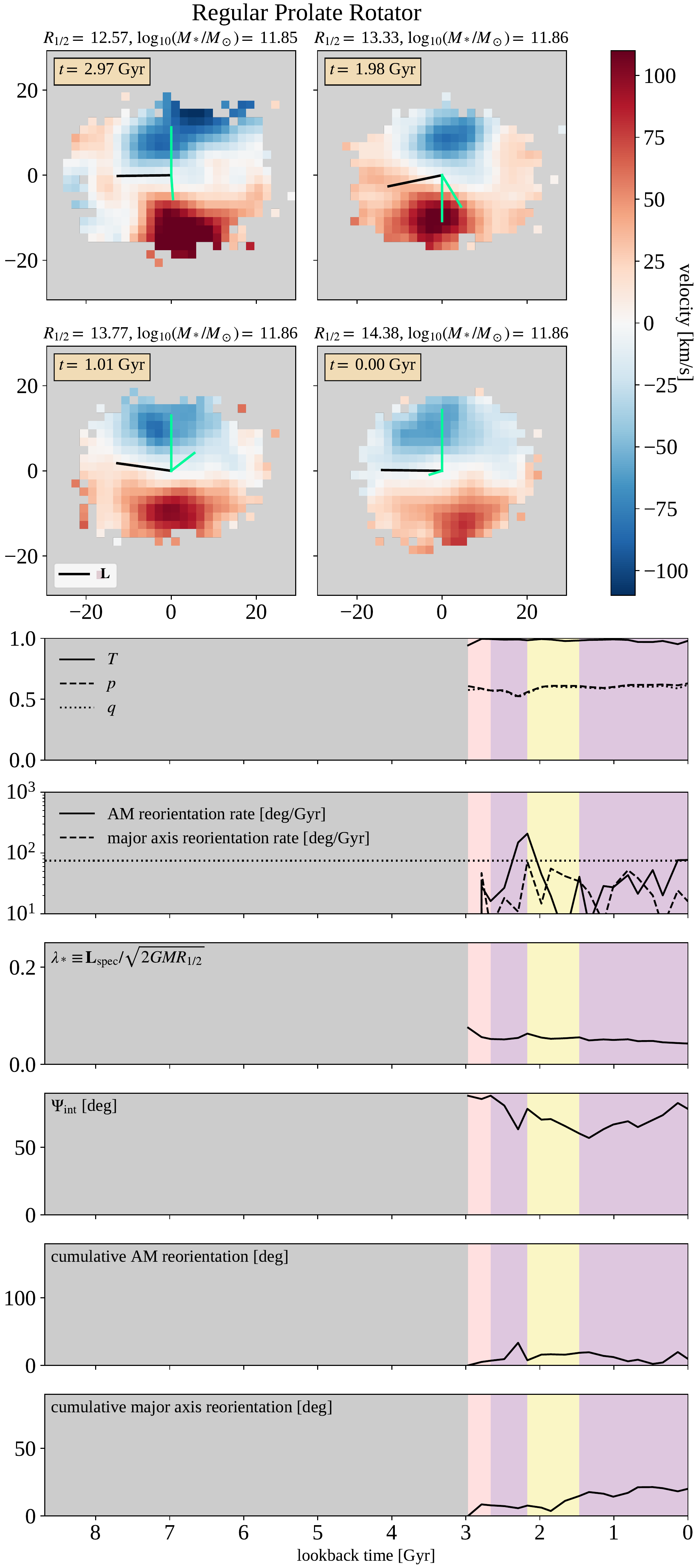} }}%
    \caption{The velocity map and associated panels for a \textbf{misaligned rapid AM reorienter} (left) and \textbf{prolate rotator} (right) galaxy. See Figure \ref{fig:prolate_slow_rotator_unsettled_map} for a description of what is presented in each panel.}
    \label{fig:prolate_rotator_PSR_mis_map}
\end{figure*}

\subsection{Definitions of the new classes}\label{ssec:def_classes}
The intuitive picture derived from those representative galaxies (in Figures \ref{fig:prolate_slow_rotator_unsettled_map}-\ref{fig:prolate_rotator_PSR_mis_map}) turns out to correspond quite well to the average behavior of a galaxy within each of the classes. 

\begin{figure*}%
    \centering
        \subfloat[The reorientation rate of the AM and major axes – the primary parameters used to define the new classes.]{{\includegraphics[scale=0.43]{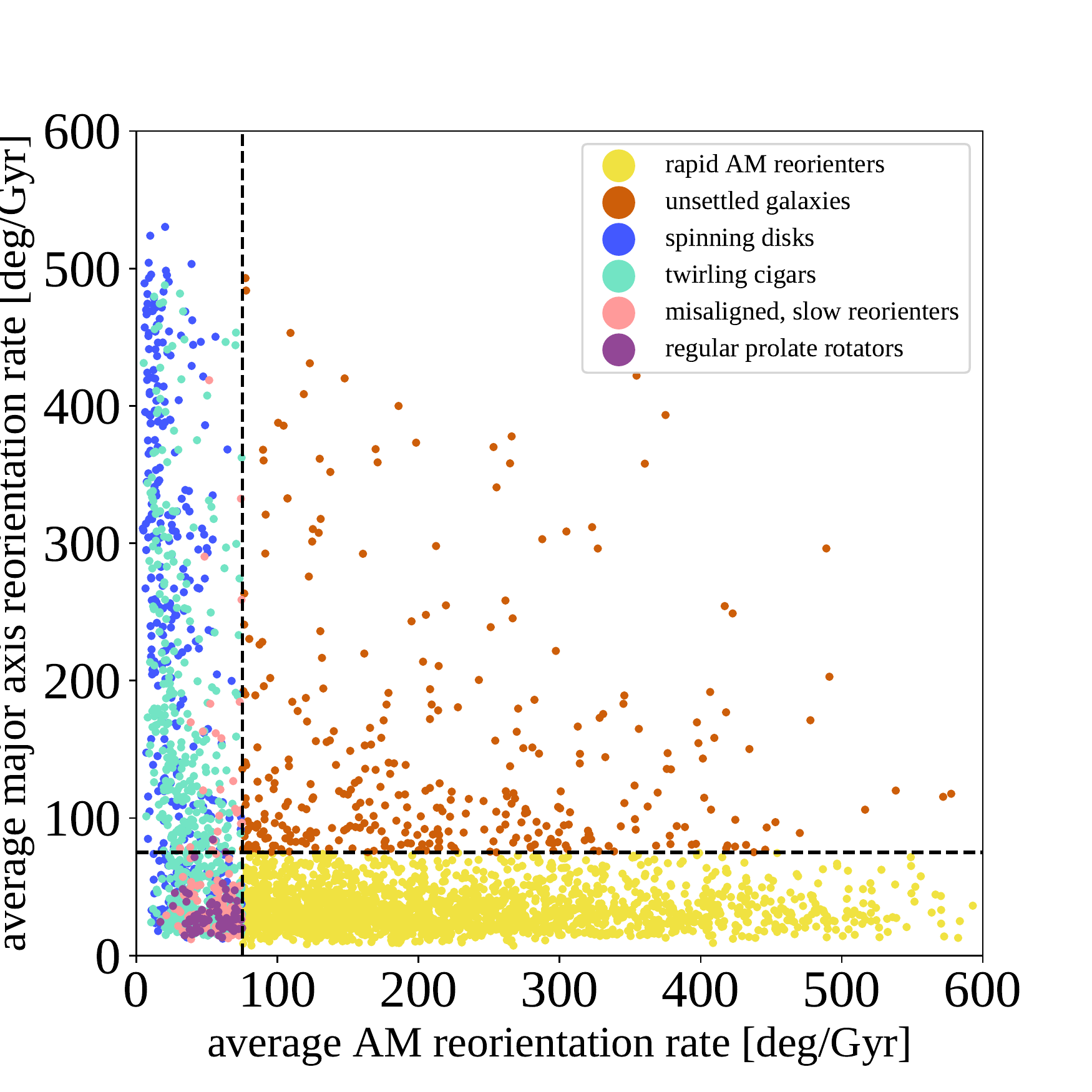} }}%
        \qquad
        \subfloat[The triaxiality ($T = \frac{1-p^2}{1-q^2}$) vs the kinematic misalignment (cf. Figure \ref{fig:morph_distribution}b). These parameters primarily define the spinning disks, twirling cigars, misaligned slow reorienters, and regular prolate rotators.]{{\includegraphics[scale=0.43]{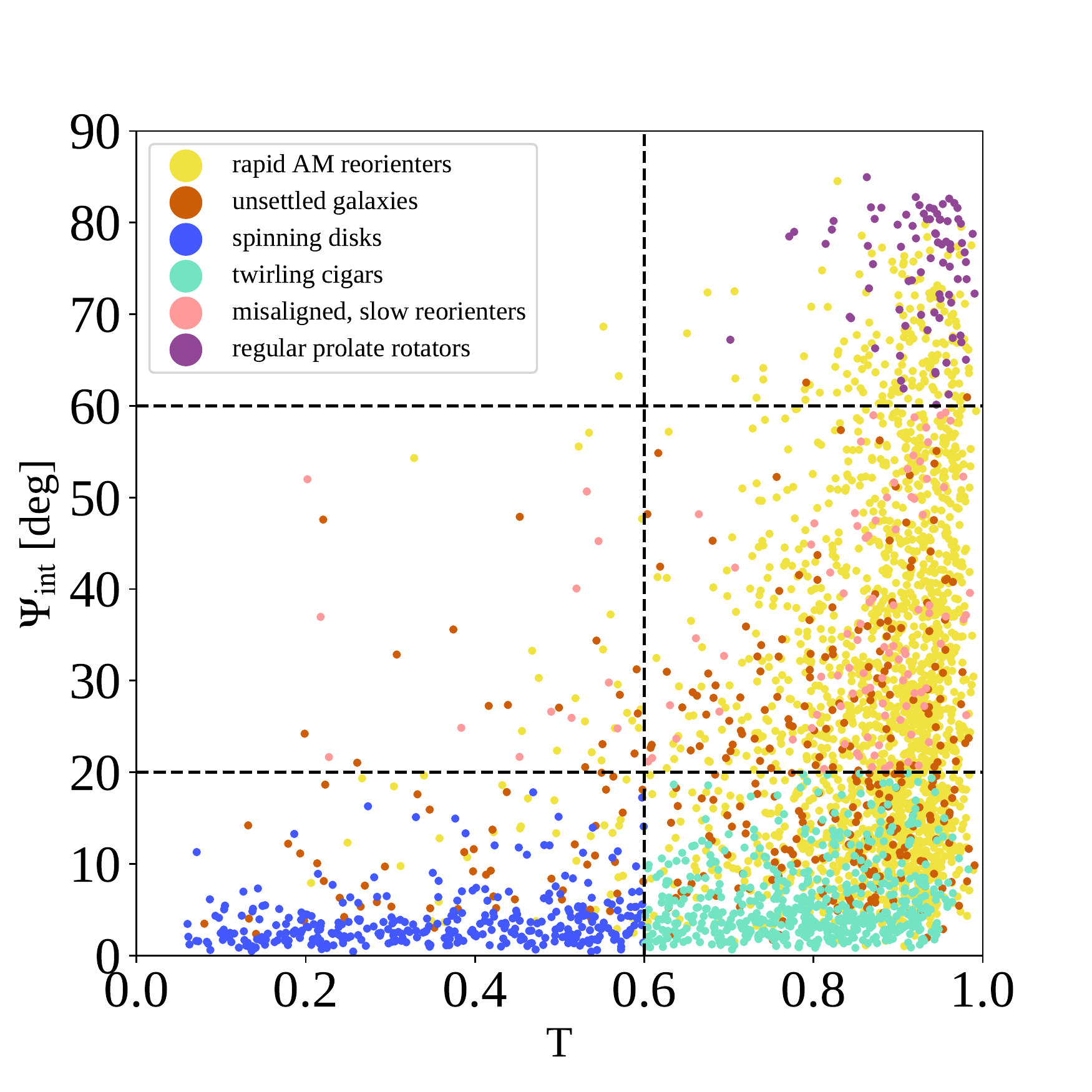} }}%
        \qquad
        \subfloat[The reorientation rate of the AM vs the kinematic misalignment $\Psi_\mathrm{int}$.]{\includegraphics[scale=0.43]{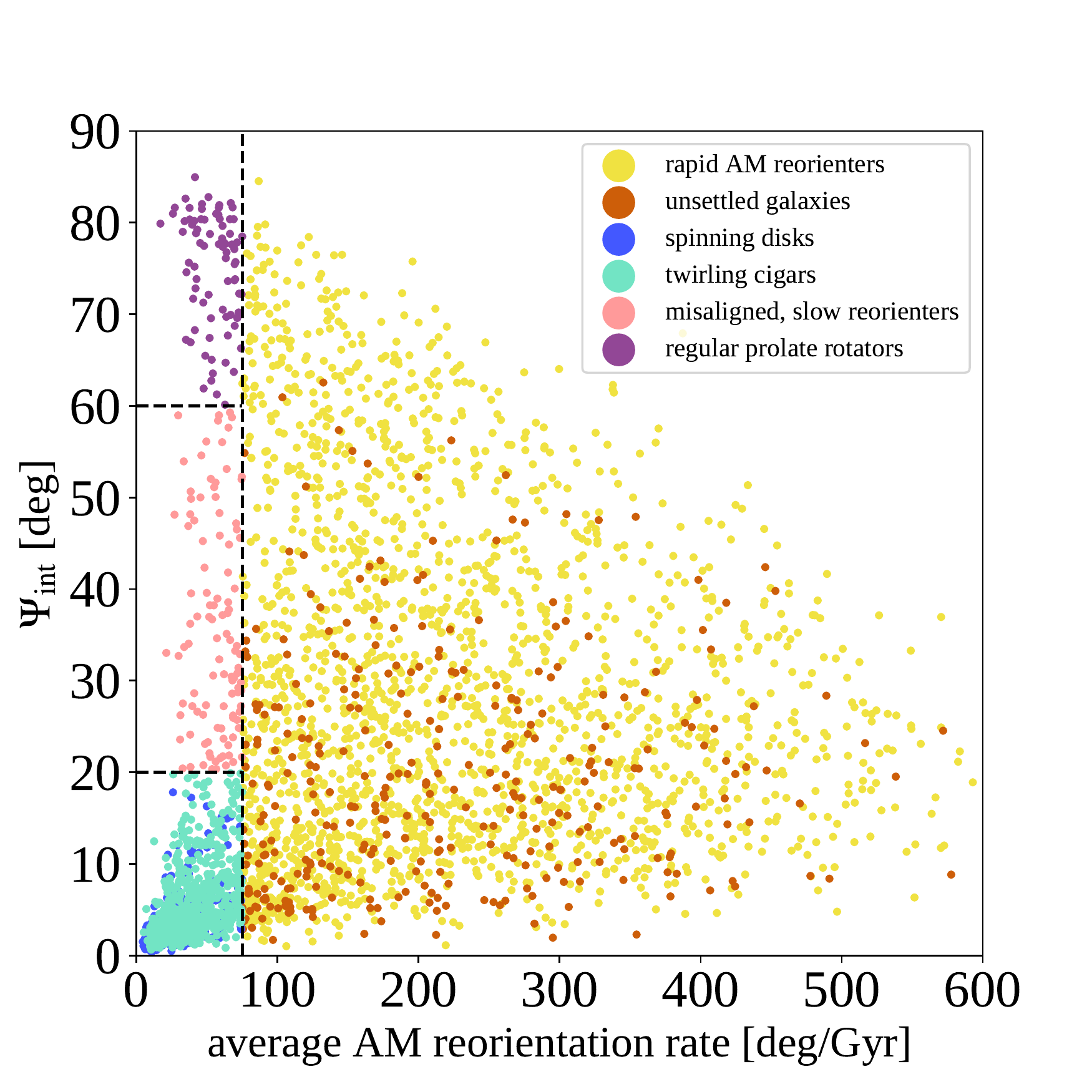}}%
        \caption{The distribution of parameters used to define the classes. These quantities are averaged over a 5 snapshot ($\sim 1$ Gyr) window. Points are colored by the classification outlined in \ref{ssec:def_classes} and black dashed lines indicate the boundaries between classes.}
        \label{fig:class_definition}%
\end{figure*}

The resulting distribution of our massive galaxy sample is given in Figure \ref{fig:class_definition}. There are several distinct regions of this figure and these correspond to observable and physical behaviors. 

Formally, we define the classes as follows:
\begin{enumerate}[label=\Roman*.]
    \item \textbf{Rapid AM Reorienters:} rapid AM and slow $\mathbf{e}_1$ reorienters ($\dot{\Phi}_L\geq 75$ deg/Gyr and $\dot{\Phi}_{\mathbf{e}_1}< 75$ deg/Gyr), \textit{58.38\%}
    \item \textbf{Unsettled galaxies:} fast reorienters in both AM and $\mathbf{e}_1$ ($\dot{\Phi}_L\geq 75$ deg/Gyr and $\dot{\Phi}_{\mathbf{e}_1}\geq 75$ deg/Gyr), \textit{9.84\%}
    \item Aligned ($\Psi_\mathrm{int} < 20^\circ$), slow AM reorienters ($\dot{\Phi}_L< 75$ deg/Gyr)
        \begin{enumerate}[label=\Alph*.]
        \item \textbf{Spinning disks:} small triaxiality (more oblate; $T< 0.6$), \textit{10.05\%}
        \item \textbf{Twirling cigars:} large triaxiality (more prolate; $T> 0.6$), \textit{16.29\%}
    \end{enumerate}
    \item Misaligned ($\Psi_\mathrm{int} \geq 20^\circ$), slow AM reorienters ($\dot{\Phi}_L< 75$ deg/Gyr)
    \begin{enumerate}[label=\Alph*.]
        \item \textbf{Misaligned, Slow Reorienters:} intermediate misalignment ($20^\circ \leq \Psi_\mathrm{int} < 60^\circ$), \textit{3.03\%}
        \item \textbf{Regular prolate rotators:} very misaligned ($\Psi_\mathrm{int} \geq 60^\circ$), \textit{2.37\%}
    \end{enumerate}
\end{enumerate}
where the italicized percentages indicate the fraction of the full sample occupied by each class at the present day. These have been grouped into subclasses based on similarity of behavior that will become more evident as we inspect their evolution. Note that the suggestive names that these have been assigned are more useful mnemonics than anything else – formally, the physical picture is the definition outlined above (e.g.~aligned, slow reorienter, etc.). From this classification it is also clear that we have omitted one class from our discussion thus far, the so-called `misaligned, slow reorienter' galaxies. These are galaxies with intermediate, unstable misalignments that do not seem to correspond to a clear physical picture in the same sense as the other galaxies and only comprise 118 of our nearly 4000 galaxies.

\subsection{Characterization of the new classes}\label{ssec:characterizingclasses}
\begin{figure*}%
    \centering
        \subfloat[The rotational support ($\lambda_*$) against the kinematic misalignment $\Psi_\mathrm{int}$.]{{\includegraphics[scale=0.28]{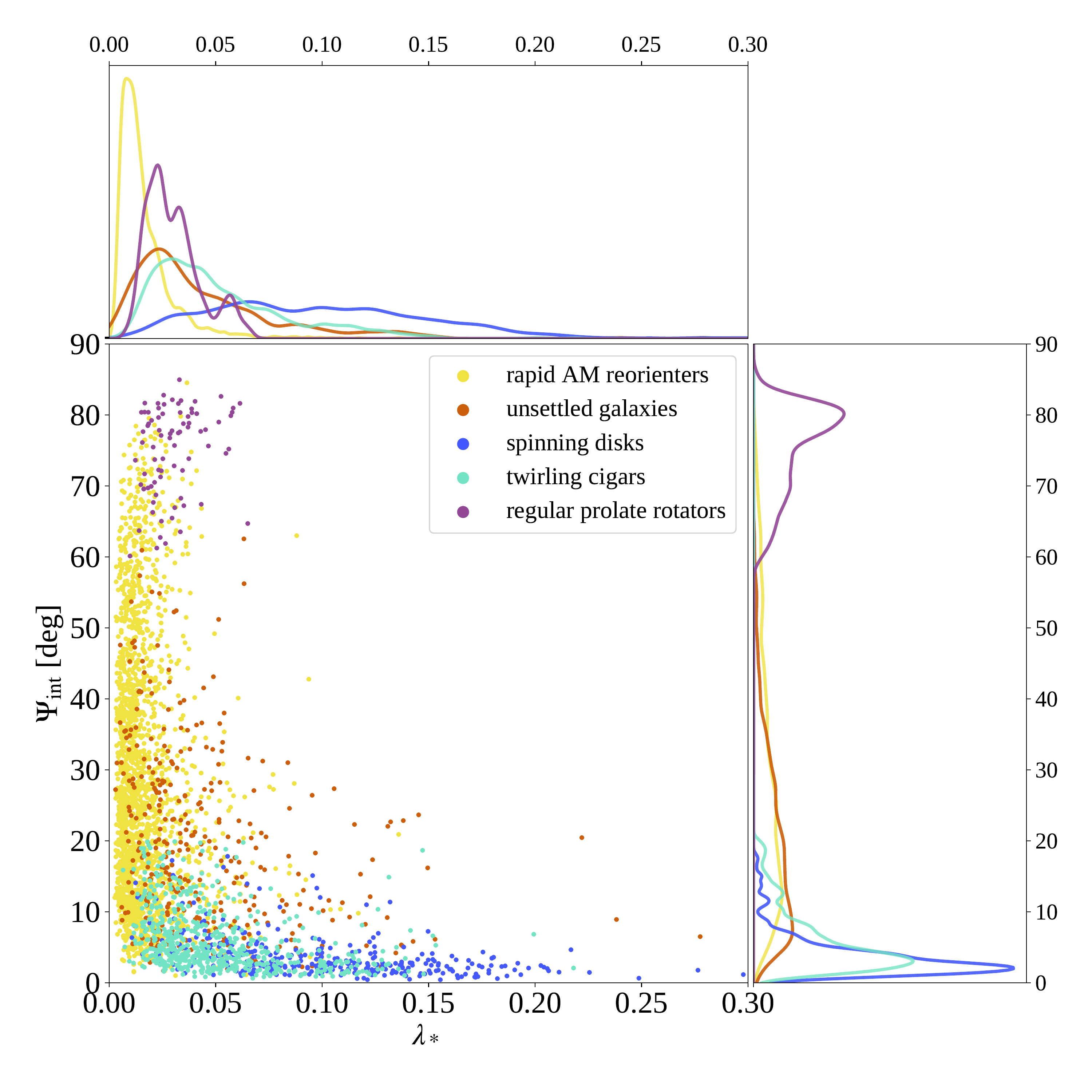} }}%
        \qquad
        \subfloat[The stellar mass against the total (baryonic and DM) mass of the group to which the galaxy belongs.]{{\includegraphics[scale=0.28]{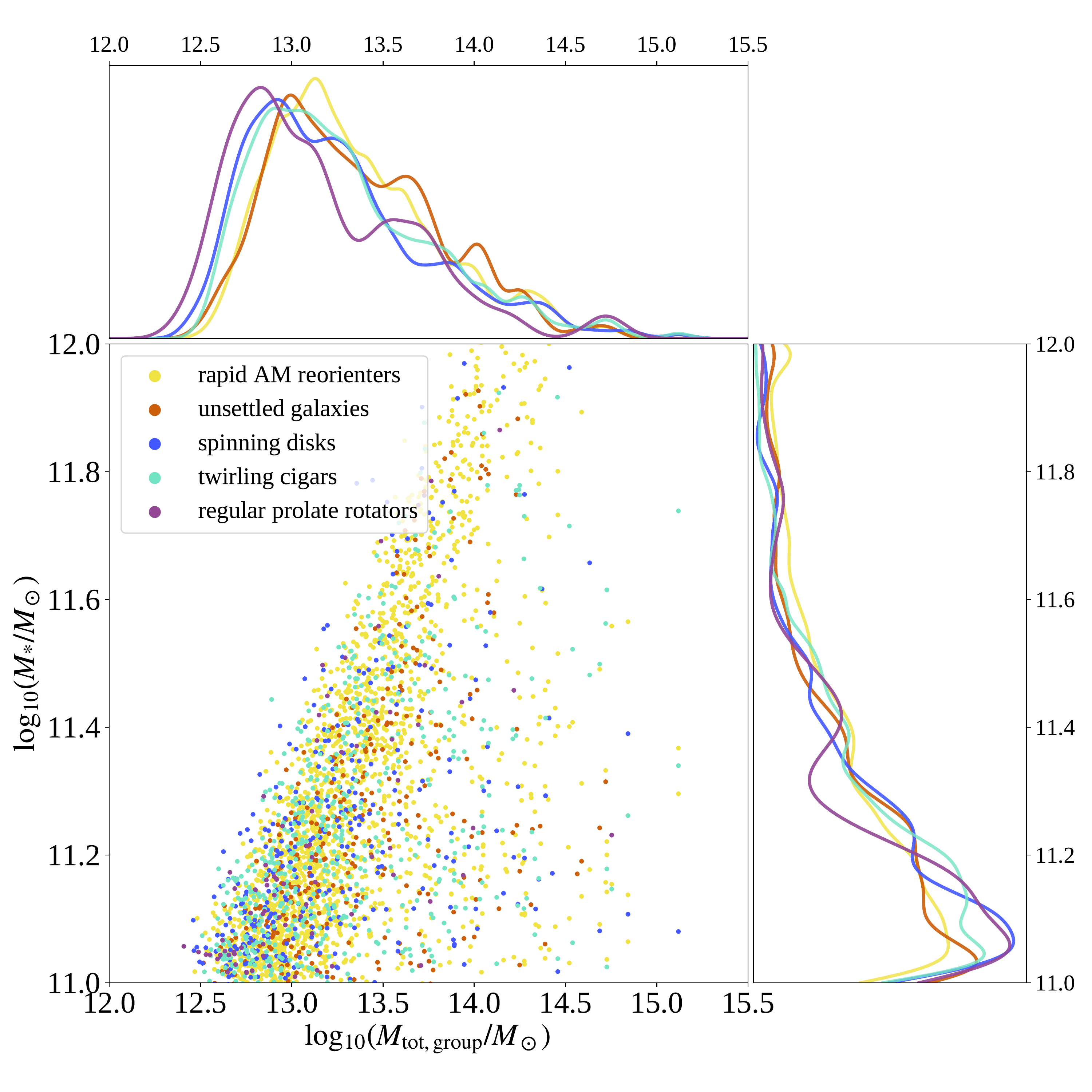} }}%
        \caption{The distribution of other parameters that characterize the classes (see Section \ref{ssec:def_classes}). Projections into 1D PDFs (using Gaussian kernel density estimation) for each quantity are shown aside each axis. Note that these distributions are normalized to the total number of galaxies within that class (so the regular prolate rotator line shows the density distribution for \textit{only} regular prolate rotators).}
        \label{fig:lam_star_M_dist}%
\end{figure*}

In this section, we investigate how galaxies with different properties are distributed amongst the classes. To do this, we analyze the aforementioned kinematic and morphological quantities in the context of our new classification. In Figure \ref{fig:lam_star_M_dist}a, we show the joint and marginal distributions of average (over a five snapshot window prior to $z=0$) stellar spin parameters and intrinsic misalignments measured for the galaxies meeting our stellar mass selection cut. Recalling the descriptions of sample galaxies from Section \ref{ssec:def_classes}, we see that broadly, the full class distributions follow many of the same trends. That is, we see that the spinning disks have the largest average rotational support, followed by the twirling cigars and regular prolate rotators. This is consistent with the understanding that low-triaxiality, flattened stellar distributions are generally supported by rapid, coherent rotation about the minor axis. We observe that the vast majority of twirling cigars and spinning disks have misalignments below $10^\circ$, again reflecting the sense of rotation about the minor axis. Though they are identified at large misalignments ($>60^\circ$) by definition, we find that the regular prolate rotators are naturally concentrated away from this misalignment boundary, with their distribution peaking near $\sim 80^\circ$. In fact, the overall misalignment has a somewhat bimodal shape – there are a significant number of galaxies at small misalignments and another, relatively smaller, group at large $\Psi_\mathrm{int}$. In the joint distribution, we see a clear correlation between the spin parameter and the misalignment. Namely, the majority of galaxies with large rotational support have misalignments below $10^\circ$. Those galaxies with low rotational support, however, exhibit a wide range of misalignments between $0^\circ$ and $90^\circ$. We find that our largest and most diverse class of galaxies, the rapid AM reorienters, are widely distributed in misalignment, instantaneously achieving states that extend to mirror the behavior of other classes. In this regard, their most defining characteristic (outside of the large AM reorientation) is the low rotational support – at all misalignments, they have smaller rotational supports than any of the other classes.

In Figure \ref{fig:lam_star_M_dist}b, we show the stellar and total halo mass distributions for our galaxy sample. We use the TNG Friends-of-Friends (FoF) group catalog to identify the group associated with each galaxy in our sample and report the total (baryonic and dark matter) mass of the group. The bounding `region' of the maximum stellar mass galaxy for a given group mass is identified with the mass of the central galaxy associated with each DM halo (i.e.~the most massive galaxy in the group). Any galaxies below that region are satellites. We find that in each of our classes, approximately 80\% of the galaxies are centrals. We see that the distributions of stellar and group masses are roughly the same for all of the classes, with the regular prolate rotators being the exception – they tend to have the smallest group and stellar masses.

\subsection{Stability of the classes}\label{sec:class_stability}
It is clear that these classes are distinguished morphologically and kinematically at the present day. This begs the question: are the class memberships stable and how do they evolve over time? To address this, we develop a measure that captures the tendency of the $z=0$ descendants of any selected subset of galaxies at a previous snapshot to have a distinct class distribution from that of the full $z=0$ sample. Specifically, we make selections based on class membership at various $z>0$, trace forward their evolution and compare the class distribution of their  $z=0$ descendants to that of our full sample.

\begin{figure*}%
    \centering
        \subfloat[rapid AM Reorienters]{{\includegraphics[scale=0.33]{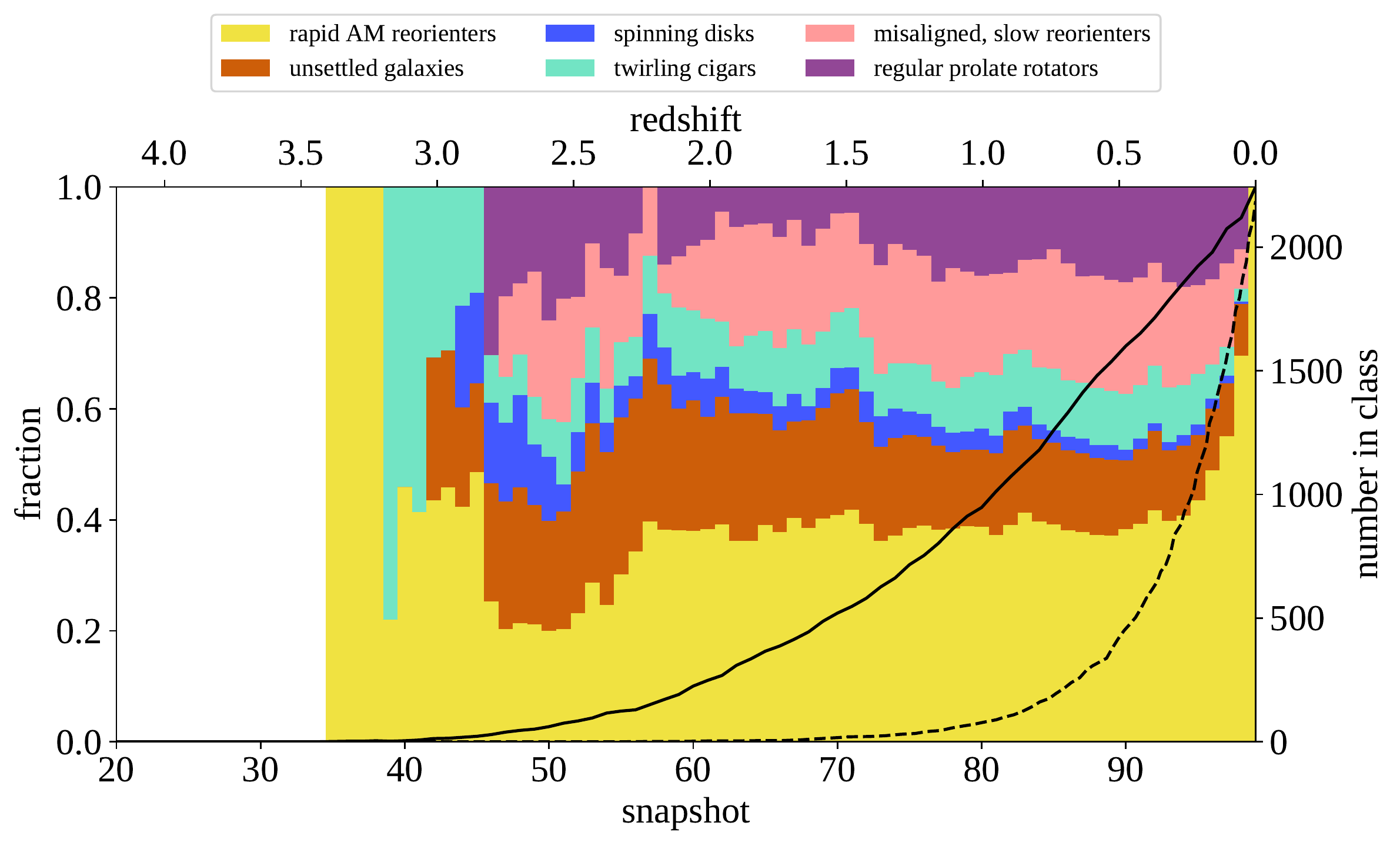} }}%
        \subfloat[Unsettled galaxies]{{\includegraphics[scale=0.33]{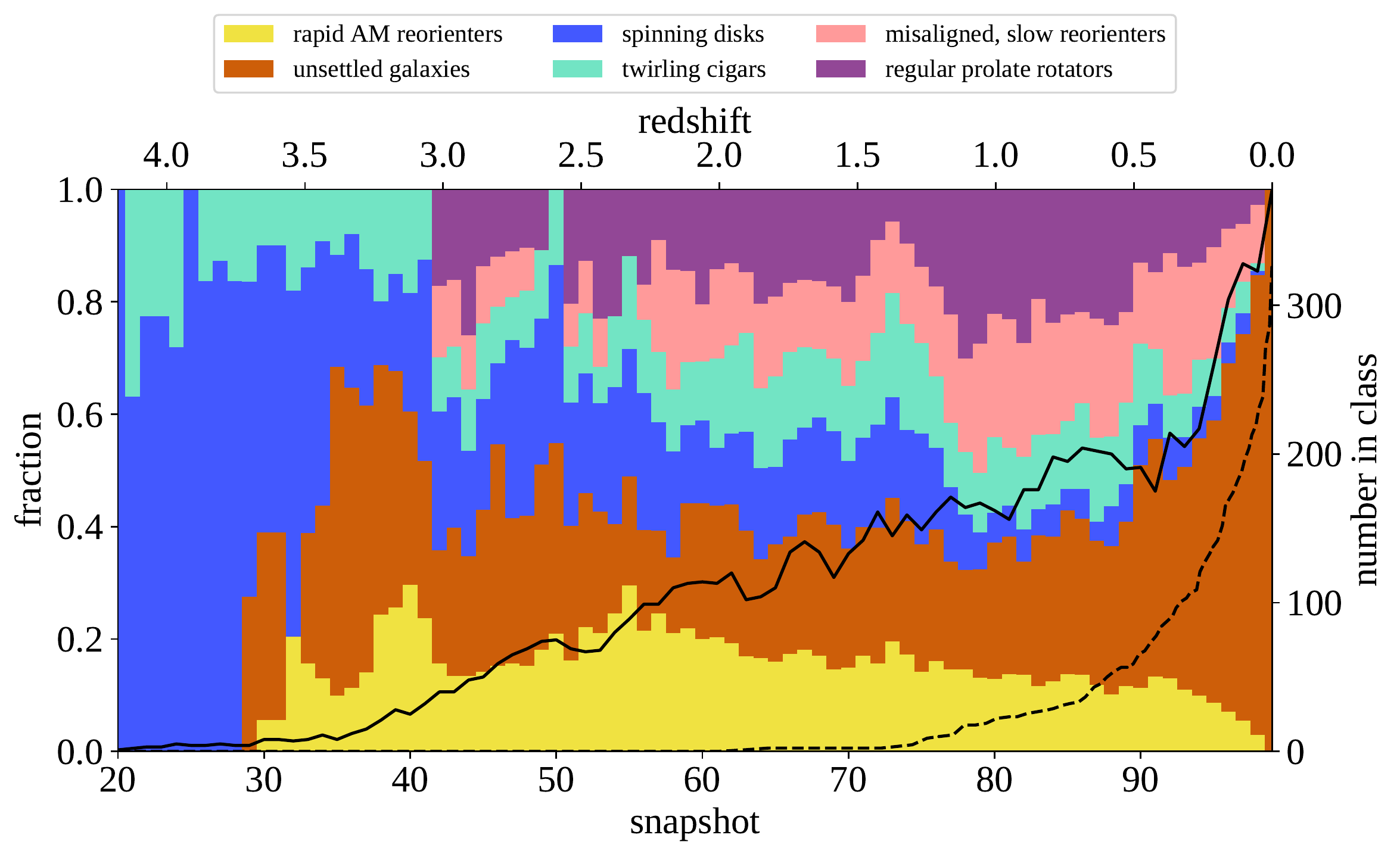} }}%
        \\
        \subfloat[Spinning disks]{{\includegraphics[scale=0.33]{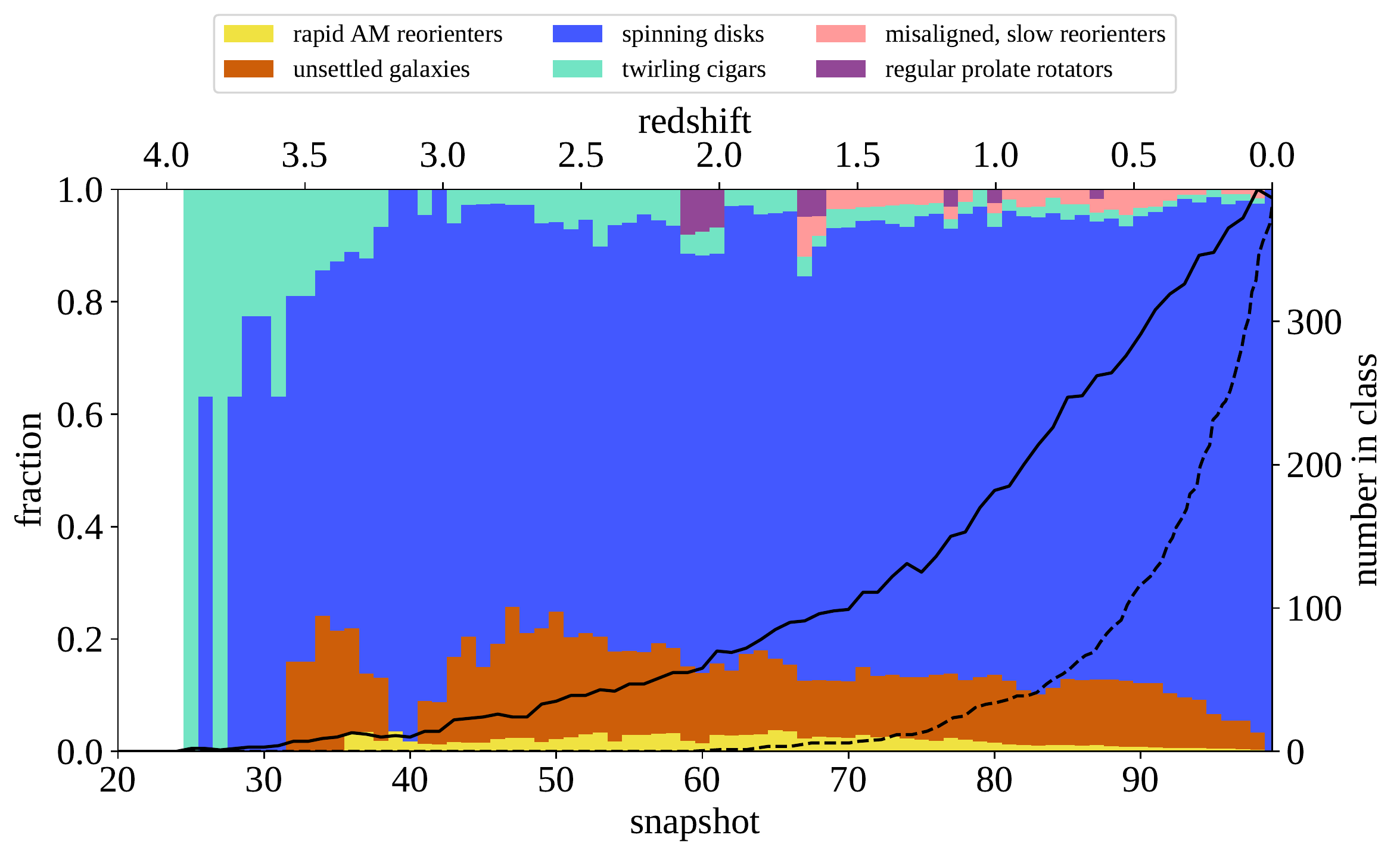} }}%
        \subfloat[Twirling cigars]{{\includegraphics[scale=0.33]{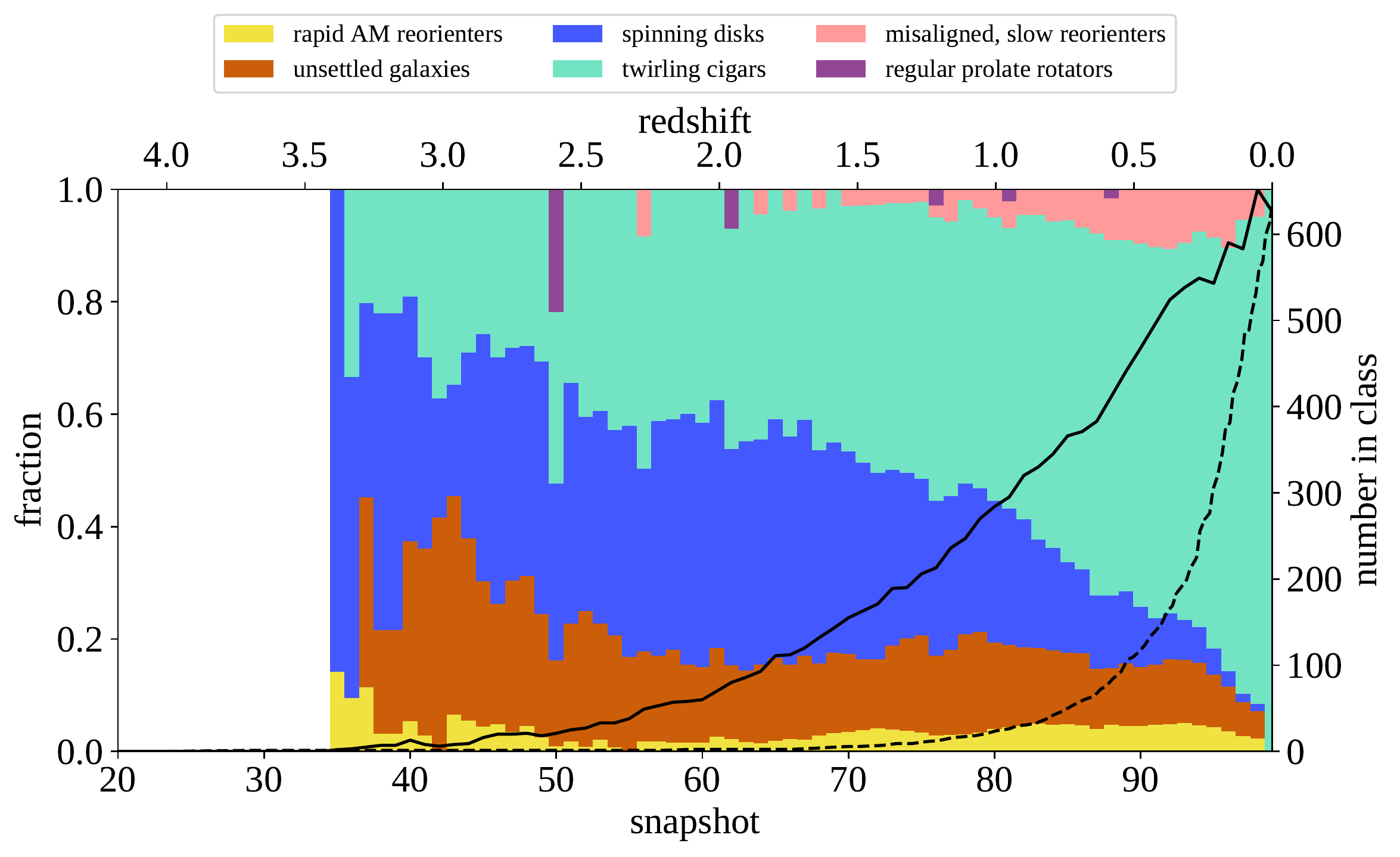} }}%
        \\
        \subfloat[Misaligned, slow reorienters]{{\includegraphics[scale=0.33]{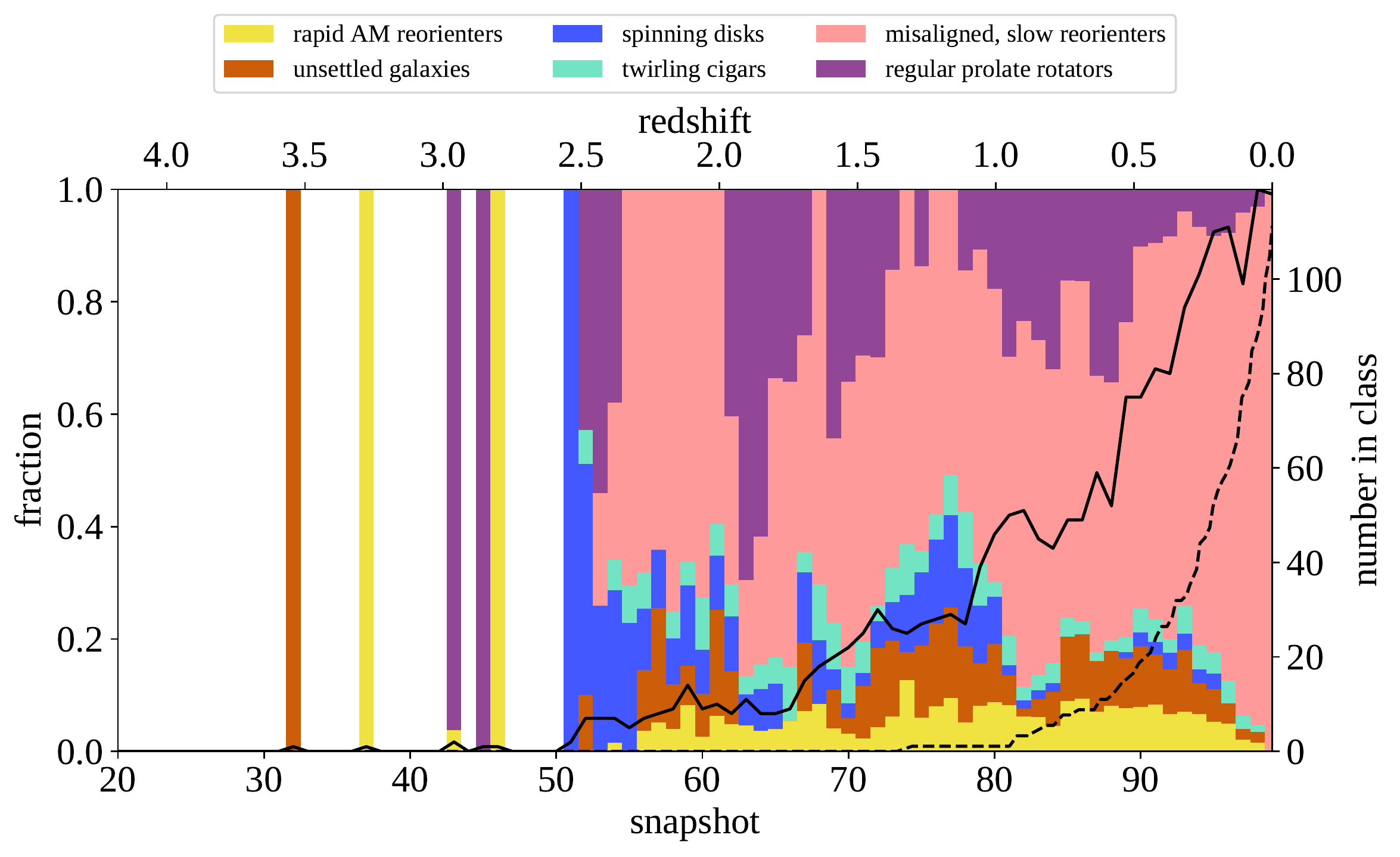} }}%
        \subfloat[Regular prolate rotators]{{\includegraphics[scale=0.33]{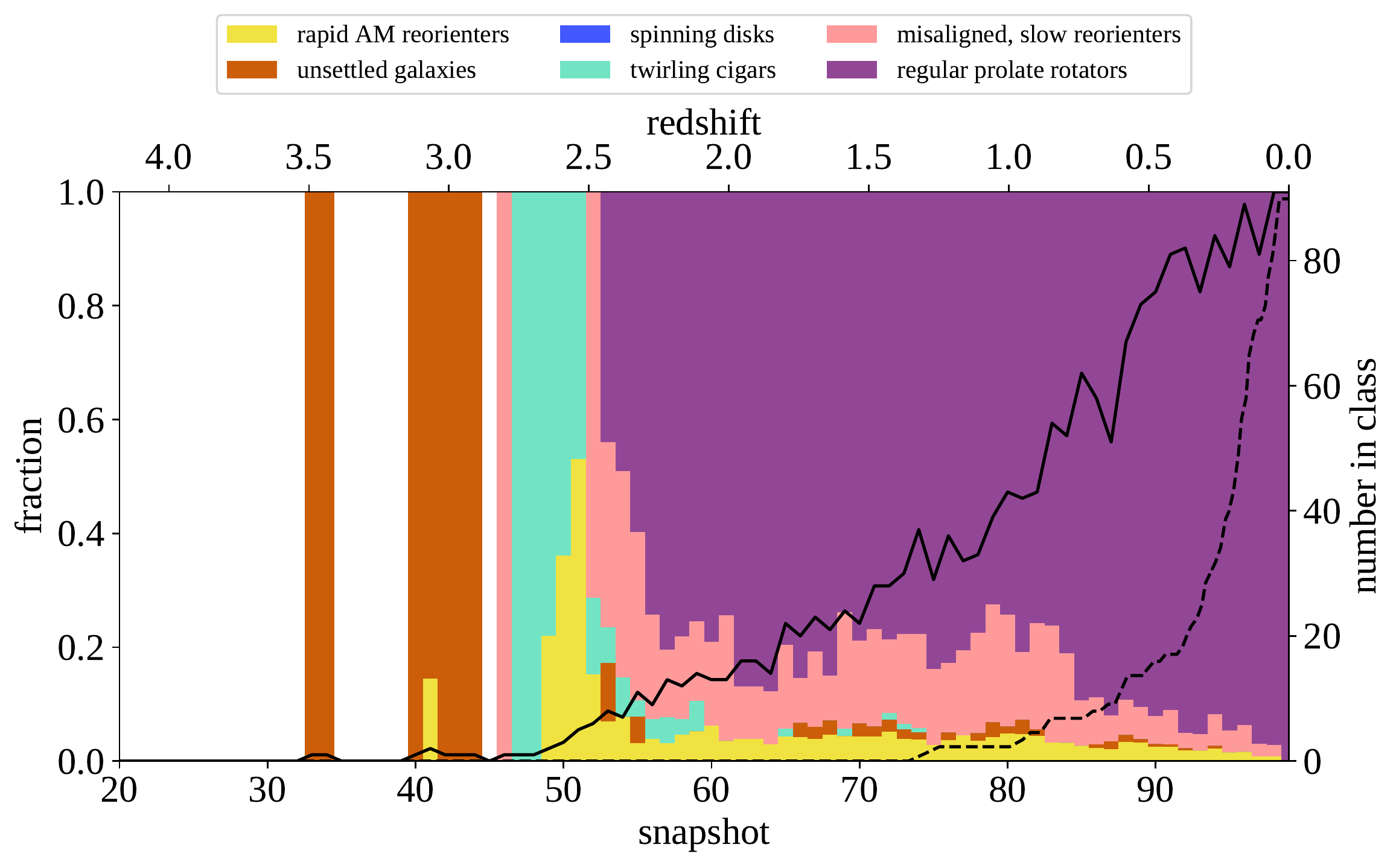} }}%
        \caption{Relative distribution of what is `observed' vs what is `expected' for galaxies of the different classes over time (snapshot; corresponding redshift is given on the top of the panels). For full procedure see Section \ref{sec:class_stability}. The black dashed curve indicates the fraction of galaxies in the current time class population that have already experienced their last major merger. The black solid curve indicates the number of galaxies that are classified as that specific class at the given snapshot (i.e.~corresponds to the number of galaxies used to generate the displayed bar). White regions correspond to times when there were no galaxies assigned that class designation. These panels display the subset of galaxies at any given time that are classified as that class \textit{and} have completed their most recent significant ($\mu>1/10$) merger.}
        \label{fig:class_bars}%
\end{figure*}

This measure can be best understood through an example, e.g.~the group of galaxies classified as regular prolate rotators (Figure~\ref{fig:class_bars}f) at $z=0.5$. With this group in hand, we look at their $z=0$ descendants and count the number of systems in each class, denoting this the `true' distribution of descendants from this subset. Then, we produce an `independent' distribution by scaling the $z=0$ distribution of galaxies (i.e.~58\% rapid AM reorienters, 9.8\% unsettled, etc.) to the total number of galaxies in the $z=0.5$ subset in question. Dividing these two sets of numbers gives us a set of six quantities (one for each class) that indicate the similarity of the `true' distribution of descendant classes to the `independent' distribution – i.e.~a ratio of unity for any given class indicates that the true fraction of descendants belonging to that class is the same as in the full $z=0$ population.
The sum of these six ratios is not necessarily 6
and to account for this we normalize by the sum of all the ratios.
Hence, values of $1/6\sim 0.17$ for this normalized ratio indicate perfect correspondence between the true distribution and the independent distribution.
This yields two extreme cases: (i) all ratios being equal and (ii) one ratio dominating. If (i) all the ratios are roughly equal (and are thus $\sim$0.17), then these galaxies are transitioning from the selected $z>0$ class to other classes with no apparent preference. If instead (ii) one ratio dominates,
then the galaxies of the selected class display a \textit{strong} preference towards transitioning to or remaining in the class with the dominant ratio. 

The results are presented in Figure~\ref{fig:class_bars}. For every snapshot, we display these ratios as a stacked set of bars colored with the same class colors as in Figure \ref{fig:class_definition}. Put simply, this means that, for a given class (say regular prolate rotators), significant swaths of the corresponding color dominance (purple) over several snapshots indicate a strong tendency to maintain that class' behavior over time (i.e.~the class is stable). If the bars display roughly equal segments for each color, then the class is relatively unstable and has a tendency to transition to the other classes.

The final step in generating the panels in Figure~\ref{fig:class_bars} is selecting which galaxies to analyze. To understand the stability of a class with this analysis, it is useful to exclude the galaxies that are yet to experience a violent, disruptive event such as a major merger. Therefore, we specifically inspect the subset of galaxies (at any given time) that have already experienced their final `significant' merger (defined as having a mass ratio $\mu > 1/10$), so that we isolate class transitions via secular transitions from those that are due to the effects of a merger. Note that the forthcoming Section~\ref{sec:merg_history} addresses the complementary selection, namely the effects of mergers.

Interpreting Figure~\ref{fig:class_bars}, the dominance of a class' bar over time indicates that the assigned class is a stable classification over time. In other words, in the absence of minor or major mergers, we are repeatedly observing a tendency for galaxies classified as a class at a given time to still be that class at $z=0$, and are more likely to be classified as such than the average galaxy randomly drawn from the distribution. This is most clearly observed for the spinning disks (panel c) – following their most recent significant merger, they tend to remain spinning disks through to the present day. This is intuitively consistent with the understanding that these flattened, rotationally supported stellar disks have regular, stable behavior over time. We see similar behavior within the twirling cigar class (panel d), with the only difference being a tendency to transition to the spinning disk class. We note that this is the only instance in which a given class displays a tendency to transition to a particular other class. This corresponds to a transition across the triaxiality boundary we use to define the classes, and physically may be caused by a flattening of the stellar distribution into a diskier shape over time. This could also indicate the presence of a bar embedded within a larger disk, with the prolate shape weakening over time. The regular prolate rotators exhibit similar stability to the spinning disks, indicating that this too is a stable configuration in the absence of mergers. We also note that a small fraction of the spinning disks and twirling cigars demonstrate a tendency to become unsettled galaxies, potentially indicative of a mini merger (mass ratio $<1:10$) or other environmental interaction driving the transition across the fast AM reorientation rate boundary.

The most inter-class transitions are observed for the unsettled galaxies, where we see a near-uniform distribution of classes over time, as we expect. That is, under the assumption that these galaxies are temporarily unsettled from their equilibrium states, by $z=0$, they should transition to another class (either they may return to the pre-merger state or `reset' and move to a new class). Given that such an unsettling is equally likely for all galaxies/classes, this produces the spread that we observe in the unsettled galaxy panel. Finally, the rapid AM reorienters show reasonable stability but also demonstrate transitions to the other classes.

In sum, we find that, generally, our new class definitions are stable in the absence of significant mergers. The key exception are the unsettled galaxies, which live up to their name and appear to be in a transition state that is a temporary deviation from the equilibrium behavior of the galaxies. Beyond that, however, class stability is most reliably seen for the spinning disks, twirling cigars, and regular prolate rotators, which all display strong tendencies to maintain their classification. 

\section{Origin of the Classes}\label{sec:merg_history}
Equipped with a characterization of the different classes at the present day and a sense of how they evolve over time, we next investigate their origin stories: how did these classes come to be? 

\subsection{Mergers set the classes}\label{ssec:merg_overview}
As a first step, we demonstrate that the most recent significant merger (with mass ratio $\mu > 1/10$) is the single most consequential event for producing the observed distinctions between the classes. We demonstrate this through a number of arguments. First, we have inspected the distribution of parameters, such as triaxiality and misalignment, prior to the most recent significant merger (measured at the time of the satellite's maximum mass, which roughly corresponds to the point just before the interaction significantly begins) and found that the distribution is roughly the same across all the classes. In other words, the progenitors of galaxies of different classes are fairly indistinguishable in their global properties before the merger. 

The second piece of evidence comes from the results of a similar procedure to that outlined in Section~\ref{sec:class_stability}. That is, we reproduced that calculation, tracing forward the evolution of galaxies in a given class from each snapshot to the present day universe, but now focusing just on the subset of galaxies that experience a significant merger between that time and $z=0$. This isolates the effects of these significant mergers on the tendency of the galaxies in each class to maintain, or change, their nature. The results of this analysis (not shown) confirm the supposition that the mergers are the most important defining characteristic of the classes: the distributions were nearly uniform (similar in appearance to Figure~\ref{fig:class_bars}b) for every class. This means that, among galaxies selected at a given time that experience a significant merger later in their evolution, any two galaxies are roughly equally likely to end up in any of the classes, irrespective of their classification at the earlier selection time. 

\begin{figure}
    \centering
    \includegraphics[scale = 0.43]{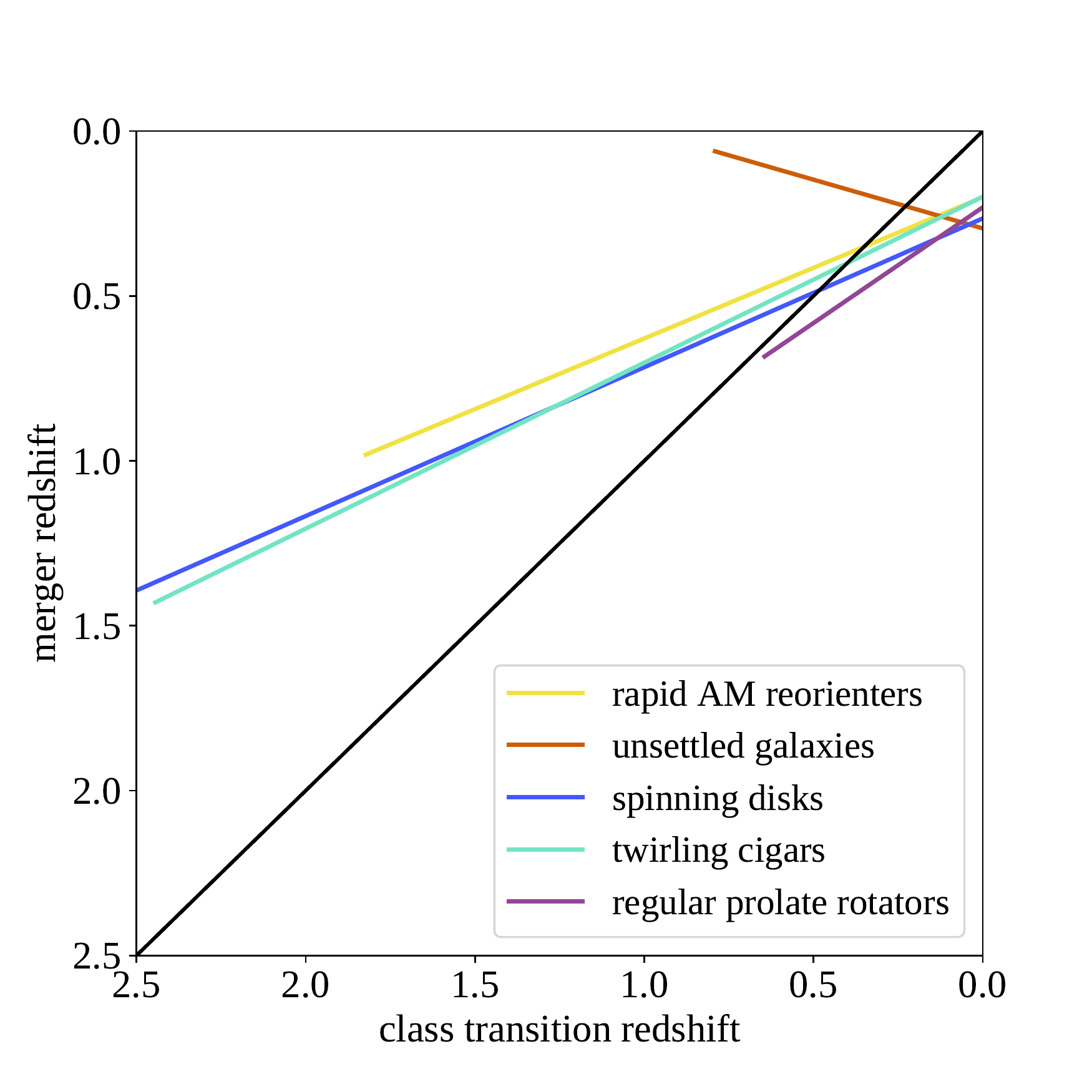}
    \caption{The time at which each class most recently achieved its $z=0$ classification against the time of the most recent significant ($\mu>1/10$) merger. The line of best fit for each class is displayed with the colors defined in Section~\ref{ssec:def_classes}. The black solid line indicates 1-1 correspondence between the times.}
    \label{fig:class_transition_time}
\end{figure}

An additional piece of evidence for this conclusion can be seen in Figure~\ref{fig:class_transition_time}, where we show the correlation between the time when each galaxy achieved its $z=0$ class and the time of the last significant merger, broken down by class. In particular, we measure the most recent time at which each galaxy transitions into its present-day classification, and display the best-fit linear correlation with the time of its most recent significant ($\mu>1/10$) merger. 
We see that the classes, with the exception of the unsettled galaxies, roughly follow the 1-1 correspondence, with the regular prolate rotators (purple) demonstrating the strongest relationship between the time of merger and transition time. 
Interestingly, the unsettled galaxies (red solid line), follow a distinctly different trend from the rest of the classes and the 1-1 line (given in black). This is evidence that these galaxies are, as we have noted previously, unsettled, in some cases due to events other than significant mergers. However, aside from this particular class, it is clear from that it is primarily some property of the merger that defines the subsequent kinematic and morphological behavior of the classes moving forward.

\begin{figure}%
    \centering
        \subfloat[The mass ratio of the merger]{{\includegraphics[scale=0.43]{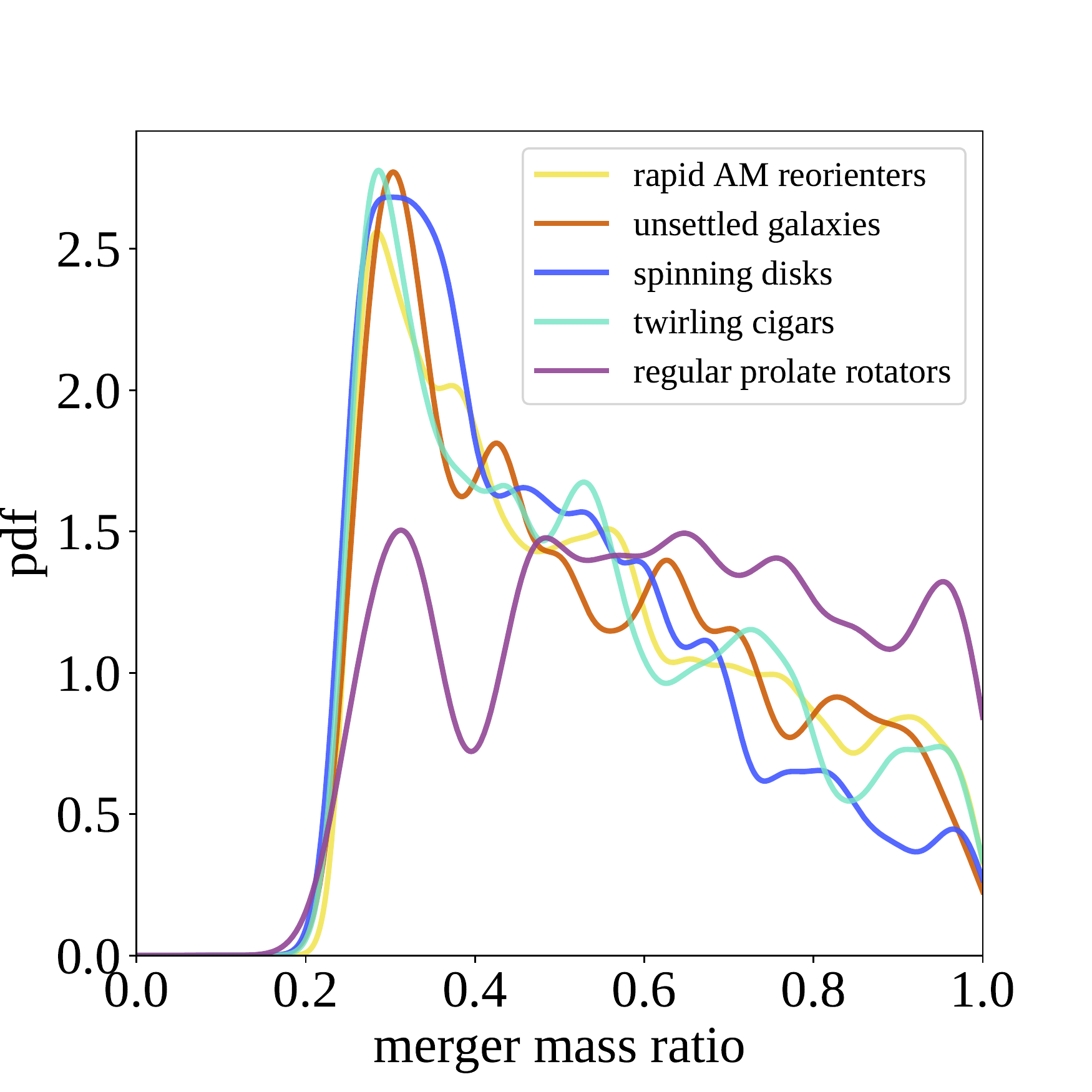} }}%
        \qquad
        \subfloat[The time of most recent significant merger.]{{\includegraphics[scale=0.43]{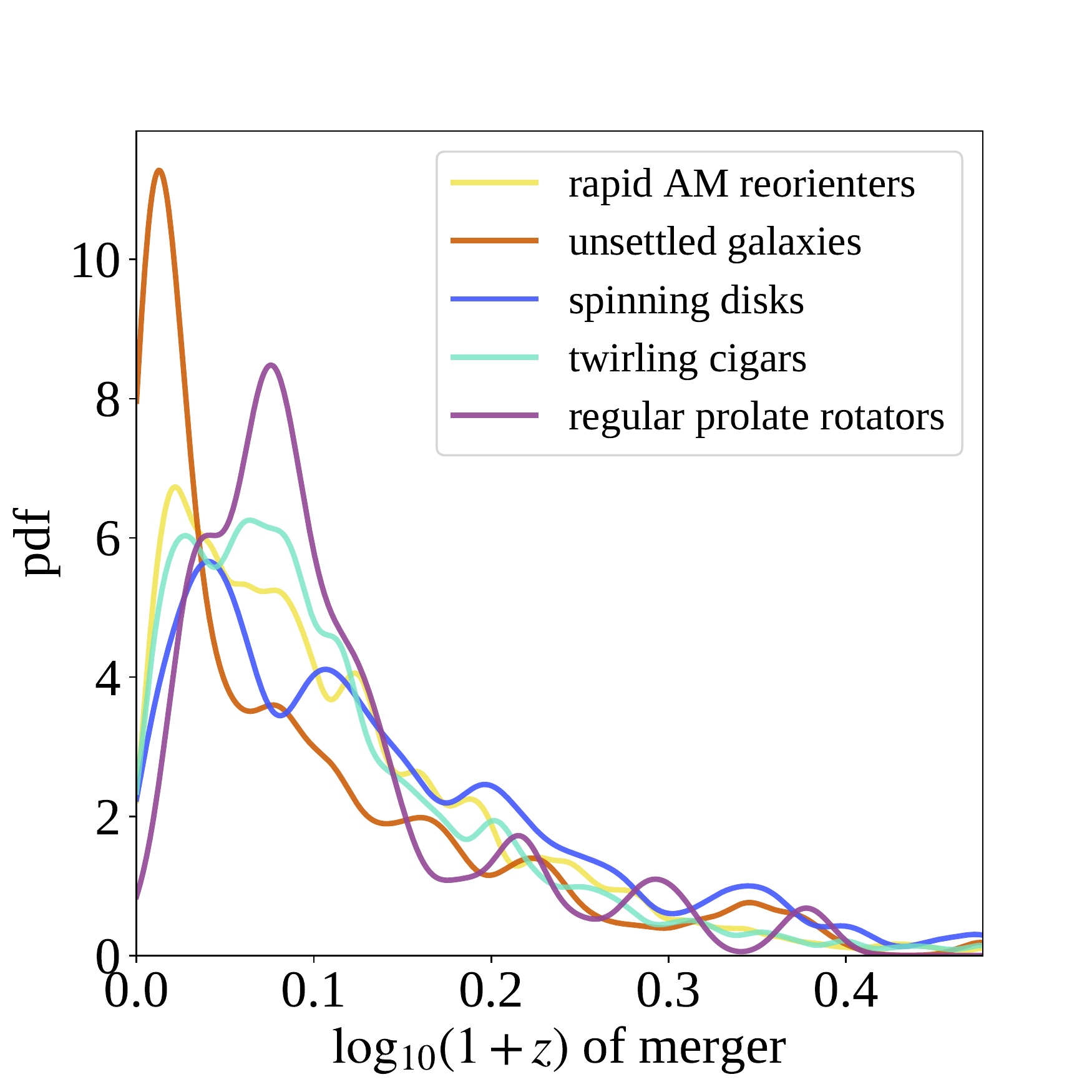} }}%
        \caption{The distribution of basic merger descriptors separated by the different classes. These 1D PDFs are generated using Gaussian kernel density estimation and are normalized to the total number of galaxies within that class.}
        \label{fig:merger_dist}%
\end{figure}

Having established the most recent merger as the primary event setting the galaxy class, we turn to a more careful examination of those mergers.
As a general summary of the merger properties, we see in Figure~\ref{fig:merger_dist} that all classes show roughly similar distributions of merger times and mass ratios, with two notable exceptions. The first exception is the tendency of the regular prolate rotators to have more major mergers, while the second is the tendency of the unsettled galaxies to have experienced their mergers more recently.

\begin{figure*}
    \centering
    \includegraphics[scale = 0.205]{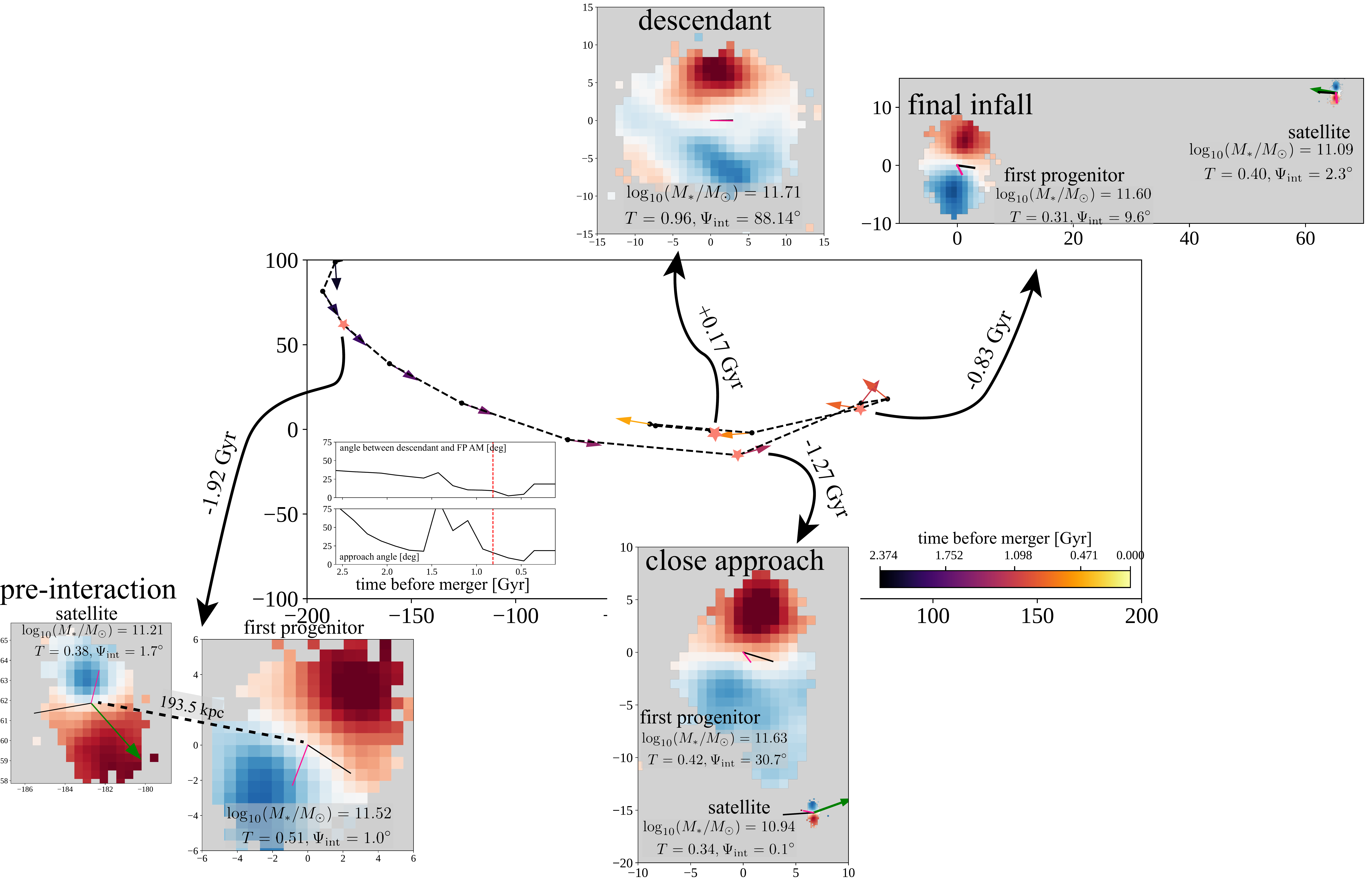}
    \caption{The evolution of the satellite's orbit over the few gigayears prior to the merger, with positions and maps oriented in the orbital frame of the merger (i.e.~with the orbital AM of the satellite pointing out of the page). Arrows at each snapshot represent the direction of the satellite's instantaneous velocity, relative to the first progenitor's velocity, and are colored according to the time before the merger of each snapshot. Each galaxy's pixel map is produced according to the procedure outlined in Section \ref{ssec:class_exemplars} and the internal AM, major axis, and velocity vector directions are given in black, pink, and green, respectively. We present the snapshots of the progenitor states at $t=1.92$ Gyr pre-merger (lower left, panels separated for visual purposes), $t=1.27$ Gyr pre-merger (lower middle), and $t=0.83$ Gyr pre-merger (upper right). We also present the velocity map for the descendant galaxy (at $t=0.17$ Gyr after the merger, leftmost panel) in this projection. These times are given relative to the time of merger and are marked along the arrows overlaid on the Figure, so negative values correspond to times before the merger and positive to after. The points along the trajectory corresponding to each of these selected snapshots are identified with pink-orange stars. Supplemental information about the progenitors is given in each panel – we provide the stellar masses $M_*$, triaxialities $T$, and misalignments $\Psi_\mathrm{int}$ of the progenitors. Inlaid in the lower-left of the central panel, we present the evolution of the angle between the internal AM direction of the descendant galaxy and instantaneous first progenitor (upper) and the approach angle $\theta$ (lower).}
    \label{fig:merg_panel_ex}
\end{figure*}

\subsection{A new approach to analyze the mergers: the final infall}\label{ssec:final_infall}
We find that direct, visual analysis of the orbits of the merging progenitor galaxies reveals a picture that is not well captured by analysis of their global properties well before the merger. To see this, it is most useful to inspect a specific case – in Figure~\ref{fig:merg_panel_ex}, we present a few selected snapshots over the 2.5 Gyr prior to the merger that produces the exemplar regular prolate rotator discussed in Section~\ref{ssec:class_exemplars} (see Figure~\ref{fig:prolate_rotator_PSR_mis_map}b). The central panel of the figure captures the trajectory of the satellite during this time relative to the position of the first progenitor (the galaxy identified on the main progenitor branch in the \texttt{SubLink} merger tree catalog). The position of the infalling satellite is presented in the orbital plane of the merger from the first displayed point and is centered by the position of the first progenitor at each time. The orbital AM vector is roughly consistent throughout the merger so this is a reasonable projection from which to visually analyze the dynamics. Within the juxtaposed panels that show the visual state of the systems at certain points in time, we identify the internal AM (black), major axis (pink), and orbital velocity (green) vectors for each of the progenitors and overlay those atop the pixelized velocity maps.

Specifically, in Figure~\ref{fig:merg_panel_ex} we present the two progenitor systems at a point relatively early in the merger (before a significant interaction but after the first fly-by; i.e.~at $t=1.92$ Gyr before the merger – the lower left pair of panels), at the time of a close approach (at $t=1.27$ Gyr before the merger – the bottom center overlaid panel), and at the start of the final infall (at $t=0.83$ Gyr before the merger – the rightmost overlaid panel). We also present the velocity map for the descendant galaxy (at $t=0.17$ Gyr after the merger – the uppermost panel). We have chosen to separate the depiction of the two progenitors into individual velocity maps for the first of these snapshots for visual purposes (i.e.~at this scale, they would be too small to see otherwise). We generate these velocity maps using the method outlined in Section~\ref{ssec:class_exemplars} and all are displayed in the orbital plane of the merger (i.e.~with the orbital AM of the satellite pointed out of the page). In the lower left of the central panel, we present two subplots that display the evolution of the angle between the internal AM of the descendant galaxy (at $t=0.17$ Gyr post-merger) and that of the instantaneous first progenitor (top) and the approach angle $\theta$ (bottom), which is the angle between the position and velocity vectors of the satellite galaxy and roughly corresponds to the circularity of the orbit (see \citealp{Zeng2021}). For the approach angle, values closer to $0^\circ$ correspond to radial orbits and values closer to $90^\circ$ correspond to more circular orbits. Within these two panels, we mark the time of final infall (the time of the second-to-last local maximum in the 3D distance between the two progenitors; $t=0.83$ Gyr pre-merger) with a red dashed line. We select the second-to-last local maximum because, generally, the last maximum occurs during the final coalescence, when the properties of the progenitors are ill-defined (e.g.~this time would be in the final $\sim 0.3$ Gyr in Figure~\ref{fig:merg_panel_ex}, when the final coalescence is clearly underway). All the angles in Figure~\ref{fig:merg_panel_ex} are reported in degrees.

At early times, well before the merger (see $t=1.92$ Gyr pre-merger), we see that the first progenitor galaxy has a strong sense of rotation and indeed this progenitor galaxy is classified as a spinning disk during this time (see Figure~\ref{fig:spinning_disk_twirling_cigar_map}a for an example of that class). Specifically, this means that this galaxy is a fast rotator, has a low triaxiality, and a very small, well-defined misalignment. However, it is not until the infalling satellite makes a close passage (at around 1.27 Gyr before the merger) that the main progenitor's angular momentum is nudged into a state that is most similar to the descendant galaxy. In this panel, we see the disturbance of the main progenitor by the proximity of the satellite and the internal AM is at a transition stage between the pre-interaction point and the descendant state. In other words, after this interaction (see first progenitor projection at $t=0.83$ Gyr before the merger; the time of final infall), we see that the internal AM of the main progenitor galaxy is oriented in roughly the same direction (approx. $15^\circ$ difference) as the ultimate internal AM of the descendant galaxy ($t=0.17$ Gyr after the merger). Indeed, this strong alignment of the two vectors (and lack of alignment between the orbital AM and the AM of the descendant galaxy) indicates that it is the strong rotational support and internal AM of the main progenitor that defines the AM direction of the resulting descendant. For the final 0.8 Gyr, the last meaningful `infall' of the satellite galaxy, we see that this interaction has also altered the shape of the satellite's orbit so that the infall is occurring on a roughly radial trajectory (with approach angles $\theta<30^\circ$ and steadily decreasing). This means that the direction of the separation vector between the two galaxies is fairly consistent over the course of those snapshots, and, we can see that it is increasingly well aligned with the major axis direction of the descendant galaxy (i.e.~compare the separation vector of the two galaxies at $t=0.83$ Gyr pre-merger with the pink line in the descendant panel, at $t=0.17$ Gyr post-merger).

With these two pieces in hand, the picture is clear: this regular prolate rotator galaxy initially formed as a spinning disk, with a high degree of rotational support and minimal misalignment ($\Psi_\mathrm{int}\sim 0^\circ$). For several gigayears prior to the merger, there was little change in this behavior and the observed early time (pre-interaction) behavior of this galaxy was not meaningfully well correlated with that of the descendant. However, following the interaction at 1.27 Gyr before the merger, the satellite was shifted into a more radial orbit and the stage was set for the regular prolate rotator to be born. From this point to final coalescence, we see that the satellite's orbit loops around the main progenitor galaxy orthogonal to its internal plane of rotation (roughly aligned with the internal AM direction). As a result, this previously disky galaxy is elongated along its minor axis direction but, because it began with a good degree of rotational support, it maintains its core sense of rotation. Therefore, it becomes a regular prolate rotator, with an ultimately well defined alignment of the major axis and internal AM directions that is steady over time. 

Inspecting a few randomly selected galaxies, we find that such behavior is not uncommon within the regular prolate rotator population, and demonstrates that it is the interaction and subsequent `final infall' (e.g.~the final 0.83 Gyr in Figure~\ref{fig:merg_panel_ex}) that are most important to the ultimate behavior of the descendant galaxy, especially for the regular prolate rotators. Therefore, we focus our analysis on the beginning of this final infall, identified as 0.83 Gyr before the merger for the depicted case. 

\begin{figure*}
    \centering
    \includegraphics[scale = 0.5]{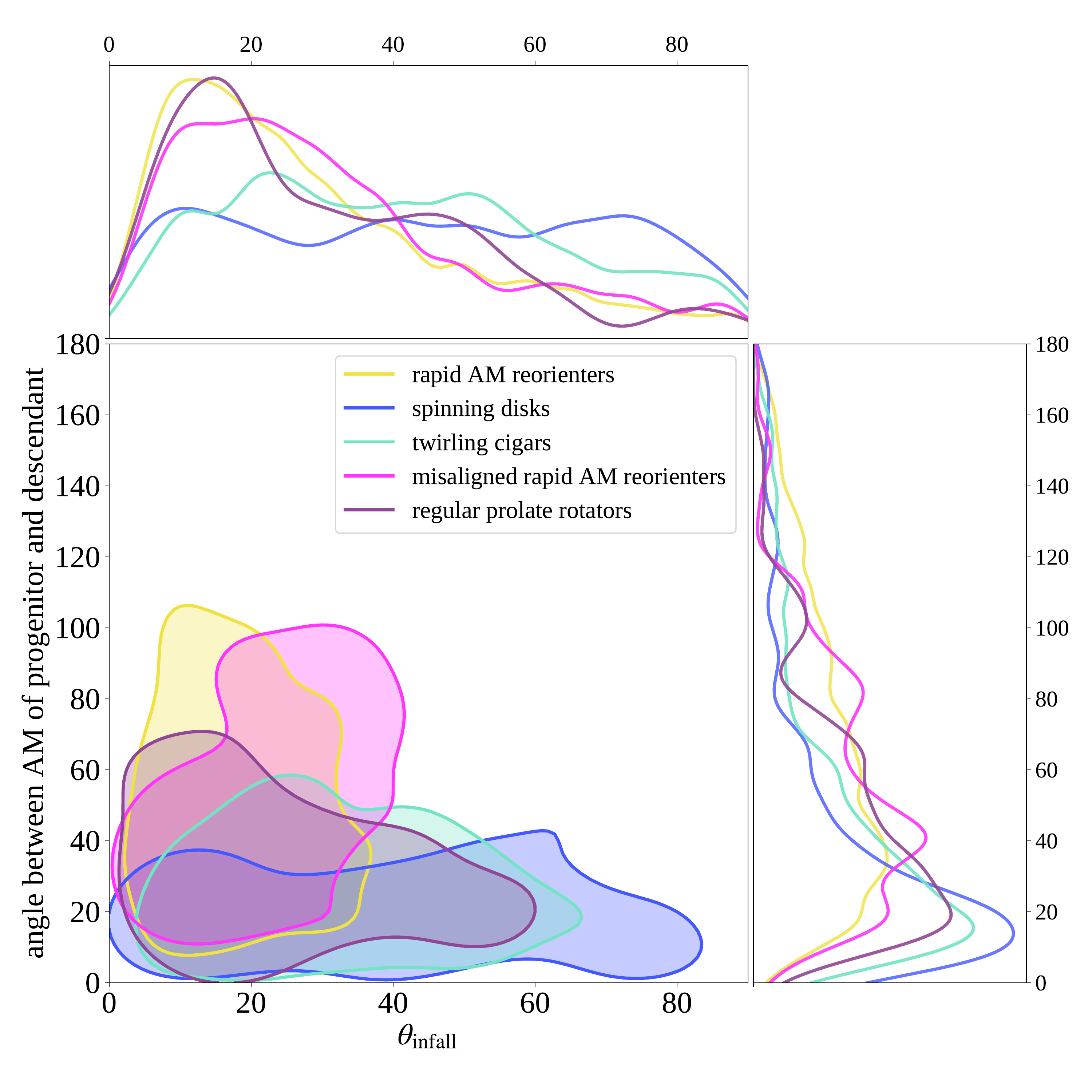}
    \caption{The approach angle against the angle between the internal AM of the first progenitor and that of the descendant, separated by class and measured at the time of final infall. Projections into 1D PDFs (using Gaussian kernel density estimation) for each quantity are shown aside each axis. Note that the pink curves display the distribution for a subclass of the rapid AM reorienters and these distributions are normalized to the total number of galaxies within that class (so the regular prolate rotator line shows the density distribution for \textit{only} regular prolate rotators). 2D PDFs are presented in the central panel and are estimated in the same way, with the displayed contour representing 85\% of that class' data.}
    \label{fig:AM_ang_app_ang_infall}
\end{figure*}

With this new perspective in hand, we can analyze some slices of the overall distribution of properties measured before and after the merger. For example, in Figure~\ref{fig:AM_ang_app_ang_infall}, we present the angle between the internal AM vectors of the first progenitor and of the descendant (along the ordinate), a quantity that should be a good indicator of the degree to which internal rotation is upset by the merger dynamics. Here, we see that, as expected, the angle is small for the spinning disks and twirling cigars, consistent with the findings that these galaxies are experiencing relatively smaller mergers and already have a strong sense of rotation. The regular prolate rotators, however, also have small angles here, with a peak near $25^\circ$, indicating that they too maintain the internal AM direction of the progenitor when this is measured relatively close to the point of merger. Along the abscissa of Figure~\ref{fig:AM_ang_app_ang_infall}, we present the approach angle (angle between the position and velocity vectors of the satellite) at this time of final infall. The distinction between classes is again clear – we see that the regular prolate rotators distribution is peaked at small angles (indicating a radial infall), while the spinning disks (and twirling cigars, to a lesser degree) have a larger spread in this distribution, corresponding to more circular orbits. Together, we see that the prolate rotators are well clustered at small values of both these angles, while the other classes tend to demonstrate extension along one of the axes.

\begin{figure}%
    \centering
        \subfloat[The magnitude of the internal AM of the first progenitor galaxy.]{{\includegraphics[scale=0.43]{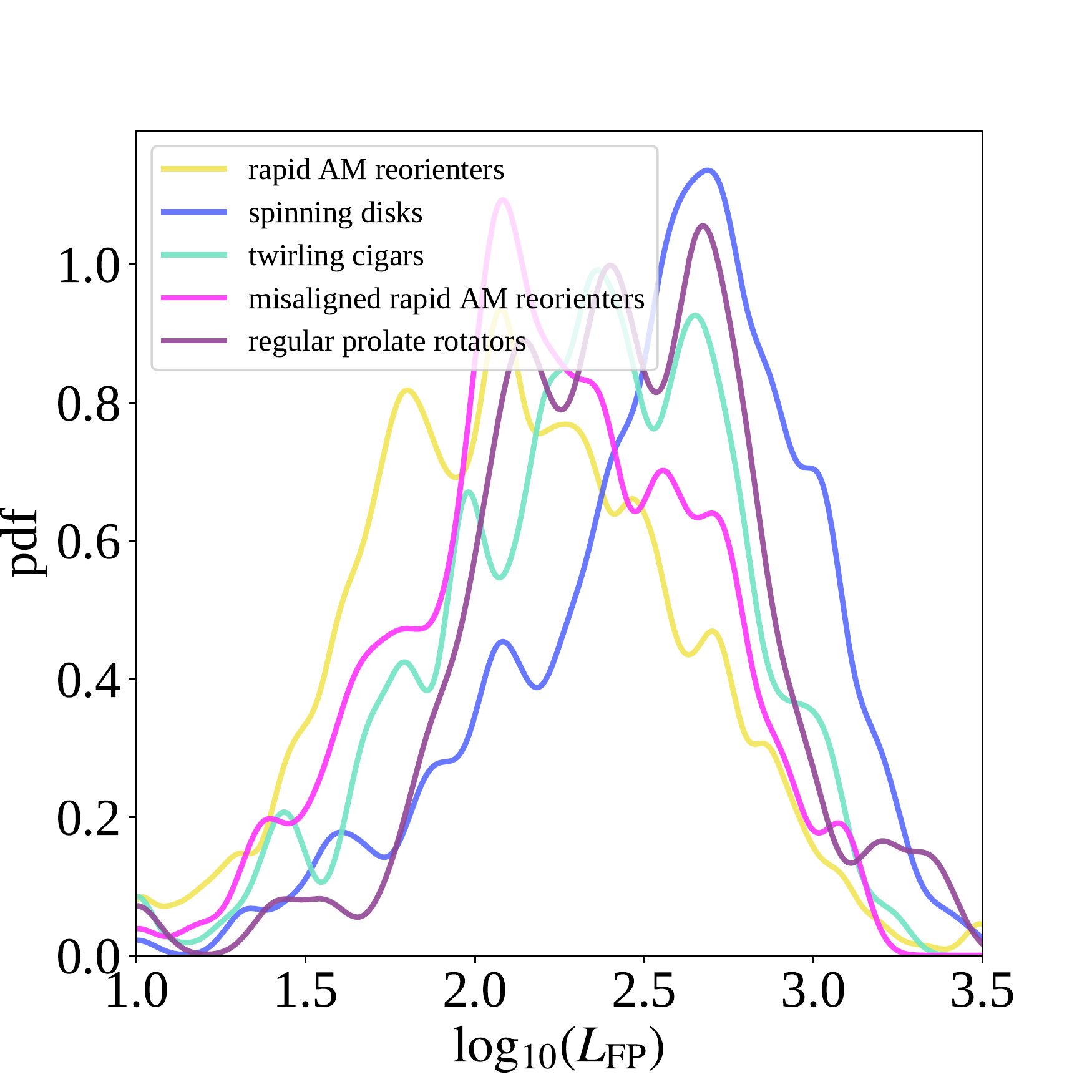} }}%
        \qquad
        \subfloat[The magnitude of the orbital AM of the merger.]{{\includegraphics[scale=0.43]{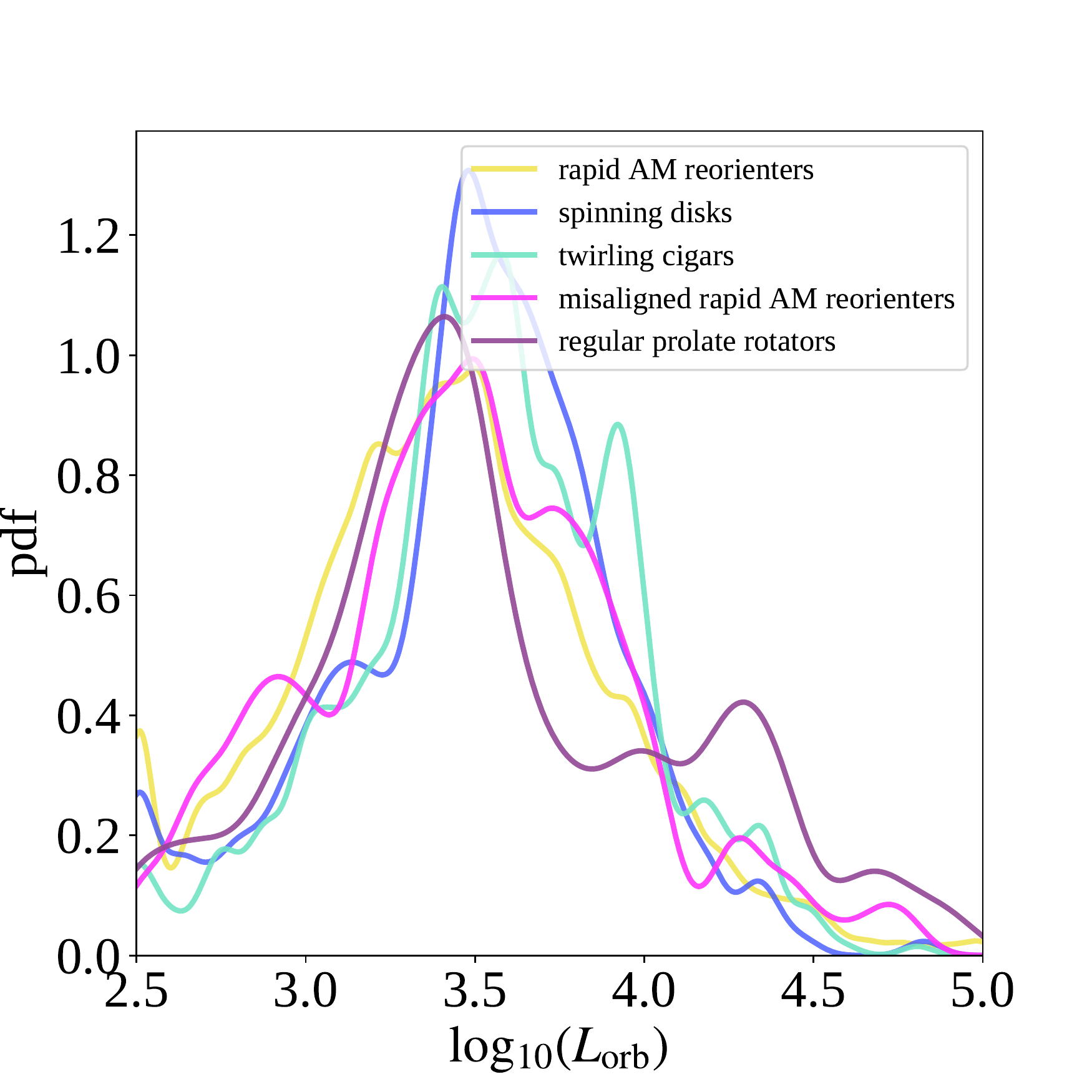} }}%
        \caption{Two slices of the distribution of pre-merger parameters measured at the time of final infall (see Section \ref{sec:merg_history}), separated by class. Note that the pink curve displays the distribution for a subclass of the rapid AM reorienters.}
        \label{fig:fp_AM_mag_orb_AM_mag}%
\end{figure}

These conclusions are further supported by Figure~\ref{fig:fp_AM_mag_orb_AM_mag}a and b, as we see that the regular prolate rotators have the smallest magnitude of orbital AM, consistent with a more radial orbit at this point in the merger. At the same time, however, we see that the regular prolate rotators also have a relatively large internal AM of the first progenitor, again consistent with that strong sense of rotation. Note that in both Figures~\ref{fig:AM_ang_app_ang_infall} and \ref{fig:fp_AM_mag_orb_AM_mag}, we have included the distribution of properties for a subclass of the rapid AM reorienters: the misaligned, rapid AM reorienters (in pink) – we will discuss these in Section~\ref{sec:misaligned_comp}.

While we are able to see these distinctions between the orbits and properties of the progenitors when measured at this time of infall, we also explored the predictive utility of measurements made well before the merger, at the time of satellite maximum mass. This time should roughly correspond to a point prior to significant interaction between the progenitors. We find that global properties measured at this time (such as combinations of alignments between the AM and major axis vectors) do not meaningfully distinguish between the classes – especially not between the regular prolate rotators – and attribute this lack of signal to the effects of a close interaction later in the orbit of the satellite as we see in Figure~\ref{fig:merg_panel_ex}.

\section{Discussion}\label{sec:discussion}

We have now completed our exploration of the new class framework, demonstrating that a galaxy's classification is remarkably robust over time, that significant mergers are the primary events which move galaxies from one class to another (with the exception of the unsettled class), and that it is the final steps of the merger pirouette that set the class membership. In the discussion, we touch on a few other topics, first briefly exploring the origin of the angular momentum reorientation that we see, before making some additional comments on the regular prolate rotator class, and finally turning to a discussion of the place of this paper to previous work.

\subsection{Quantifying the Effects of Torques}\label{ssec:tidaltorques}
Having investigated the long-term behavior of these systems, an important question remains – how much of the evolution is driven by processes such as gravitational torques induced by nearby systems, and how much can be attributed to the merger-induced behavior that we hope to describe? To answer this, we carry out an investigation of the effect of external torques on the galaxies in our sample. To this end, we follow the analysis outlined in \citealp{Danovich2015}, wherein the authors describe the use of tidal torque theory to compute the time derivative of the angular momentum induced by external torques, directly computed from derivatives of the potential. Broadly, we use this approach to estimate the effect of tidal torques on the AM and compare the results to our measured reorientation rates, to evaluate their importance to the galaxies’ evolution.

Specifically, leveraging the particle potentials reported in the TNG database, we fit a second-degree polynomial to a sample of the particles at the scale of the galaxy half-mass radius, the relevant scale of our reorientation rate computations. We use this to compute the second derivatives of the potential and construct the tidal tensor (see equation 3 in \citealp{Danovich2015}). The antisymmetric tensor product of this tensor with the inertia tensor (Eq.~\ref{eq:inertia}) yields the time derivative of the AM due to external torques, $\mathbf{\dot{J}}$. We find that this measurement is roughly constant between snapshots, so we multiply this by $t = 5$ snapshots to estimate $\mathbf{\Delta J}$. We then add this to the measured AM and compute the angle between $\mathbf{J}$ and $\mathbf{J}+\mathbf{\Delta J}$. Finally, dividing this angle by the time window $\Delta t$ yields a measurement analogous to the reorientation rate in [deg/Gyr].

\begin{figure}
    \centering
    \includegraphics[scale = 0.28]{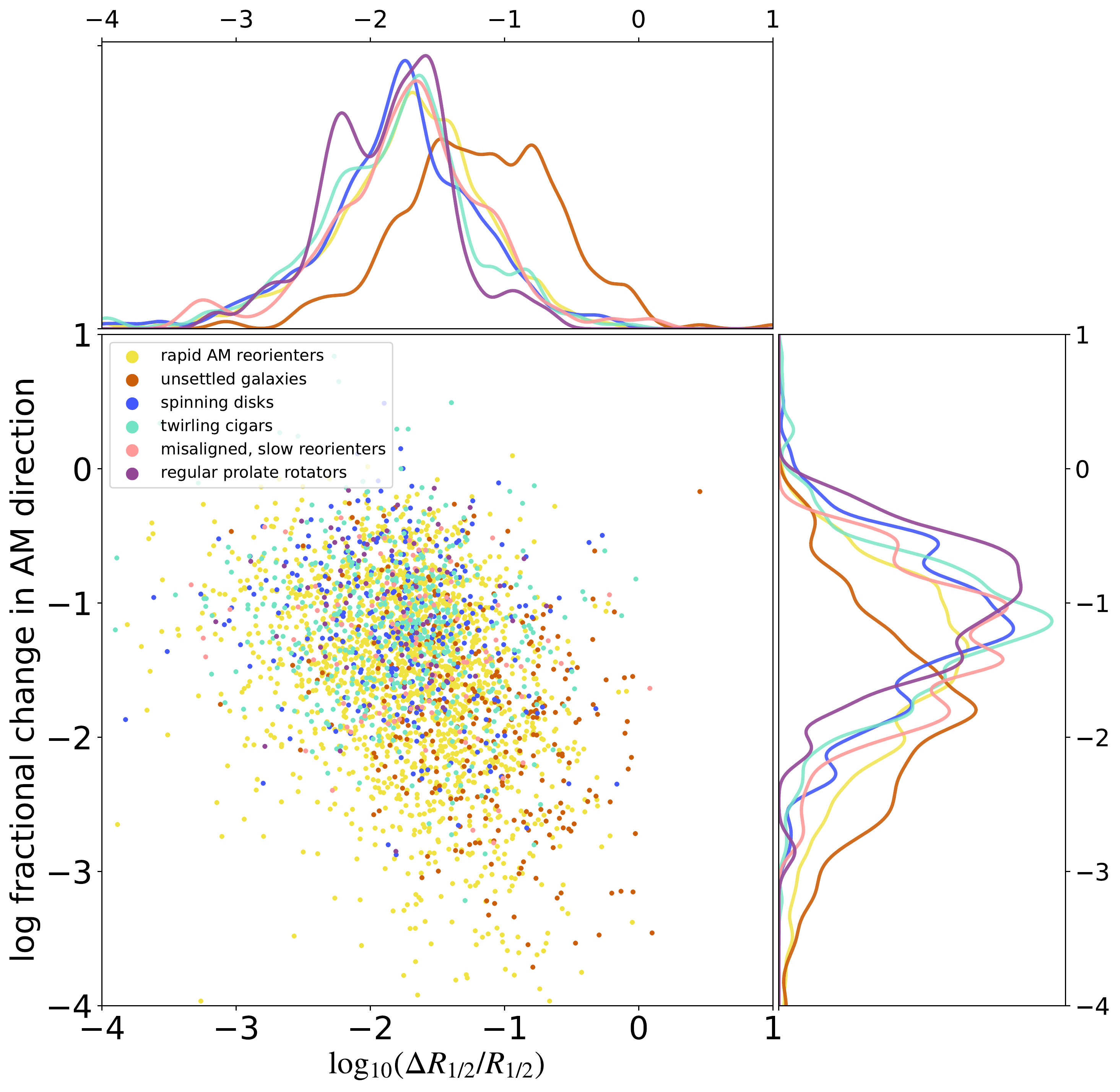}
    \caption{The growth of the galaxy's half-mass radius $R_{1/2}$ against the fractional change in the AM direction due to external torques (see Section \ref{ssec:tidaltorques}). Projections into 1D PDFs (using Gaussian kernel density estimation)
    for each quantity are shown aside each axis. Note that these distributions are normalized to the total number of galaxies within that class (so the regular prolate rotator
    line shows the density distribution for only regular prolate rotators). The reorientation is most closely connected to galaxy growth rather than external gravitational torques.}
    \label{fig:delR_J_ang_frac}
\end{figure}

We compare these two reorientation rates – the rate predicted by tidal torque theory and the measured rate we have discussed thus far – by computing their ratio. For almost all galaxies, this ratio is less than unity and for many, the reorientation rate due to external torques is an order of magnitude or more smaller than the measured reorientation rate.  The distribution of this ration can be seen in the PDF along the vertical axis in Figure \ref{fig:delR_J_ang_frac}, which captures the ratio of these two reorientation rates.  

This, then, begs the question of where the reorientation rates come from. To begin answering this, we examine the connection between reorientation and galaxy growth. In the central scatterplot of Figure \ref{fig:delR_J_ang_frac}, we plot the aforementioned ratio against the normalized change in the size ($R_{1/2}$) of the galaxy, which is directly connected to the galaxy’s growth rate. Here, we visually see an anticorrelation between the galaxy’s growth rate and the effect of external torques, as we would expect if angular momentum addition from galaxy growth were primarily responsible for reorientation. We verify this visual conclusion by computing the correlation coefficient between these quantities and find a statistically significant value of -0.2-0.3 for most classes. This indicates that for galaxies with more significant growth rates, the effect of external torques is subdominant, and other processes instead drive the reorientation rate.

In sum, this brief investigation demonstrates that, in general, external torques are not the dominant force driving the reorientation rates we discuss here.

\subsection{All prolate rotators are misaligned, but are all misaligned galaxies prolate rotators?}\label{sec:misaligned_comp}
With these results in hand, we can revisit one of our classes – the regular prolate rotators – and identify the features that distinguish them from galaxies that may be simply be misaligned. For a rough idea of the distinction at hand, recall Figure~\ref{fig:prolate_rotator_PSR_mis_map}, where we see that the misaligned galaxy is visually and quantitatively distinct from the regular prolate rotator – most notably in its lack of rotational support. We can also see an example of another type of misalignment in Figure~\ref{fig:prolate_slow_rotator_unsettled_map}a, where at approximately 1 Gyr before the present day, that galaxy achieves a significant degree of misalignment. This is the case more commonly observed in our sample and is observationally difficult to distinguish from consistent misalignment. In other words, because these galaxies are instantaneously misaligned at a given time and have a prolate shape, they could be classified as regular prolate rotators, but we argue that they are actually distinct.

\begin{figure*}
    \centering
        \subfloat[Misaligned rapid AM reorienters.]{{\includegraphics[scale=0.46]{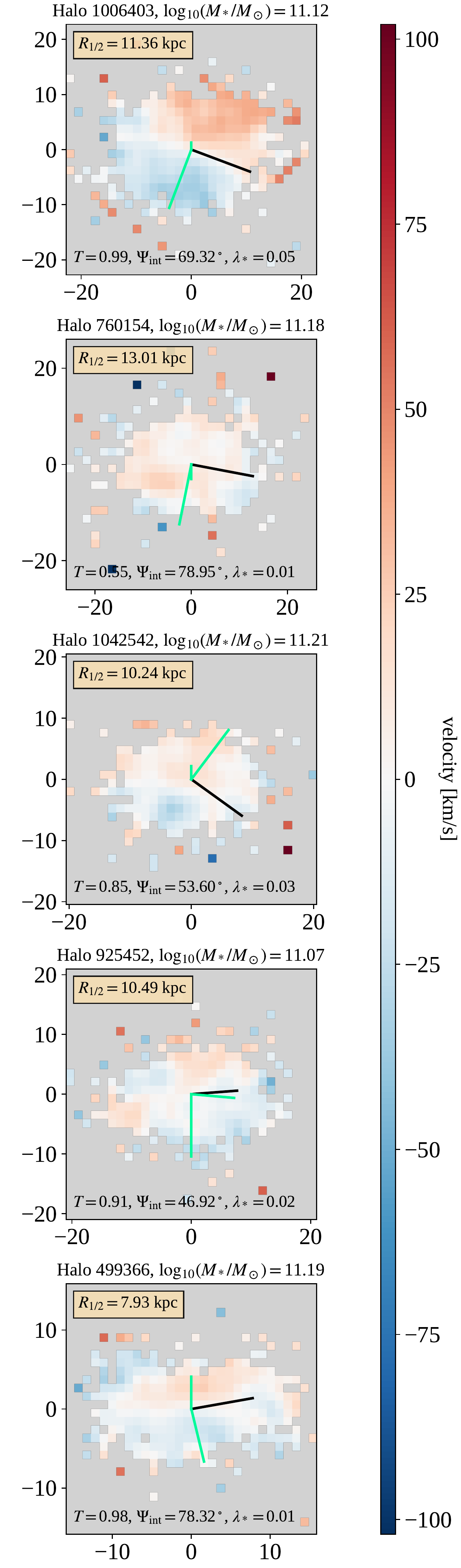} }}%
        \qquad
        \subfloat[Regular prolate rotators.]{{\includegraphics[scale=0.46]{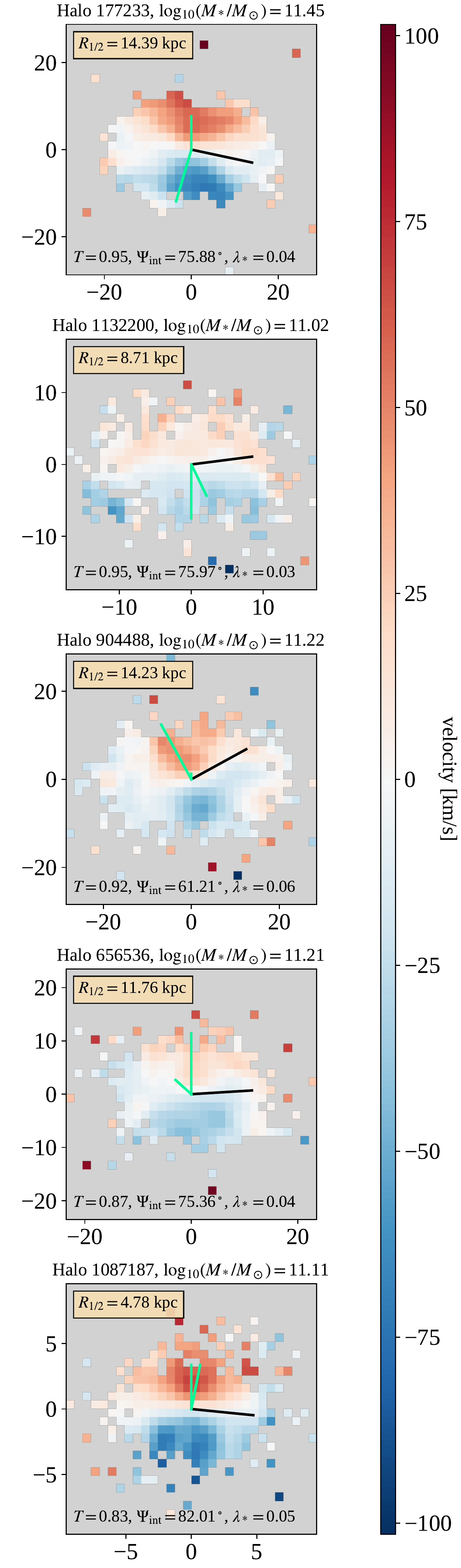} }}%
    \caption{Five velocity maps for \textbf{misaligned, rapid AM reorienters} (left) and \textbf{regular prolate rotators} (right) at $z=0$. See Section \ref{ssec:class_exemplars} for a description of how these maps were generated. These maps are oriented with the major axis of the galaxy along the horizontal and the minor axis as the vertical and we indicate the internal AM direction with a black line (accompanied by two orthogonal green lines for visual purposes). The stellar mass $M_*$, half-mass radius $R_{1/2}$, triaxiality $T$, intrinsic misalignment $\Psi$, and stellar spin parameter $\lambda_*$ are given in each panel.}
    \label{fig:PR_PSR_mis_examples}
\end{figure*}

For more examples of the distinction, we present velocity maps of randomly selected galaxies from each class – either regular prolate rotators or misaligned, rapid AM reorienters – viewed at $z=0$ in Figure~\ref{fig:PR_PSR_mis_examples}. We orient these velocity maps with the major axis of the galaxy along the horizontal and the minor axis along the vertical and overlay the internal AM direction with a black line, so comparing deviations of the black from the horizontal visually gives an idea of the misalignment we report. In addition to the misalignment angle, we present the stellar mass of the galaxy, the half-mass radius (in kpc), the triaxiality, and the stellar spin parameter. With these examples, the diversity of the classes and the distinctions between the two sets of misaligned galaxies are even more explicit. On the left, we generally have a less obvious sense of rotation in the velocity maps than those on the right (evidenced both visually and in the reported values for the stellar spin parameter $\lambda_*$). This is indicative of the latter observation made previously – that is, some galaxies are misaligned for only brief periods in their evolution and do not stably exhibit this behavior. 

Across the whole population, the distinction between these groups is most clearly seen from their definitions, as we identify the misaligned rapid AM reorienters as galaxies that have a more volatile AM direction. We find that as a result of this, their misalignments, while large, tend to be somewhat smaller than those of the regular prolate rotators ($\Psi_\mathrm{int, PSR}\sim 60^\circ-70^\circ$; cf. $\Psi_\mathrm{int, PR}\gtrapprox 70^\circ$) and also demonstrate larger variations. Similarly, we see that the degree of rotational support in misaligned rapid AM reorienters is, on average, smaller than that of the regular prolate rotators. These observations are not pictured but can be extrapolated from the trends observed in the reported properties of regular prolate rotators vs. misaligned, rapid AM reorienters in Figure~\ref{fig:PR_PSR_mis_examples}.

\begin{figure}
    \centering
    \includegraphics[scale=0.43]{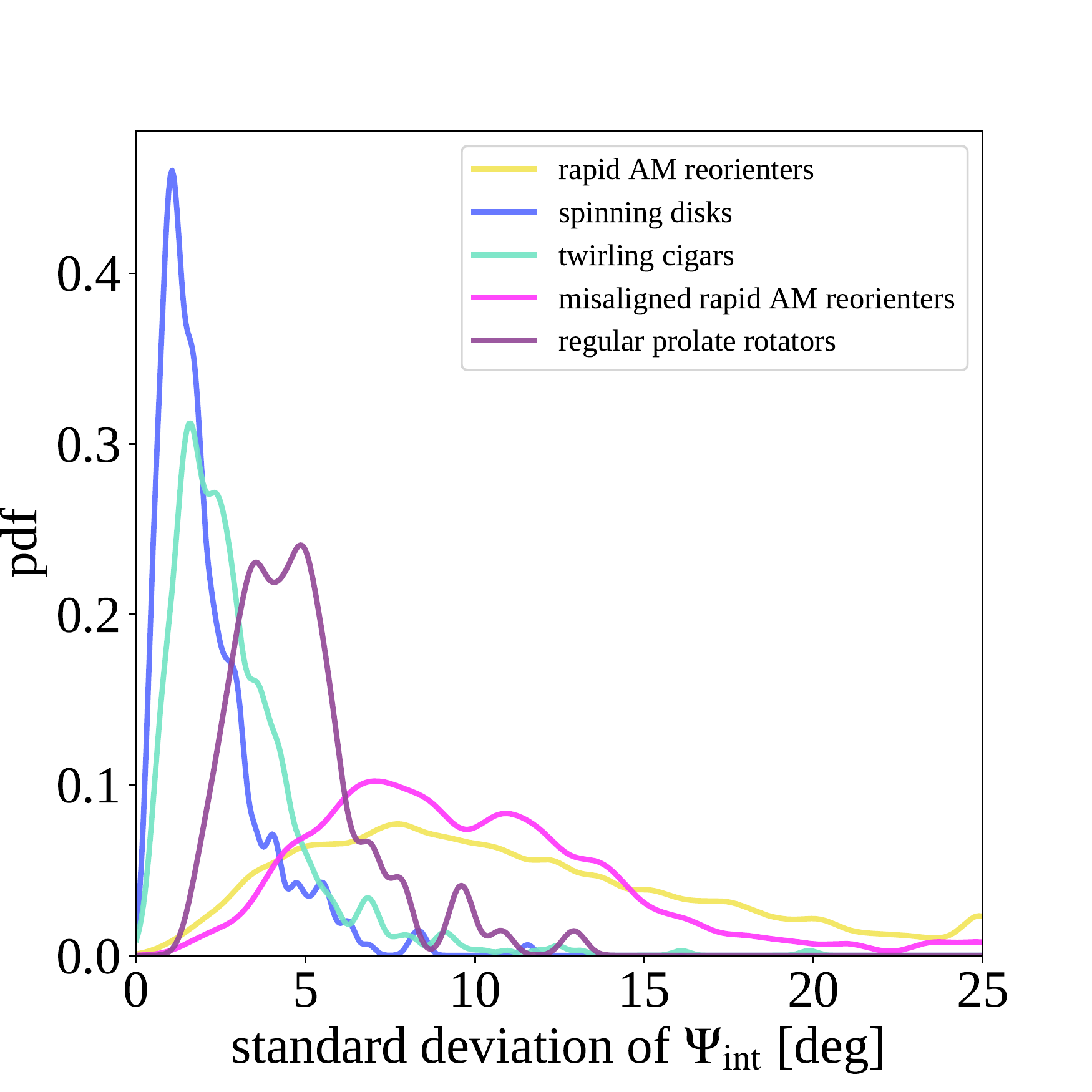}
    \caption{The standard deviation of the measured misalignment over a 5 snapshot window prior to $z=0$, separated by class. Note that the pink curve displays the distribution for a subclass of the rapid AM reorienters.}
    \label{fig:psi_std_grp}
\end{figure}

In Figure~\ref{fig:psi_std_grp}, we present the standard deviation of the misalignment in a five snapshot window prior to $z=0$, separated by class. This is a measure of the variability in the misalignment measurement of a given galaxy and the results are consistent with the description we have outlined thus far. That is, the spinning disks and twirling cigars have extremely well-defined measurements of their misalignments, and this agrees with the picture that we have of disk and prolate galaxies, respectively, stably rotating around their minor axes. The rapid AM reorienters, on the other hand, demonstrate a large spread in this plot, with a peak around $8-10^\circ$, which is also consistent with the exemplar picture given in Figure \ref{fig:prolate_slow_rotator_unsettled_map}a, where we see significant variability in the misalignment over time for this galaxy. The regular prolate rotators have relatively well defined misalignments, with a slightly larger spread than the spinning disks and twirling cigars. Finally, the misaligned, rapid AM reorienters demonstrate almost the largest average variations in the misalignment, indicating that, on average, misalignment measurements for these galaxies are relatively volatile and inconsistent. We do not explore this in this work, but note that for a potentially more direct connection to observable distinctions between the subgroups of misaligned galaxies, one could analyze the radial variation of the quantities we measure (i.e. measurements of the spread of misalignments as a function of radius could potentially be correlated with classifications of galaxies as true prolate rotators vs simply being misaligned).

The merger investigation, as discussed in Section \ref{sec:merg_history}, helps shed light on these observed distinctions. In particular, the progenitors of regular prolate rotator galaxies tend to have lower triaxialities and stronger degrees of rotational support so, using our classification system, these galaxies would be classified as spinning disks or twirling cigars prior to the merger. Evidently, as can be seen from the similarity in the distributions of misaligned, rapid AM reorienters and regular prolate rotators in Figure \ref{fig:AM_ang_app_ang_infall}, the dynamics of the merger – specifically the radial infall of the satellite along the direction of the internal AM of the progenitor – are sufficient to generate misalignment in a galaxy. It is then the kinematic and morphological state of the progenitor that sets apart the regular prolate rotators from their more unstably misaligned counterparts.

\begin{figure*}
    \centering
    \includegraphics[scale = 0.5]{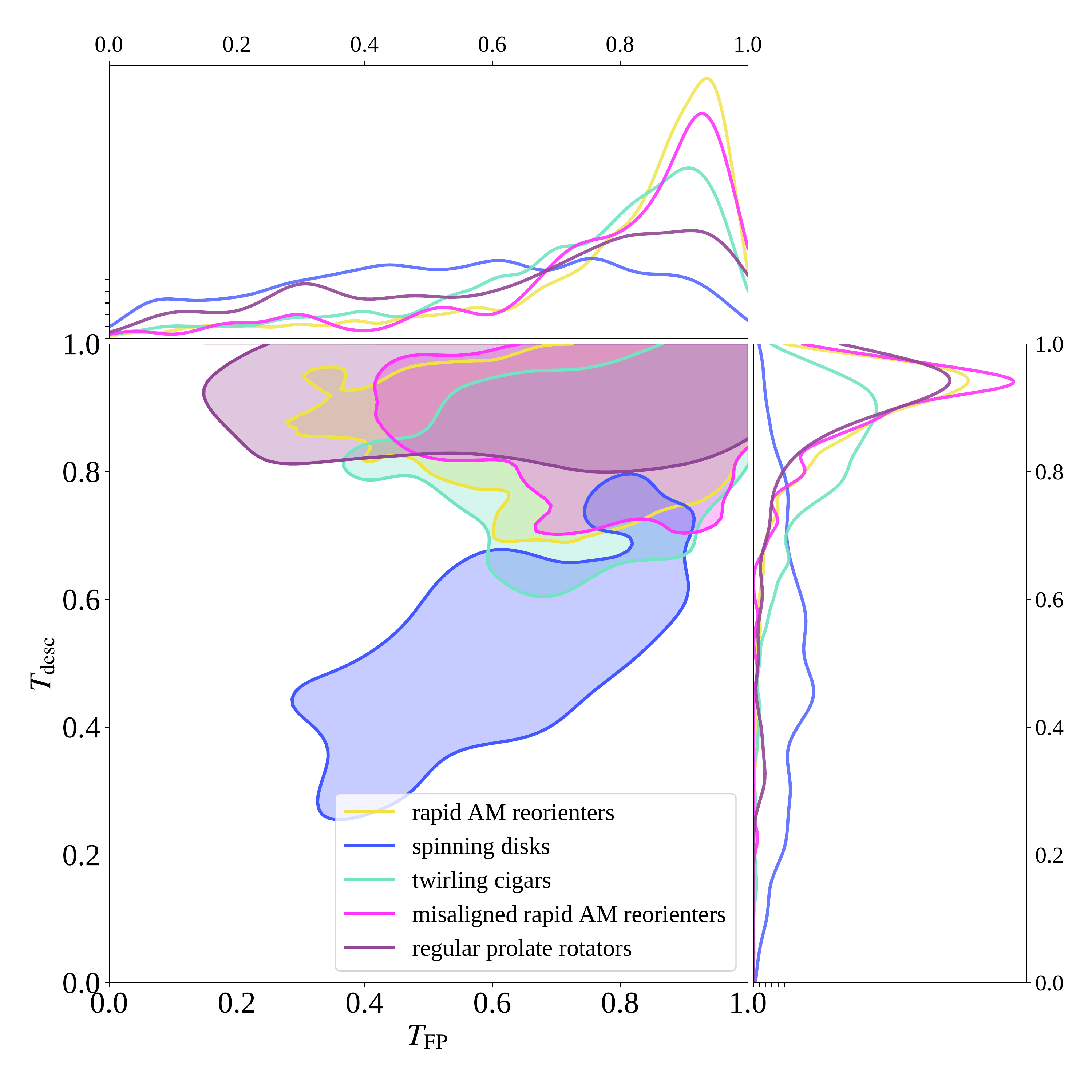}
    \caption{The triaxiality of the progenitor galaxy plotted against the triaxiality of the descendant galaxy, separated by class. Quantities prior to the merger are measured at the time of final infall (see Section \ref{sec:merg_history}). Projections into 1D PDFs (using Gaussian kernel density estimation) for each quantity are shown aside each axis. Note that the pink curve displays the distribution for a subclass of the rapid AM reorienters and these distributions are normalized to the total number of galaxies within that class (so the regular prolate rotator line shows the density distribution for \textit{only} regular prolate rotators). 2D PDFs are presented in the central panel and are estimated the same way, with the displayed contour representing 85\% of that class' data.}
    \label{fig:T_fp_T_desc}
\end{figure*}

This difference is evident in Figure \ref{fig:T_fp_T_desc}, where we display the triaxiality of the progenitor (measured at the time of final infall, see Section \ref{sec:merg_history}) against the triaxiality of the descendant. From this, it seems that the regular prolate rotators demonstrate the most significant transition in shape as a result of the merger. Prior to the merger, the regular prolate rotators are fairly widely distributed, with a slight peak at large triaxialities, but a broad range – almost a combination of the spinning disk and twirling cigar distributions. Following the merger, however, we see that the regular prolate rotators experience a marked transition and are clustered at large triaxialities, a shift that we know to continue over time (see Figure \ref{fig:class_definition}b). In the joint probability distribution, we see that the regular prolate rotators are spread along the top, while many of the other classes, most notably the spinning disks and twirling cigars, have a significant component of their joint distribution clustered along the diagonal. This indicates that for these classes, the merger induces relatively minimal transitions in shape. The misaligned rapid AM reorienters are one such case and are approximately clustered around the high triaxiality diagonal (i.e.~they maintain roughly the same shape before and after the merger), with a slight spread in progenitor triaxialities. In kind, revisiting Figure \ref{fig:fp_AM_mag_orb_AM_mag}, the modulus of the orbital AM for the misaligned, rapid AM reorienters is somewhat larger than that of the regular prolate rotators while the modulus of the internal AM demonstrates the opposite trend, supporting the hypothesis that it is this sense of rotation in the progenitor that determines the ultimate state of prolate rotation. The joint distribution similarly demonstrates that the regular prolate rotators occupy regions of relatively low orbital AM (more radial orbits) and large internal AM (strong degree of rotational support).

Returning to the panels in Figure \ref{fig:AM_ang_app_ang_infall}, now focusing on the two classes of misaligned galaxies – the misaligned rapid AM reorienters and regular prolate rotators – it seems that the misaligned rapid AM reorienters are almost an intermediate class between the regular prolate rotators and the other classes, but do demonstrate some similarities to the regular prolate rotators. Evidently, it is this configuration of the merger – the more radial orbits along the internal AM direction of the main progenitors – that produces the misalignment in the descendant and the progenitor state (shape and rotation) that specifically distinguishes the regular prolate rotators from their generally misaligned counterparts.

\begin{figure*}
    \centering
        \subfloat[Misaligned rapid AM reorienters.]{{\includegraphics[scale=0.33]{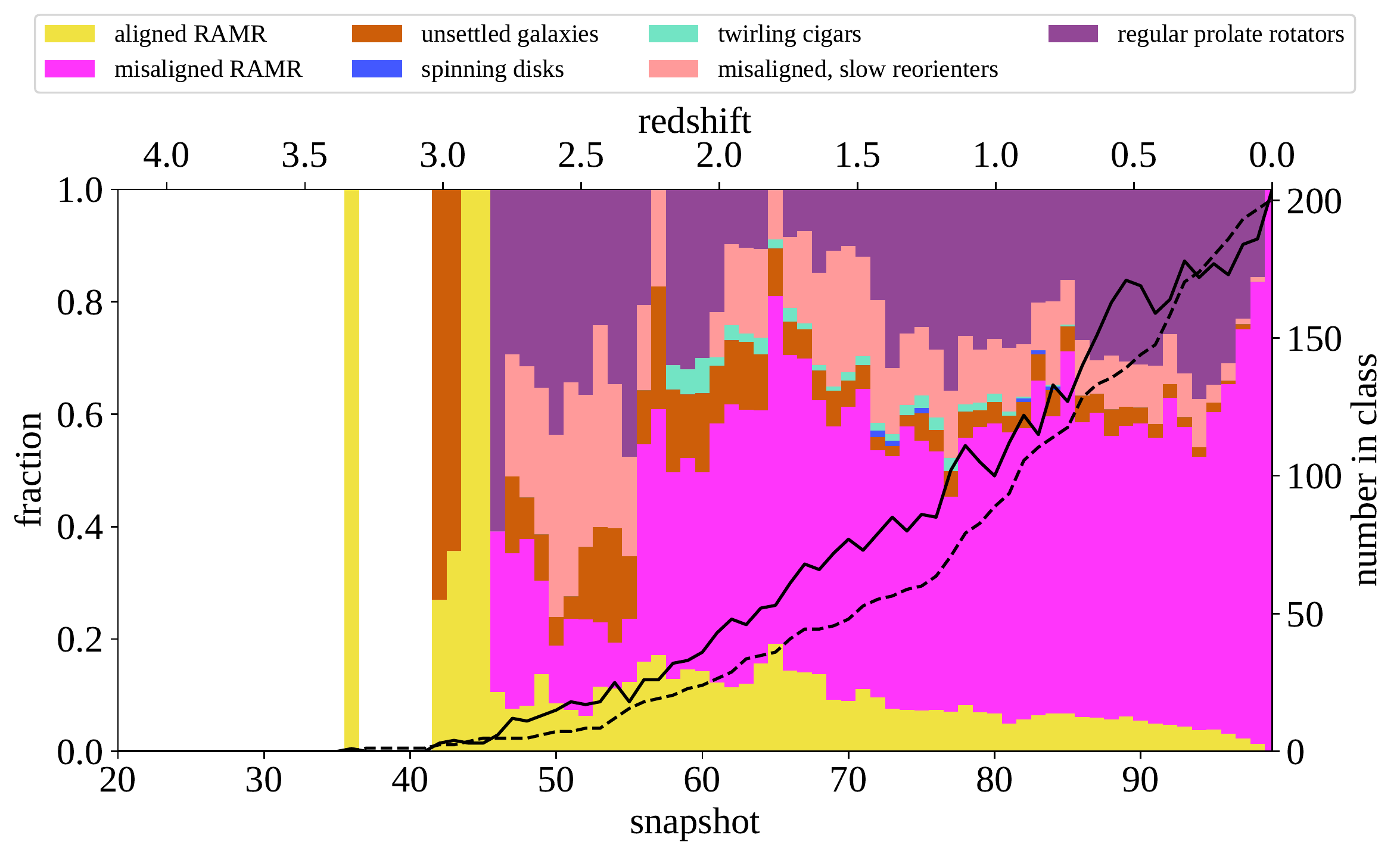} }}%
        \qquad
        \subfloat[Regular prolate rotators.]{{\includegraphics[scale=0.33]{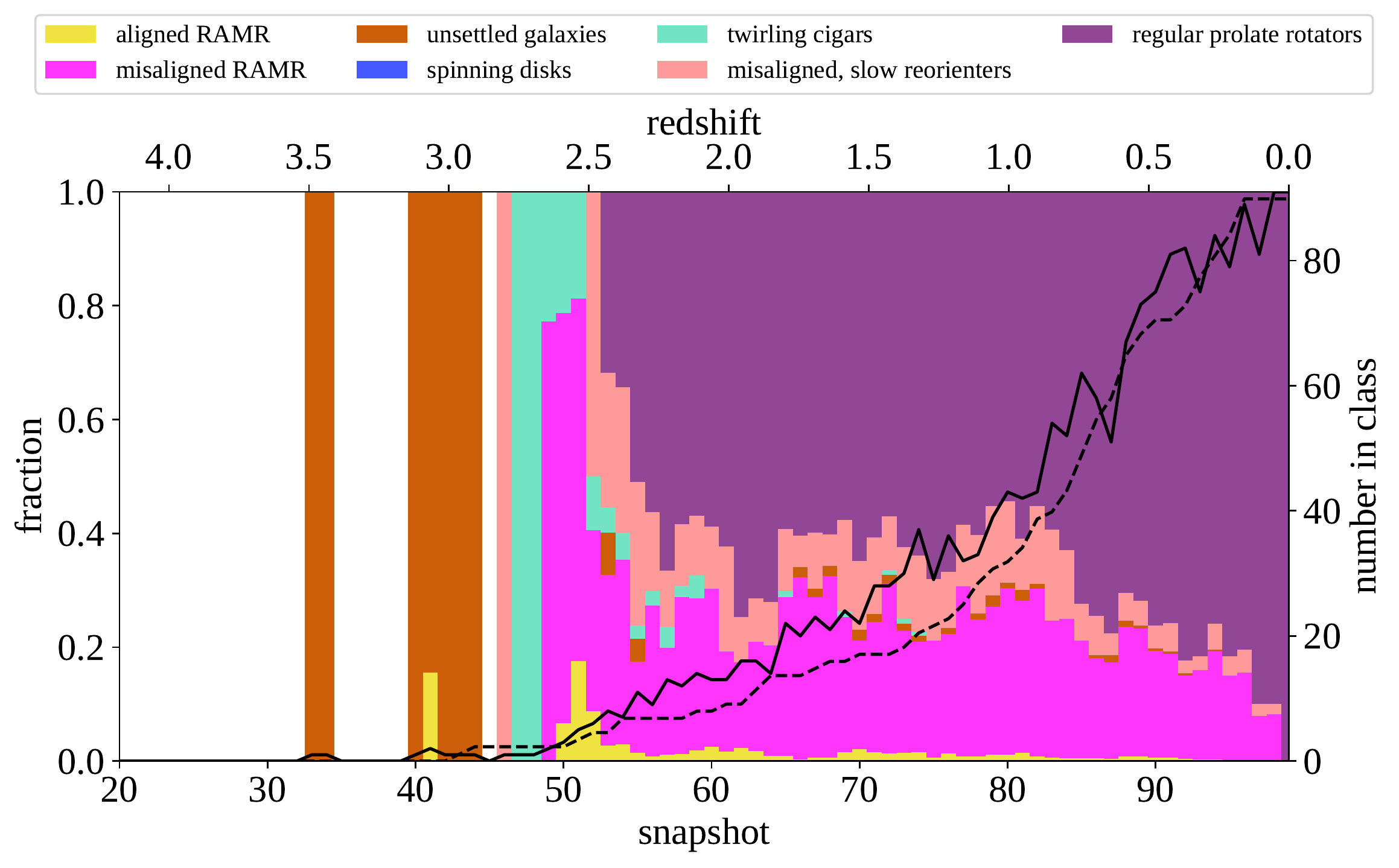} }}%
    \caption{An analogous measurement to that of Figure \ref{fig:class_bars}, now focusing on comparing the misaligned subclass – the misaligned rapid AM reorienters – to the regular prolate rotators (see Section \ref{sec:class_stability} for details on the computation). Note that we abbreviate the rapid AM reorienters as `RAMR' in the legend.}
    \label{fig:group_bar_ev}
\end{figure*}

We also carry out the same investigation as Section \ref{sec:class_stability} (see Figure \ref{fig:group_bar_ev}), now dividing the rapid AM reorienters (denoted RAMR in Figure \ref{fig:group_bar_ev}) into two subclasses – those that are aligned ($\Psi_\mathrm{int}<60^\circ$) and those that are misaligned (the complementary group). This allows us to directly explore the connection between the regular prolate rotators and misaligned rapid AM reorienters. In particular, both panels demonstrate that, while the two groups have distinct formation pathways (as is discussed above), there is a subset of both groups that displays a tendency to transition to the other group. This can be understood given that the two groups differ by a hard boundary in the class definitions – specifically, the reorientation rate of the AM axis – so a tendency to cross that boundary is a reasonable observation (and is a similar one to that observed between the spinning disks and twirling cigars in panels c and d of Figure \ref{fig:class_bars}). In other words, these groups demonstrate some similarities, and the distinctions between them are more general observations than strict, physical differences – these group definitions are not binary. This overlap is demonstrated in the first panel in Figure \ref{fig:PR_PSR_mis_examples}a – one of the randomly selected misaligned, rapid AM reorienters looks quite similar to a prolate rotator (i.e. shares many of the standard properties) and is only classified differently because it went through a period of rapid AM reorientation two snapshots prior to $z=0$.

\subsection{Connections to Previous Work}\label{sec:prev_work}

The most clear and direct connection to previous studies, in terms of focus and methodology, is with \citet{EL2017}, wherein the authors discuss their analysis of prolate rotation in the Illustris simulation (the precursor to the TNG simulations used in this work). They employ a somewhat different approach in terms of selecting the sample, choosing to analyze galaxies within a radius defined by a preselected limiting density and selecting regular prolate rotators by a combination of their triaxiality and the ratio of the AM in the major axis direction to the total AM of the galaxy. In addition, the smaller volume of Illustris compared to TNG300 forced them to draw their sample from a lower-mass population than the galaxies analyzed here. Regardless, their selection results in a sample of 59 regular prolate rotators and they conduct analysis of the global properties of these galaxies' mergers at the point of \textit{first} infall. In our analysis, we found that attempts to analyze the merger state at such a point (prior to any meaningful interaction between the progenitor galaxies) did not result in any clear prediction of the post-merger state of the galaxies. Indeed, as they find in their work, there is little significant signal in the parameters they use to summarize the merger properties at this selected time (and they too conclude that, instead, the final stages of the merger are seemingly the most relevant for formation of prolate rotation). This conclusion on the importance of the `last stretch' of the orbit mirrors that of Section 6.4 of \citet{Pop2018}, where they discuss that orbits of satellites in high mass-ratio mergers are quickly radialized. 

Returning to \citet{EL2017}, if we look to their Figure 6, we can see that they too present a series of frames tracing the final infall of a selected regular prolate rotator galaxy (cf. our Figure \ref{fig:merg_panel_ex}). Interestingly, with the discussion outlined in Section \ref{sec:merg_history} in mind, the picture of this prolate rotator's merger is fairly consistent with our results. That is, we can see that the satellite approaches the first progenitor galaxy on a roughly radial trajectory that is both aligned with the internal AM of the galaxy and the major axis of the resulting descendant. In fact, if we look to their Figure 5 which describes the evolution of triaxiality for this same galaxy, we see that this galaxy transitions from a state of low triaxiality (oblate shape) and minimal rotation about the major axis (disk-like rotation) before the merger to one with prolate rotation afterwards (with descendant internal AM aligned with the internal AM of the progenitor galaxy). Therefore, this combination of progenitor shape/kinematics and merger dynamics meets all the criteria we have identified for the production of prolate rotation and, indeed, we see that the resulting system is what they term a `golden' example of a prolate rotator.

Another similarly related work is that of \citet{BF2019}. The authors of that study similarly use the original Illustris simulation to identify relationships between galaxy intrinsic shapes and kinematic properties and it is from this work that we borrow many of the definitions we use for quantities, such as the intrinsic misalignment and 3D shape. Much as with \citet{EL2017}, that study focuses on galaxies at a lower mass range than employed in our work, and they attempt to draw a more direct connection to observational results by constructing mock photometric and kinematic maps of the galaxies in their sample and measuring mock observables from these. Our Figure \ref{fig:class_definition}b is a direct comparison with their Figure 4 and demonstrates how the higher mass cut in this work yields a distinctly different distribution of galaxies for our sample. They primarily limit their analysis to a shape-based classification of galaxies and we are able to reproduce many of their results within our sample as well. However, as explained in Section \ref{sec:new_classes}, these simple shape distinctions are not sufficient to capture the diversity of our higher-mass sample of galaxies. Finally, in \citet{BF2019}, they also connect the shape transitions to the merger history of their galaxies but do not look very closely at the details of the interactions, arguing instead that the snapshot separation of Illustris is too coarse to truly track the dynamics of the merger. In this work, we take this analysis further and develop a more extensive classification system that yields more insight into the merger history of these high-mass galaxies.

For an observational connection, we can look at \citet{Ebrova2021}, wherein the authors analyze optical images of 19 prolate rotators compiled from the literature and document signs of interactions in all the galaxies in the sample. In this – and other similar – observational works, prolate rotation is largely identified solely on the basis of the observed instantaneous misalignment, as is evident from Table 1 in that paper. In this paper, we argue that this single criterion is potentially not sufficient for identifying \textit{regular} prolate rotation as a long-lived and stable kinematic class of galaxies (see Section \ref{sec:misaligned_comp}). In addition, we see that the spread of misalignments measured in the sample is distinct from that which we measure in TNG. That is, we see (in Figure \ref{fig:lam_star_M_dist}a, for example) that regular prolate rotators are clustered at large misalignments, with a strong peak near $\Psi_\mathrm{int}\sim 80^\circ$, while they report misalignments ranging from $45^\circ$ to $80^\circ$. In identifying signs of interactions in their observed sample, they also find a connection between prolate rotation and the presence of multiple shells in these galaxies, which are believed to result from nearly radial minor and major mergers, an understanding of the mergers consistent with the results that we have found in our analysis of regular prolate rotators here with TNG.

There are a number of other works that motivated some of the analysis and discussion carried out in this study, albeit less prominently than the three discussed thus far. In particular, it was the discussion in \citet{Zeng2021} and \citet{Vasiliev2021} that motivated some of our detailed analysis of merger dynamics. Both of these works focused on the evolution of the circularity of the satellite's orbit, \citet{Zeng2021} with an eye towards disk galaxies and \citet{Vasiliev2021} with a focus on radialization of satellite orbits over time. The latter discussion is especially relevant for the regular prolate rotators we have analyzed in this study as we see close interactions and radialization of the orbit as a prerequisite for the development of a misaligned state. As is argued in \citet{Vasiliev2021}, higher mass satellites tend to experience stronger radialization, and we indeed see that the galaxies with the most radial progenitor orbits (Figure \ref{fig:AM_ang_app_ang_infall}b), the regular prolate rotators, tend to have the largest merger mass ratios (Figure \ref{fig:merger_dist}a). 

Another related work is that of \citet{cardona-barrero2021}, wherein the authors analyze one identified case of prolate rotation from a sample of cosmologically simulated dwarf galaxies. In this case, they find that the prolate rotation arose after a previously oblate system experienced a major merger, but they do not speculate on the configuration of the merger that led to this state. That being said, the information provided is consistent with our picture of regular prolate rotator formation – indeed, they observe a transition from a state of oblate, disk-like rotation to one of prolate rotation and note that the prolate rotation behavior is maintained from the point of merger to $z=0$ (approx. 6 Gyr of stability).  

\section{Conclusions}\label{sec:conclusions}
In this work, we have examined the shapes and kinematics of the most massive galaxies generated by the IllustrisTNG cosmological simulation suite. Building from past studies (e.g.~\citealp{EL2017}, \citealp{BF2019}, \citealp{Pulsoni2020}), we aim to characterize the distribution of global properties that can be measured for these galaxies with an eye towards developing a robust mechanism for reliably identifying and understanding the behavior of prolate galaxies that demonstrate ordered rotation about their major axis. To this end, we develop a new classification system based on the kinematic \textit{and} morphological properties of the galaxies in our sample and analyze these classes in the context of the most recent significant ($\mu > 1/10$) mergers.

We find that the appearance of these kinematic classes can be well connected to the merger history of the galaxies and our findings are summarized as follows:
\begin{itemize}
    \item Taking time-averaged measurements of the directions of the kinematic and morphological axes of these galaxies provides a useful new perspective that can be connected to a clear physical picture of their behavior. This picture is relatively stable over time and, as a result, allows for reliable analysis of these galaxies throughout their evolution.  
    \item A significant fraction ($\sim 70\%$) of our galaxies are prolate, but within this designation, display a variety of behaviors. We denote these as rapid AM reorienters, twirling cigars, and regular prolate rotators, and each of these classes roughly corresponds to a different range of kinematic misalignment.
    \item The classifications we develop – rapid AM reorienters, unsettled galaxies, spinning disks, twirling cigars, misaligned, slow reorienters, and regular prolate rotators – are fairly stable over time (after their most recent significant mergers) and display a variety of unique behaviors.
    \item The transitions between classes are well correlated with the presence of significant mergers, most notably in the case of regular prolate rotators, though the specifics of the mergers that generate each class are unique.
    \item The mergers that produce regular prolate rotators generally seem to follow a standard narrative: a rotationally supported spinning disk/twirling cigar experiences a close interaction with an infalling satellite that sends the satellite onto a roughly radial orbit for its final infall. This orbit, when approximately aligned with the internal AM of the progenitor causes an elongation of the stellar distribution of that galaxy in the direction of the internal AM and produces a descendant galaxy with alignment of the AM and major axis directions – a prolate rotator. 
    \item The subset of galaxies that are misaligned ($\Psi_\mathrm{int}> 60^\circ$) but are not regular prolate rotators – i.e.~the misaligned, rapid AM reorienters – exist as an almost intermediate state between the regular prolate rotators and the rapid AM reorienters. Their behavior demonstrates that a suitably configured radial merger is sufficient to produce misalignment in galaxies but it is the progenitor state (rotationally supported with low triaxiality or not) that determines the stability of this misalignment over time.
    \item The reorientation of the angular momentum direction in our galaxies is more correlated with galaxy size growth than external gravitational torques.
\end{itemize}

Broadly, we have presented a set of kinematic classes that offer a new perspective on the evolution of massive galaxies. While the direct connection to observational results is not immediately apparent, a comparison of our identified galaxy properties (through mock observations, as in \citet{BF2019} for example) and signatures of the mergers (as in \citet{Ebrova2021}) to those in observed massive galaxies would be invaluable. The study of massive galaxies has often settled on identifying two main classes – the slow and fast rotators – but our work demonstrates that there ought to be further nuance and thought given to such classification. As we have shown, a majority of the galaxies in our sample are classified as rapid AM reorienters, an inherently kinematically unstable and disordered class of galaxies – that is, a class of non-rotators. With this in mind, the traditional `slow rotator' moniker that these would have earned by virtue of their small spin parameters is somewhat misleading and motivates more careful consideration of this distinction. In kind, we demonstrate that highly misaligned galaxies display more nuance than previously believed and that regular prolate rotation is not a trivial identification.

\acknowledgements
GLB acknowledges support from the NSF (OAC-1835509, AST-2108470, and computational resources through XSEDE), a NASA TCAN award, and the Simons Foundation. The Flatiron Institute is supported by the Simons Foundation

\bibliographystyle{aasjournal}
\bibliography{biblio}

\begin{thebibliography}{}
\expandafter\ifx\csname natexlab\endcsname\relax\def\natexlab#1{#1}\fi
\providecommand{\url}[1]{\href{#1}{#1}}
\providecommand{\dodoi}[1]{doi:~\href{http://doi.org/#1}{\nolinkurl{#1}}}
\providecommand{\doeprint}[1]{\href{http://ascl.net/#1}{\nolinkurl{http://ascl.net/#1}}}
\providecommand{\doarXiv}[1]{\href{https://arxiv.org/abs/#1}{\nolinkurl{https://arxiv.org/abs/#1}}}

\bibitem[{{Bacon} {et~al.}(2001){Bacon}, {Copin}, {Monnet}, {Miller},
  {Allington-Smith}, {Bureau}, {Carollo}, {Davies}, {Emsellem}, {Kuntschner},
  {Peletier}, {Verolme}, \& {de Zeeuw}}]{Bacon2001}
{Bacon}, R., {Copin}, Y., {Monnet}, G., {et~al.} 2001, \mnras, 326, 23,
  \dodoi{10.1046/j.1365-8711.2001.04612.x}

\bibitem[{{Barrera-Ballesteros} {et~al.}(2015){Barrera-Ballesteros},
  {Garc{\'\i}a-Lorenzo}, {Falc{\'o}n-Barroso}, {van de Ven}, {Lyubenova},
  {Wild}, {M{\'e}ndez-Abreu}, {S{\'a}nchez}, {Marquez}, {Masegosa},
  {Monreal-Ibero}, {Ziegler}, {del Olmo}, {Verdes-Montenegro},
  {Garc{\'\i}a-Benito}, {Husemann}, {Mast}, {Kehrig}, {Iglesias-Paramo},
  {Marino}, {Aguerri}, {Walcher}, {V{\'\i}lchez}, {Bomans}, {Cortijo-Ferrero},
  {Gonz{\'a}lez Delgado}, {Bland-Hawthorn}, {McIntosh}, \&
  {Bekerait{\.{e}}}}]{barreraballesteros2015}
{Barrera-Ballesteros}, J.~K., {Garc{\'\i}a-Lorenzo}, B., {Falc{\'o}n-Barroso},
  J., {et~al.} 2015, \aap, 582, A21, \dodoi{10.1051/0004-6361/201424935}

\bibitem[{{Bassett} \& {Foster}(2019)}]{BF2019}
{Bassett}, R., \& {Foster}, C. 2019, \mnras, 487, 2354,
  \dodoi{10.1093/mnras/stz1440}

\bibitem[{{Bender} {et~al.}(1989){Bender}, {Surma}, {Doebereiner},
  {Moellenhoff}, \& {Madejsky}}]{Bender1989}
{Bender}, R., {Surma}, P., {Doebereiner}, S., {Moellenhoff}, C., \& {Madejsky},
  R. 1989, \aap, 217, 35

\bibitem[{{Cappellari}(2016)}]{Cappellari2016}
{Cappellari}, M. 2016, \araa, 54, 597,
  \dodoi{10.1146/annurev-astro-082214-122432}

\bibitem[{{Cappellari} {et~al.}(2011){Cappellari}, {Emsellem}, {Krajnovi{\'c}},
  {McDermid}, {Scott}, {Verdoes Kleijn}, {Young}, {Alatalo}, {Bacon}, {Blitz},
  {Bois}, {Bournaud}, {Bureau}, {Davies}, {Davis}, {de Zeeuw}, {Duc},
  {Khochfar}, {Kuntschner}, {Lablanche}, {Morganti}, {Naab}, {Oosterloo},
  {Sarzi}, {Serra}, \& {Weijmans}}]{Cappellari2011}
{Cappellari}, M., {Emsellem}, E., {Krajnovi{\'c}}, D., {et~al.} 2011, \mnras,
  413, 813, \dodoi{10.1111/j.1365-2966.2010.18174.x}

\bibitem[{{Cardona-Barrero} {et~al.}(2021){Cardona-Barrero}, {Battaglia}, {Di
  Cintio}, {Revaz}, \& {Jablonka}}]{cardona-barrero2021}
{Cardona-Barrero}, S., {Battaglia}, G., {Di Cintio}, A., {Revaz}, Y., \&
  {Jablonka}, P. 2021, \mnras, 505, L100, \dodoi{10.1093/mnrasl/slab059}

\bibitem[{{Danovich} {et~al.}(2015){Danovich}, {Dekel}, {Hahn}, {Ceverino}, \&
  {Primack}}]{Danovich2015}
{Danovich}, M., {Dekel}, A., {Hahn}, O., {Ceverino}, D., \& {Primack}, J. 2015,
  \mnras, 449, 2087, \dodoi{10.1093/mnras/stv270}

\bibitem[{{de Zeeuw} {et~al.}(2002){de Zeeuw}, {Bureau}, {Emsellem}, {Bacon},
  {Carollo}, {Copin}, {Davies}, {Kuntschner}, {Miller}, {Monnet}, {Peletier},
  \& {Verolme}}]{deZeeuw2002}
{de Zeeuw}, P.~T., {Bureau}, M., {Emsellem}, E., {et~al.} 2002, \mnras, 329,
  513, \dodoi{10.1046/j.1365-8711.2002.05059.x}

\bibitem[{{Dolfi} {et~al.}(2021){Dolfi}, {Forbes}, {Couch}, {Bekki},
  {Ferr{\'e}-Mateu}, {Romanowsky}, \& {Brodie}}]{Dolfi2021}
{Dolfi}, A., {Forbes}, D.~A., {Couch}, W.~J., {et~al.} 2021, \mnras, 504, 4923,
  \dodoi{10.1093/mnras/stab1023}

\bibitem[{{Ebrov{\'a}} {et~al.}(2021){Ebrov{\'a}}, {B{\'\i}lek},
  {Vudragovi{\'c}}, {Y{\i}ld{\i}z}, \& {Duc}}]{Ebrova2021}
{Ebrov{\'a}}, I., {B{\'\i}lek}, M., {Vudragovi{\'c}}, A., {Y{\i}ld{\i}z},
  M.~K., \& {Duc}, P.-A. 2021, \aap, 650, A50,
  \dodoi{10.1051/0004-6361/202140588}

\bibitem[{{Ebrov{\'a}} \& {{\L}okas}(2015)}]{Ebrova2015}
{Ebrov{\'a}}, I., \& {{\L}okas}, E.~L. 2015, \apj, 813, 10,
  \dodoi{10.1088/0004-637X/813/1/10}

\bibitem[{{Ebrov{\'a}} \& {{\L}okas}(2017)}]{EL2017}
---. 2017, \apj, 850, 144, \dodoi{10.3847/1538-4357/aa96ff}

\bibitem[{{Emsellem} {et~al.}(2007){Emsellem}, {Cappellari}, {Krajnovi{\'c}},
  {van de Ven}, {Bacon}, {Bureau}, {Davies}, {de Zeeuw}, {Falc{\'o}n-Barroso},
  {Kuntschner}, {McDermid}, {Peletier}, \& {Sarzi}}]{Emsellem2007}
{Emsellem}, E., {Cappellari}, M., {Krajnovi{\'c}}, D., {et~al.} 2007, \mnras,
  379, 401, \dodoi{10.1111/j.1365-2966.2007.11752.x}

\bibitem[{{Emsellem} {et~al.}(2011){Emsellem}, {Cappellari}, {Krajnovi{\'c}},
  {Alatalo}, {Blitz}, {Bois}, {Bournaud}, {Bureau}, {Davies}, {Davis}, {de
  Zeeuw}, {Khochfar}, {Kuntschner}, {Lablanche}, {McDermid}, {Morganti},
  {Naab}, {Oosterloo}, {Sarzi}, {Scott}, {Serra}, {van de Ven}, {Weijmans}, \&
  {Young}}]{Emsellem2011}
---. 2011, \mnras, 414, 888, \dodoi{10.1111/j.1365-2966.2011.18496.x}

\bibitem[{Faber {et~al.}(1997)Faber, Tremaine, Ajhar, Byun, Dressler, Gebhardt,
  Grillmair, Kormendy, Lauer, \& Richstone}]{Faber1997}
Faber, S., Tremaine, S., Ajhar, E., {et~al.} 1997, Astronomical Journal, 114,
  1771, \dodoi{10.1086/118606}

\bibitem[{Ferrarese {et~al.}(2006)Ferrarese, Cote, Jordan, Peng, Blakeslee,
  Piatek, Mei, Merritt, Milosavljevi{\'{c}}, Tonry, \& West}]{Ferrarese2006}
Ferrarese, L., Cote, P., Jordan, A., {et~al.} 2006, 164, 334,
  \dodoi{10.1086/501350}

\bibitem[{{Genel} {et~al.}(2018){Genel}, {Nelson}, {Pillepich}, {Springel},
  {Pakmor}, {Weinberger}, {Hernquist}, {Naiman}, {Vogelsberger}, {Marinacci},
  \& {Torrey}}]{Genel2018}
{Genel}, S., {Nelson}, D., {Pillepich}, A., {et~al.} 2018, \mnras, 474, 3976,
  \dodoi{10.1093/mnras/stx3078}

\bibitem[{{Graham} {et~al.}(2018){Graham}, {Cappellari}, {Li}, {Mao},
  {Bershady}, {Bizyaev}, {Brinkmann}, {Brownstein}, {Bundy}, {Drory}, {Law},
  {Pan}, {Thomas}, {Wake}, {Weijmans}, {Westfall}, \& {Yan}}]{Graham2018}
{Graham}, M.~T., {Cappellari}, M., {Li}, H., {et~al.} 2018, \mnras, 477, 4711,
  \dodoi{10.1093/mnras/sty504}

\bibitem[{{Greene} {et~al.}(2015){Greene}, {Janish}, {Ma}, {McConnell},
  {Blakeslee}, {Thomas}, \& {Murphy}}]{Greene2015}
{Greene}, J.~E., {Janish}, R., {Ma}, C.-P., {et~al.} 2015, \apj, 807, 11,
  \dodoi{10.1088/0004-637X/807/1/11}

\bibitem[{{Kormendy} \& {Bender}(1996)}]{Kormendy1996}
{Kormendy}, J., \& {Bender}, R. 1996, \apjl, 464, L119, \dodoi{10.1086/310095}

\bibitem[{{Krajnovi{\'c}} {et~al.}(2006){Krajnovi{\'c}}, {Cappellari}, {de
  Zeeuw}, \& {Copin}}]{Krajinovic2006}
{Krajnovi{\'c}}, D., {Cappellari}, M., {de Zeeuw}, P.~T., \& {Copin}, Y. 2006,
  \mnras, 366, 787, \dodoi{10.1111/j.1365-2966.2005.09902.x}

\bibitem[{{Krajnovi{\'c}} {et~al.}(2018){Krajnovi{\'c}}, {Emsellem}, {den
  Brok}, {Marino}, {Schmidt}, {Steinmetz}, \& {Weilbacher}}]{Krajinovic2018}
{Krajnovi{\'c}}, D., {Emsellem}, E., {den Brok}, M., {et~al.} 2018, \mnras,
  477, 5327, \dodoi{10.1093/mnras/sty1031}

\bibitem[{{Krajnovi{\'c}} {et~al.}(2008){Krajnovi{\'c}}, {Bacon}, {Cappellari},
  {Davies}, {de Zeeuw}, {Emsellem}, {Falc{\'o}n-Barroso}, {Kuntschner},
  {McDermid}, {Peletier}, {Sarzi}, {van den Bosch}, \& {van de
  Ven}}]{Krajinovic2008}
{Krajnovi{\'c}}, D., {Bacon}, R., {Cappellari}, M., {et~al.} 2008, \mnras, 390,
  93, \dodoi{10.1111/j.1365-2966.2008.13712.x}

\bibitem[{{Krajnovi{\'c}} {et~al.}(2011){Krajnovi{\'c}}, {Emsellem},
  {Cappellari}, {Alatalo}, {Blitz}, {Bois}, {Bournaud}, {Bureau}, {Davies},
  {Davis}, {de Zeeuw}, {Khochfar}, {Kuntschner}, {Lablanche}, {McDermid},
  {Morganti}, {Naab}, {Oosterloo}, {Sarzi}, {Scott}, {Serra}, {Weijmans}, \&
  {Young}}]{Krajinovic2011}
{Krajnovi{\'c}}, D., {Emsellem}, E., {Cappellari}, M., {et~al.} 2011, \mnras,
  414, 2923, \dodoi{10.1111/j.1365-2966.2011.18560.x}

\bibitem[{{Lagos} {et~al.}(2020){Lagos}, {Emsellem}, {van de Sande},
  {Harborne}, {Cortese}, {Davison}, {Foster}, \& {Wright}}]{Lagos2020}
{Lagos}, C. d.~P., {Emsellem}, E., {van de Sande}, J., {et~al.} 2020, arXiv
  e-prints, arXiv:2012.08060.
\newblock \doarXiv{2012.08060}

\bibitem[{{Li} {et~al.}(2018){Li}, {Mao}, {Cappellari}, {Graham}, {Emsellem},
  \& {Long}}]{Li2018}
{Li}, H., {Mao}, S., {Cappellari}, M., {et~al.} 2018, \apjl, 863, L19,
  \dodoi{10.3847/2041-8213/aad54b}

\bibitem[{{Marinacci} {et~al.}(2018){Marinacci}, {Vogelsberger}, {Pakmor},
  {Torrey}, {Springel}, {Hernquist}, {Nelson}, {Weinberger}, {Pillepich},
  {Naiman}, \& {Genel}}]{TNG5}
{Marinacci}, F., {Vogelsberger}, M., {Pakmor}, R., {et~al.} 2018, \mnras, 480,
  5113, \dodoi{10.1093/mnras/sty2206}

\bibitem[{{Naab} {et~al.}(2014){Naab}, {Oser}, {Emsellem}, {Cappellari},
  {Krajnovi{\'c}}, {McDermid}, {Alatalo}, {Bayet}, {Blitz}, {Bois}, {Bournaud},
  {Bureau}, {Crocker}, {Davies}, {Davis}, {de Zeeuw}, {Duc}, {Hirschmann},
  {Johansson}, {Khochfar}, {Kuntschner}, {Morganti}, {Oosterloo}, {Sarzi},
  {Scott}, {Serra}, {van de Ven}, {Weijmans}, \& {Young}}]{Naab2014}
{Naab}, T., {Oser}, L., {Emsellem}, E., {et~al.} 2014, \mnras, 444, 3357,
  \dodoi{10.1093/mnras/stt1919}

\bibitem[{{Naiman} {et~al.}(2018){Naiman}, {Pillepich}, {Springel},
  {Ramirez-Ruiz}, {Torrey}, {Vogelsberger}, {Pakmor}, {Nelson}, {Marinacci},
  {Hernquist}, {Weinberger}, \& {Genel}}]{TNG3}
{Naiman}, J.~P., {Pillepich}, A., {Springel}, V., {et~al.} 2018, \mnras, 477,
  1206, \dodoi{10.1093/mnras/sty618}

\bibitem[{{Nelson} {et~al.}(2018){Nelson}, {Pillepich}, {Springel},
  {Weinberger}, {Hernquist}, {Pakmor}, {Genel}, {Torrey}, {Vogelsberger},
  {Kauffmann}, {Marinacci}, \& {Naiman}}]{TNG2}
{Nelson}, D., {Pillepich}, A., {Springel}, V., {et~al.} 2018, \mnras, 475, 624,
  \dodoi{10.1093/mnras/stx3040}

\bibitem[{{Nevin} {et~al.}(2021){Nevin}, {Blecha}, {Comerford}, {Greene},
  {Law}, {Stark}, {Westfall}, {Vazquez-Mata}, {Smethurst},
  {Argudo-Fern{\'a}ndez}, {Brownstein}, \& {Drory}}]{Nevin2021}
{Nevin}, R., {Blecha}, L., {Comerford}, J., {et~al.} 2021, \apj, 912, 45,
  \dodoi{10.3847/1538-4357/abe2a9}

\bibitem[{{Oser} {et~al.}(2010){Oser}, {Ostriker}, {Naab}, {Johansson}, \&
  {Burkert}}]{Oser2010}
{Oser}, L., {Ostriker}, J.~P., {Naab}, T., {Johansson}, P.~H., \& {Burkert}, A.
  2010, \apj, 725, 2312, \dodoi{10.1088/0004-637X/725/2/2312}

\bibitem[{{Penoyre} {et~al.}(2017){Penoyre}, {Moster}, {Sijacki}, \&
  {Genel}}]{Penoyre2017}
{Penoyre}, Z., {Moster}, B.~P., {Sijacki}, D., \& {Genel}, S. 2017, \mnras,
  468, 3883, \dodoi{10.1093/mnras/stx762}

\bibitem[{{Pillepich} {et~al.}(2018){Pillepich}, {Nelson}, {Hernquist},
  {Springel}, {Pakmor}, {Torrey}, {Weinberger}, {Genel}, {Naiman}, {Marinacci},
  \& {Vogelsberger}}]{TNG1}
{Pillepich}, A., {Nelson}, D., {Hernquist}, L., {et~al.} 2018, \mnras, 475,
  648, \dodoi{10.1093/mnras/stx3112}

\bibitem[{{Pop} {et~al.}(2018){Pop}, {Pillepich}, {Amorisco}, \&
  {Hernquist}}]{Pop2018}
{Pop}, A.-R., {Pillepich}, A., {Amorisco}, N.~C., \& {Hernquist}, L. 2018,
  \mnras, 480, 1715, \dodoi{10.1093/mnras/sty1932}

\bibitem[{{Pulsoni} {et~al.}(2020){Pulsoni}, {Gerhard}, {Arnaboldi},
  {Pillepich}, {Nelson}, {Hernquist}, \& {Springel}}]{Pulsoni2020}
{Pulsoni}, C., {Gerhard}, O., {Arnaboldi}, M., {et~al.} 2020, \aap, 641, A60,
  \dodoi{10.1051/0004-6361/202038253}

\bibitem[{{Rodriguez-Gomez} {et~al.}(2019){Rodriguez-Gomez}, {Snyder}, {Lotz},
  {Nelson}, {Pillepich}, {Springel}, {Genel}, {Weinberger}, {Tacchella},
  {Pakmor}, {Torrey}, {Marinacci}, {Vogelsberger}, {Hernquist}, \&
  {Thilker}}]{RodriguezGomez2019}
{Rodriguez-Gomez}, V., {Snyder}, G.~F., {Lotz}, J.~M., {et~al.} 2019, \mnras,
  483, 4140, \dodoi{10.1093/mnras/sty3345}

\bibitem[{{Sersic}(1968)}]{Sersic1968}
{Sersic}, J.~L. 1968, {Atlas de Galaxias Australes}

\bibitem[{{Springel}(2010)}]{Springel2010}
{Springel}, V. 2010, \mnras, 401, 791, \dodoi{10.1111/j.1365-2966.2009.15715.x}

\bibitem[{{Springel} {et~al.}(2018){Springel}, {Pakmor}, {Pillepich},
  {Weinberger}, {Nelson}, {Hernquist}, {Vogelsberger}, {Genel}, {Torrey},
  {Marinacci}, \& {Naiman}}]{TNG4}
{Springel}, V., {Pakmor}, R., {Pillepich}, A., {et~al.} 2018, \mnras, 475, 676,
  \dodoi{10.1093/mnras/stx3304}

\bibitem[{{Tsatsi} {et~al.}(2017){Tsatsi}, {Lyubenova}, {van de Ven}, {Chang},
  {Aguerri}, {Falc{\'o}n-Barroso}, \& {Macci{\`o}}}]{Tsatsi2017}
{Tsatsi}, A., {Lyubenova}, M., {van de Ven}, G., {et~al.} 2017, \aap, 606, A62,
  \dodoi{10.1051/0004-6361/201630218}

\bibitem[{{Vasiliev} {et~al.}(2021){Vasiliev}, {Belokurov}, \&
  {Evans}}]{Vasiliev2021}
{Vasiliev}, E., {Belokurov}, V., \& {Evans}, W. 2021, arXiv e-prints,
  arXiv:2108.00010.
\newblock \doarXiv{2108.00010}

\bibitem[{{Wilkinson} {et~al.}(2015){Wilkinson}, {Maraston}, {Thomas},
  {Coccato}, {Tojeiro}, {Cappellari}, {Belfiore}, {Bershady}, {Blanton},
  {Bundy}, {Cales}, {Cherinka}, {Drory}, {Emsellem}, {Fu}, {Law}, {Li},
  {Maiolino}, {Masters}, {Tremonti}, {Wake}, {Wang}, {Weijmans}, {Xiao}, {Yan},
  {Zhang}, {Bizyaev}, {Brinkmann}, {Kinemuchi}, {Malanushenko}, {Malanushenko},
  {Oravetz}, {Pan}, \& {Simmons}}]{Wilkinson2015}
{Wilkinson}, D.~M., {Maraston}, C., {Thomas}, D., {et~al.} 2015, \mnras, 449,
  328, \dodoi{10.1093/mnras/stv301}

\bibitem[{{Zeng} {et~al.}(2021){Zeng}, {Wang}, \& {Gao}}]{Zeng2021}
{Zeng}, G., {Wang}, L., \& {Gao}, L. 2021, \mnras, 507, 3301,
  \dodoi{10.1093/mnras/stab2294}

\end{thebibliography}

\end{document}